\documentclass[preprintnumbers,amsmath,nofootinbib,secnumarabic]{revtex4}
\usepackage{graphicx}
\usepackage{mathrsfs}
\usepackage{dsfont}
\usepackage{amssymb,amsmath,amsbsy}
\usepackage{dcolumn}
\usepackage{axodraw}
\usepackage{pstricks}
\usepackage{color}
\usepackage{amssymb,amsmath,amsbsy,bbold}
\usepackage{latexsym}

\numberwithin{equation}{section}
\newcommand\T{\rule{0pt}{2.6ex}}
\newcommand\Bot{\rule[-1.2ex]{0pt}{0pt}}
\def\lsim{\mathrel{\raise.3ex\hbox{$<$\kern-.75em\lower1ex\hbox{$\sim$}}}}
\def\gsim{\mathrel{\raise.3ex\hbox{$>$\kern-.75em\lower1ex\hbox{$\sim$}}}}

\def\anti{\overline}

\def\ur{U_R}
\def\dr{D_R}
\def\mud{M_U}
\def\mdd{M_D}
\def\abar{{\bar a}}
\def\bbar{{\bar b}}
\def\cbar{{\bar c}}
\def\dbar{{\bar d}}
\def\ebar{{\bar e}}
\def\fbar{{\bar f}}
\def\gbar{{\bar g}}
\def\hb{{\bar h}}
\def\qcal{\mathcal{Q}}
\def\mud{M_U}
\def\mdd{M_D}
\def\hl{{h^0}}
\def\hh{{H^0}}
\def\ha{A^0}

\def\mha{m_{\ha}}
\def\mhl{m_{\hl}}
\def\mhh{m_{\hh}}
\def\cbma{\cos(\beta-\alpha)}
\def\sbma{\sin(\beta-\alpha)}
\def\sb  {\sin\beta}
\def\cb  {\cos\beta}
\def\mc{m_{H^\pm}}

\def\phm{\phantom{-}}

\def\gc{\frac{g^2}{16\pi^2c_W^2}}
\def\hi{h_i \hspace{2mm} (i=1,2,3)}
\def\Ref#1{ref.~\cite{#1}}
\def\refs#1#2{refs.~\cite{#1} and \cite{#2}}

\def\eq#1{eq.~(\ref{#1})}
\def\eqs#1#2{eqs.~(\ref{#1}) and (\ref{#2})}

\def\eqst#1#2{eqs.~(\ref{#1})--(\ref{#2})}
\def\eqthree#1#2#3{eqs.~(\ref{#1}), (\ref{#2}) and (\ref{#3})}
\def\Eq#1{Eq.~(\ref{#1})}
\def\Eqst#1#2{Eqs.~(\ref{#1})--(\ref{#2})}
\def\Eqs#1#2{Eqs.~(\ref{#1}) and (\ref{#2})}
\def\Eqst#1#2{Eqs.~(\ref{#1})--(\ref{#2})}
\def\sec#1{Section~\ref{#1}}

\def\ifmath#1{\relax\ifmmode #1\else $#1$\fi}
\def\ls#1{\ifmath{_{\lower1.5pt\hbox{$\scriptstyle #1$}}}}
\def\lss#1{\ifmath{^{\,\lower2.5pt\hbox{$\scriptstyle #1$}}}}
\def\lsup#1{^{\lower 6pt\hbox{$\scriptstyle#1$}}}
\def\llsup#1{^{\lower 3pt\hbox{$\scriptstyle#1$}}}
\def\lasup#1{^{\lower 2pt\hbox{$\scriptstyle#1$}}}
\def\nicefrac#1#2{\hbox{$\frac{#1}{#2}$}}

\def\half{\tfrac{1}{2}}

\def\quarter{\tfrac{1}{4}}
\def\eighth{\tfrac{1}{8}}

\def\LL{{\Lambda^2}}
\def\zsix{Z_6 e^{-i \theta_{23}}}
\def\zfive{Z_5 e^{-2 i \theta_{23}}}
\def\zfivei{\Im(Z_5 \,e^{-2i\theta_{23}})v^2}

\def\vev#1{\langle #1 \rangle}
\def\lsup#1{^{\lower 6pt\hbox{$\scriptstyle#1$}}}
\def\phaa{\phantom{AA}}
\def\zsixr{\Re(Z_6 e^{-i \theta_{23}})}
\def\zsixi{\Im(Z_6 e^{-i \theta_{23}})}
\def\zfiver{\Re(Z_5 \,e^{-2i\theta_{23}})}
\def\ddel{\!\!\mathrel{\raise1.5ex\hbox{$\leftrightarrow$\kern-.85em
\lower1.7ex\hbox{$\partial$}}}}

\def\thet{e^{i \theta_{23}}}

\def\thetdoub{e^{-2 i \theta_{23}}}

\def\Tr{{\rm Tr}}
\def\gtwo{\frac{g^2}{16\pi^2}}
\def\iso{\mathchoice{\cong}{\cong}{\isoS}{\cong}}
\def\isoS{\vbox{\baselineskip 0pt  \lineskip 0.5pt
    \ialign{$ \mathsurround=0pt  \scriptstyle \hfil ## \hfil $\crcr
        \sim \crcr = \crcr}}}
\newcommand{\fourpoint}[4]{\fcolorbox{white}{white}{
  \begin{picture}(190,60) (78,-47)
    \SetWidth{0.5}
    \SetColor{Black}
    \DashCArc(105,-20)(18,117,477){4}
    \Photon(68,-42)(143,-42){4}{6.5}
    \Vertex(106,-38){1.6}
    \Text(69,-35)[lb]{\Black{$#1$}}
    \Text(130,-35)[lb]{\Black{$#2$}}
    \Text(100,0)[lb]{\Black{$#3$}}
 \Text(157,-20)[lb]{\Black{$#4$}}

\end{picture} }
}

\newcommand{\gaugeprop}[5]{\fcolorbox{white}{white}{

 \begin{picture}(194,50) (88,-2)
    \SetWidth{0.5}
    \SetColor{Black}
    \Photon(102,9)(138,9){4}{3}
    \DashCArc(123,12)(18,0,-180){4}
    \Photon(80,9)(102,9){4}{2.3}
    \Photon(138,9)(160,9){4}{2.3}

    \Text(80,16)[lb]{\Black{$#1$}}
    \Text(145,16)[lb]{\Black{$#2$}}
    \Text(115,-3)[lb]{\Black{$#3$}}
    \Text(115,30)[lb]{\Black{$#4$}}
 \Text(167,9)[lb]{\Black{$#5$}}

\end{picture} }}
\newcommand{\Gaugeprop}[5]{\fcolorbox{white}{white}{

 \begin{picture}(250,40) (86,-2)
    \SetWidth{0.5}
    \SetColor{Black}
    \Photon(102,9)(138,9){4}{3}
    \DashCArc(123,12)(18,0,-180){4}
    \Photon(80,9)(102,9){4}{2.3}
    \Photon(138,9)(160,9){4}{2.3}

    \Text(80,16)[lb]{\Black{$#1$}}
    \Text(148,16)[lb]{\Black{$#2$}}
    \Text(115,-3)[lb]{\Black{$#3$}}
    \Text(115,30)[lb]{\Black{$#4$}}
 \Text(167,9)[lb]{\Black{$#5$}}

\end{picture} }}

\newcommand{\loopgraph}[6]{\fcolorbox{white}{white}{
  \begin{picture}(200,53) (86,-12)
    \SetWidth{0.5}
    \SetColor{Black}
    \DashArrowArc(120,9)(18,0,-180){4}
    \DashArrowArc(120,9)(18,180,-0){4}
    \Photon(80,9)(102,9){-4}{2.3}
    \Photon(138,9)(160,9){4}{2.3}

    \Text(80,16)[lb]{\Black{$#1$}}
    \Text(148,16)[lb]{\Black{$#2$}}
    \Text(115,-6)[lb]{\Black{$#3$}}
    \Text(115,30)[lb]{\Black{$#4$}}
 \Text(170,9)[lb]{\Black{$#5$}}
 \Text(170,-9)[lb]{\Black{$#6$}}
\end{picture} }
}
\newcommand{\Loopgraph}[6]{\fcolorbox{white}{white}{
  \begin{picture}(250,58) (85,-12)
    \SetWidth{0.5}
    \SetColor{Black}
    \DashArrowArc(120,9)(18,0,-180){4}
    \DashArrowArc(120,9)(18,180,-0){4}
    \Photon(80,9)(102,9){-4}{2.3}
    \Photon(138,9)(160,9){4}{2.3}

    \Text(80,16)[lb]{\Black{$#1$}}
    \Text(148,16)[lb]{\Black{$#2$}}
    \Text(115,-6)[lb]{\Black{$#3$}}
    \Text(115,30)[lb]{\Black{$#4$}}
 \Text(170,9)[lb]{\Black{$#5$}}
 \Text(170,-9)[lb]{\Black{$#6$}}
\end{picture} }
}
\newcommand{\feynrule}[4]{\fcolorbox{white}{white}{
  \begin{picture}(180,48) (48,-42)

    \SetWidth{0.5}
    \SetColor{Black}
    \Photon(45,-20)(70,-20){3.5}{3}
    \DashArrowLine(90,-5)(70,-20){4}
    \DashArrowLine(70,-20)(90,-35){4}
  \Text(50,-12)[lb]{\Black{$#1^\mu$}}
 \Text(92,-7)[lb]{\Black{$#2$}}
 \Text(92,-42)[lb]{\Black{$#3$}}

 \Text(160,-25)[b]{$ #4 (p_{1} + p_{2})^\mu $}
  \end{picture}}
}
\newcommand{\feynruleVV}[4]{\fcolorbox{white}{white}{
  \begin{picture}(170,50) (31,-43)
    \SetWidth{0.5}
    \SetColor{Black}
    \Photon(40,-3)(70,-20){3}{3.5}
    \Photon(40,-37)(70,-20){-3}{3.5}
    \DashLine(100,-20)(70,-20){4}
  \Text(21,-5)[lb]{\Black{$#1^\mu$}}
  \Text(21,-45)[lb]{\Black{$#1^\mu$}}
 \Text(90,-18)[lb]{\Black{$#3$}}

 \Text(150,-25)[b]{$#4 \hspace{1mm}g^{\mu\nu} $}
  \end{picture}}
}

\newcommand{\feynrulefour}[5]{\fcolorbox{white}{white}{
  \begin{picture}(170,58) (42,-47)
    \SetWidth{0.5}
    \SetColor{Black}
    \Photon(40,-3)(70,-20){3}{3.5}
    \Photon(40,-37)(70,-20){-3}{3.5}
    \DashArrowLine(90,-3)(70,-20){4}
    \DashArrowLine(70,-20)(90,-37){4}
  \Text(35,0)[lb]{\Black{$#1^\mu$}}
  \Text(35,-50)[lb]{\Black{$#1^\mu$}}
 \Text(92,0)[lb]{\Black{$#2$}}
 \Text(92,-48)[lb]{\Black{$#3$}}

 \Text(160,-20)[b]{$#4 g^{\mu\nu} $}
 \Text(140,-39)[b]{$#5$}
  \end{picture}}
}
\def\beq{\begin{equation}}
\def\eeq{\end{equation}}
\newenvironment{Eqnarray}%
     {\arraycolsep 0.14em\begin{eqnarray}}{\end{eqnarray}}
\def\beqa{\begin{Eqnarray}}
\def\eeqa{\end{Eqnarray}}
\def\bea{\begin{Eqnarray*}}
\def\eea{\end{Eqnarray*}}
\renewcommand{\Re}{{\rm Re}}
\renewcommand{\Im}{{\rm Im}}
\begin{document}
\preprint{SCIPP 10/18}
\title{Basis-independent methods for the two-Higgs-doublet model III:
The CP-conserving limit,
custodial symmetry, and the oblique parameters $\boldsymbol{S}$,
$\boldsymbol{T}$, $\boldsymbol{U}$}
\author{Howard E. Haber}
\author{Deva O'Neil\footnote{Present address: Physics Department,
Bridgewater College, Bridgewater, VA 22812.}}
\affiliation{Santa Cruz Institute for Particle Physics  \\
University of California, Santa Cruz, CA 95064, U.S.A. \\
\vspace{1cm}}

\begin{abstract}

In the Standard Model, custodial symmetry is violated
by the hypercharge U(1) gauge interactions and the Yukawa couplings,
while being preserved by the Higgs scalar potential.  In the
two-Higgs doublet model
(2HDM), the generic scalar potential introduces new sources of
custodial symmetry breaking.  We obtain a basis-independent
expression for the constraints that impose custodial symmetry
on the 2HDM scalar potential.  These constraints impose
CP-conservation on the scalar potential and vacuum, and in addition
add one extra constraint on the scalar potential parameters.
We clarify the mass degeneracies of the 2HDM that arise as a
consequence of the custodial symmetry.  We also provide
a computation of the ``oblique'' parameters ($S$, $T$, and $U$)
for the most general CP-violating 2HDM
in the basis-independent formalism.  We demonstrate that the 2HDM
contributions to $T$ and $U$ vanish in the custodial symmetry
limit, as expected.  Using the experimental bounds on $S$ and $T$ from
precision electroweak data, we examine the resulting constraints
on the general 2HDM parameter space.
\end{abstract}

\maketitle

\section{Introduction: The CP-Violating Two Higgs Doublet Model (2HDM)}
\label{sec:intro}

In the most general two-Higgs-doublet extension of the Standard Model
(2HDM), the two hypercharge-one Higgs doublet fields $\Phi_1$ and
$\Phi_2$ are indistinguishable.
Consequently, all physical observables must be independent of
a change in the scalar basis, which corresponds to a redefinition of
the scalar doublets by a global U(2) transformation, $\Phi_a \rightarrow
U_{a\bar{b}}\Phi_b$.  In refs.~\cite{davidson} and \cite{haberoneil},
a basis-independent formalism for the 2HDM was introduced and
developed.\footnote{There is an alternative approach, which we do not employ 
in this paper, that emphasizes the role 
of gauge-invariant scalar field bilinears.  For further details, see 
e.g. refs.~\cite{Maniatis:2006fs,Nishi:2006tg, Ivanov:2006yq,Maniatis:2007vn}.
\label{bilinears}}
In particular, a basis-independent form for the
most general 2HDM interactions was obtained in ref.~\cite{haberoneil}.
A recap of the basis-independent formalism for the 2HDM is provided
in Section 2 in order to make this paper self-contained.

However, the most general form of the 2HDM is certainly not realized
in nature.  For example, for generic 2HDM parameters, one expects
large flavor-changing neutral currents and a significant violation of
custodial symmetry, in conflict with experimental observations.
These problems are ameliorated in restricted parameter regimes of the
2HDM.  These restricted regions are either fine-tuned or can be
implemented by imposing additional symmetries (e.g. discrete
symmetries or supersymmetry) on the 2HDM scalar potential.  Such
additional symmetries would in general distinguish between the
two-Higgs doublet fields, and thereby choose a preferred basis.
If 2HDM phenomena are observed in nature, one important goal of
experimental Higgs studies at future colliders will be to determine
the nature of the additional symmetry structures (if present) that
restrict the 2HDM parameters, and the associated preferred scalar basis.
However, prior to determining whether such a preferred scalar basis exists,
the basis-independent techniques will be critical for exploring the
phenomenological profile of the 2HDM and determining its
theoretical structure.

In this paper, we provide a basis-independent formulation of custodial
symmetry for the most general 2HDM.  If custodial symmetry were exact,
then there would be no Higgs sector corrections to the tree-level
relation $m_W^2=m_Z^2\cos^2\theta_W$ to all orders in perturbation
theory.  Of course, custodial symmetry is not an exact symmetry of the
Standard Model, as it is violated by the hypercharge gauge
interactions and the Higgs--fermion Yukawa interactions.  The precision
measurements of electroweak observables by LEP and the Tevatron
suggest that additional sources of custodial symmetry breaking beyond
that which is contained in the Standard Model must be small.  This
imposes interesting constraints on the most general 2HDM.

The custodial symmetric 2HDM scalar potential must be CP-conserving.
Thus, in Section 3 we first review the basis-independent conditions
for a CP-conserving 2HDM potential.  We then establish the
basis-independent conditions for a custodial symmetric 2HDM scalar
potential in Section~4.  These results clarify the significance of the
conditions for custodial symmetry in the 2HDM obtained previously in
the literature~\cite{pomarol}.  The effects of the custodial
symmetry-violating terms on the 2HDM have phenomenological
consequences.  In particular, these terms would lead to shifts in the
Peskin-Takeuchi $T$ and $U$ parameters~\cite{peskin}.  In contrast,
shifts in the $S$ parameter~\cite{peskin} can be generated even in the
presence of an exact custodial symmetry.  In Section 5, we have obtained
basis-independent expressions for the 2HDM contributions to the
oblique parameters $S$, $T$ and $U$.  Using these results, we present
in Section 6 a numerical study of the size of the 2HDM contributions
to the oblique parameters as a function of the 2HDM parameter space.
By comparing these results to the experimental bounds on $S$ and $T$,
we determine some of the features of the constraints on the 2HDM
parameter space.  Conclusions are given in Section 7.

Some details have been relegated to the appendices.  In Appendix A,
we reproduce the cubic and quartic bosonic couplings of the 2HDM
obtained in \Ref{haberoneil}.  These couplings are critical for
determining the CP-quantum numbers of the neutral Higgs states
for the CP-conserving scalar potential.
In Appendix B, we record some useful expressions involving
the neutral Higgs masses and invariant mixing angles.
In Appendix C we
summarize the basis-independent treatment of the CP-conserving 2HDM.
This appendix also examines a number of special cases in which some
of the neutral Higgs scalars are mass-degenerate. Appendix D provides
details of the computation of the 2HDM contributions to $S$, $T$ and $U$,
along with the relevant Feynman rules (in the basis-independent formalism)
and one-loop graphs.  Appendix E summarizes the key features of the
decoupling limit of the 2HDM.  Finally, in Appendix F, we examine the
tree-level unitarity bounds on the scalar potential parameters
in the basis-independent formalism.
These bounds are implemented in the parameter space scans of
Section 6.

\section{Recap of the basis-independent formalism for the 2HDM}

The scalar potential may be written in a basis-independent
form as~\cite{davidson,branco}
\beq
\label{genericpot} \mathcal{V}=Y_{a\bbar}\Phi_\abar^\dagger\Phi_b
+\half Z_{a\bbar c\dbar}(\Phi_\abar^\dagger\Phi_b)
(\Phi_\cbar^\dagger\Phi_d)\,,
\eeq
where $Z_{a\bbar c\dbar}=Z_{c\dbar a\bbar}$ and hermiticity implies
$Y_{a \bbar}= (Y_{b \abar})^\ast$ and
$Z_{a\bbar c\dbar}= (Z_{b\abar d\cbar})^\ast$.
The indices $a$, $b$, $c$ and
$d$ label the two Higgs doublets, and there is an implicit sum over
unbarred--barred index pairs.   The barred indices help keep track
of which indices transform with
$U$ and which transform with $U^\dagger$. For example, under a global
U(2) transformation, the parameters of the scalar potential change
according to
\beq
Y_{a\bbar}\to U_{a\cbar}Y_{c\dbar}U^\dagger_{d\bbar}\qquad {\rm and}
\qquad Z_{a\bbar c\dbar}\to U_{a\ebar}U^\dagger_{f\bbar}U_{c\gbar}
U^\dagger_{h\dbar} Z_{e\fbar g\hb}\,.
\eeq

The vacuum expectation values of the two Higgs fields
can be parametrized as
\beq \label{emvev} \langle
\Phi_a \rangle={\frac{v}{\sqrt{2}}} \left(
\begin{array}{c} 0\\ \widehat v_a \end{array}\right)\,,\qquad
{\rm with}\qquad
\widehat v_a \equiv e^{i\eta}\left(
\begin{array}{c} \cb\,\\ \sb\,e^{i\xi} \end{array}\right)
\,,
\eeq
where $v=246$~GeV and $\eta$ is an arbitrary phase.
The unit vector $\widehat v_a$ satisfies $ \widehat v_a \widehat
v^*_{\bar a} = 1$, where
$\widehat{v}^*_{\bar a}\equiv (\widehat{v}_a)^*$.
If we define the hermitian matrix
$V_{a\bbar}\equiv \hat v_a \hat v^*_{\bbar}$, then the
scalar potential minimum condition is given by the invariant
condition:
\beq \label{VY}
\Tr(VY)+\half v^2 Z_{a\bbar c\dbar} V_{b\abar}V_{d\cbar}=0\,.
\eeq
The orthonormal eigenvectors of $V_{a\bbar}$ are $\hat v_b$ and
\beq \label{wdef}
\widehat w_b \equiv \widehat
v_{\bar{a}}^\ast\epsilon_{ab}=e^{-i\eta}\left(
\begin{array}{c} -\sb\,e^{-i\xi}\\ \cb\end{array}\right)\,,\
\eeq
where $\epsilon_{12}=-\epsilon_{21}=1$ and $\epsilon_{11}=\epsilon_{22}=0$.
 Under the U(2) transformation,
$\widehat v_a\to U_{a\bbar}\hat v_b$, whereas
$$\widehat w_a\to ({\rm det}~U)^{-1}\,U_{a\bbar\,} \widehat w_b\,,
$$
where ${\rm det}~U\equiv e^{i\chi}$ is a pure phase.
That is, $\widehat w_a$ is a pseudo-vector with respect to
global U(2) transformations.
One can use $\widehat w_a$ and $\widehat w^*_{\abar}
\equiv (\widehat w_a)^*$
to construct a proper second-rank tensor,
$W_{a\bbar}\equiv \widehat w_a \widehat w^*_{\bbar}
\equiv\delta_{a\bbar}-V_{a\bbar}$.

One can always
define the so-called Higgs basis in which
only one of the two Higgs doublets has a neutral component with
a non-zero vacuum expectation value~\cite{branco,cpx2}.  The Higgs basis fields are
given by
\beq
H_1= \widehat{v}^*_{\bar{a}} \Phi_a\,,\qquad\qquad
H_2= \widehat{w}^*_{\bar{a}} \Phi_a\,.
\label{hconv}
\eeq
Since $\widehat v_a$ and $\widehat w_a$ are orthonormal vectors,
it follows that
\beq \label{higgsvevs}
\vev{H_1^0}=\frac{v}{\sqrt{2}}\,,\qquad\qquad \vev{H_2^0}=0\,.
\eeq

Note that $H_1$ is an invariant field whereas $H_2\to ({\rm det}U)H_2$
is a pseudo-invariant field under the global U(2) transformation.
The scalar potential can
then be expressed using the Higgs basis fields as follows:
\beqa \label{hbasispot}
\mathcal{V}&=& Y_1 H_1^\dagger H_1+ Y_2 H_2^\dagger H_2
+[Y_3 H_1^\dagger H_2+{\rm h.c.}]\nonumber \\[5pt]
&&\quad
+\half Z_1(H_1^\dagger H_1)^2 +\half Z_2(H_2^\dagger H_2)^2
+Z_3(H_1^\dagger H_1)(H_2^\dagger H_2)
+Z_4( H_1^\dagger H_2)(H_2^\dagger H_1) \nonumber \\[5pt]
&&\quad +\left\{\half Z_5 (H_1^\dagger H_2)^2
+\big[Z_6 (H_1^\dagger H_1)
+Z_7 (H_2^\dagger H_2)\big]
H_1^\dagger H_2+{\rm h.c.}\right\}\,,
\eeqa
where $Y_1$, $Y_2$ and $Z_{1,2,3,4}$  are real-valued U(2)-invariants,
\beqa
Y_1 &\equiv& \Tr(YV)\,,\qquad\qquad\qquad\,\,\,\, Y_2 \equiv \Tr(YW)\,,\\
Z_1 &\equiv& Z_{a\bbar c\dbar}\,V_{b\abar}V_{d\cbar}\,,\qquad\qquad\,\,\,\,
Z_2 \equiv Z_{a\bbar c\dbar}\,W_{b\abar}W_{d\cbar}\,,\qquad\qquad \\
Z_3 &\equiv& Z_{a\bbar
c\dbar}\,V_{b\abar}W_{d\cbar}\,,\qquad\qquad\,\,\, Z_4 \equiv
Z_{a\bbar c\dbar}\,V_{b\cbar}W_{d\abar}
\eeqa
and
$Y_3$ and $Z_{5,6,7}$ are complex ``pseudoinvariants,''
\beqa
\vspace{-0.3in}
\hspace{-0.6in} Y_3 &\equiv&
Y_{a\bbar}\,\widehat v_\abar^\ast\, \widehat w_b\,,\qquad\qquad\qquad
\,Z_5 \equiv Z_{a\bbar c\dbar}\,\widehat v_\abar^\ast\, \widehat w_b\,
\widehat v_\cbar^\ast\, \widehat w_d\,,\\
\hspace{-0.6in}
Z_6 &\equiv& Z_{a\bbar c\dbar}\,\widehat v_\abar^\ast\,\widehat v_b\,
\widehat v_\cbar^\ast\, \widehat w_d\,,\qquad\quad
Z_7 \equiv Z_{a\bbar c\dbar}\,\widehat v_\abar^\ast\, \widehat w_b\,
\widehat w_\cbar^\ast\,\widehat w_d\,.
\eeqa %
which transform as
\beq \label{tpseudo}
[Y_3, Z_6, Z_7]\to (\det U)^{-1}[Y_3, Z_6, Z_7]\qquad {\rm and} \qquad
Z_5\to  (\det U)^{-2} Z_5 \,.
\eeq
The scalar potential minimum condition [\eq{VY}] fixes
\beq\label{Y3}
Y_1=-\half Z_1 v^2\,,\qquad\qquad Y_3=-\half Z_6 v^2\,.
\eeq

The three physical neutral Higgs boson mass-eigenstates
can be determined by diagonalizing a $3\times 3$ squared-mass
matrix in the Higgs basis.  The diagonalizing matrix is a $3\times 3$
real orthogonal matrix that depends on three angles:
$\theta_{12}$, $\theta_{13}$ and $\theta_{23}$.
As shown in ref.~\cite{haberoneil},
under a U(2) transformation,
\beq
\theta_{12}\,,\, \theta_{13}~{\hbox{\text{are invariant}}}\quad {\rm and}\quad
e^{i\theta_{23}}\to (\det U)^{-1} e^{i\theta_{23}}\,.
\eeq
In particular, with respect to the \textit{invariant}
Higgs basis neutral fields
$\{{\rm Re}~\overline{H}_1\lsup{0}\,,\,{\rm Re}(e^{i\theta_{23}}H_2^0)\,,\,
{\rm Im}(e^{i\theta_{23}}H_2^0)\}$, where $\overline{H}_1\lsup{0}\equiv H_1^0-
(v/\sqrt{2})$,
the neutral Higgs squared-mass matrix is given by:
\beq \label{mtilmatrix}
\mathcal{M}^2=
v^2\left( \begin{array}{ccc}
Z_1&\,\, \Re(Z_6 \, e^{-i\theta_{23}}) &\,\, -\Im(Z_6 \, e^{-i\theta_{23}})\\
\Re(Z_6 e^{-i\theta_{23}}) &\,\, A^2/v^2+\Re(Z_5 \,e^{-2i\theta_{23}}) & \,\,
- \half \Im(Z_5 \,e^{-2i\theta_{23}})\\ -\Im(Z_6 \,e^{-i\theta_{23}})
 &\,\, - \half \Im(Z_5\, e^{-2i\theta_{23}}) &\,\, A^2/v^2\end{array}\right)\,,
\eeq
where
\beq
A^2\equiv Y_2+\half[Z_3+Z_4-\Re(Z_5
e^{-2i\theta_{23}})]v^2\,.
\eeq
Note that $\mathcal{M}^2$ is manifestly basis-independent, in which
case the neutral Higgs mass eigenstates are invariant fields with
respect to U(2) transformations.
Diagonalizing the neutral Higgs squared-mass matrix then gives
\beq \label{diagtil}
R\,\mathcal{M}^2\,R^T=\mathcal{M}^2_D={\rm
 diag}(m_1^2\,,\,m_2^2\,,\,m_3^2)\,,
\eeq
where  $m_1,$ $m_2$ and $m_3$ are the neutral Higgs boson masses and
\beq \label{rtil}
R=\left(\begin{array}{ccc}c_{12}c_{13} & \quad -s_{12} &
\quad -c_{12}s_{13} \\ s_{12}c_{13} & \quad \phm c_{12} & \quad-s_{12}s_{13}\\
s_{13} & \quad 0 & \quad c_{13}\end{array}\right)\,,
\eeq
where $c_{ij}\equiv\cos\theta_{ij}$ and $s_{ij}\equiv\sin\theta_{ij}$.
As shown in \Ref{haberoneil}, one can choose a convention
(without loss of generality) where
$-\half\pi\leq\theta_{12}\,,\,\theta_{13}<\half\pi$.
The neutral Goldstone boson is identified as $G^0\equiv{\rm Im}~H_1^0$.
One can express the mass eigenstate
neutral Higgs bosons, $h_k$ ($k=1,2,3$) and the neutral
Goldstone boson ($h_4\equiv G^0$)
directly in terms of the original shifted neutral
fields,  $\overline\Phi_a\lsup{0}\equiv \Phi_a^0-v\widehat
v_a/\sqrt{2}$,
\beq \label{hmassinv}
h_k=\frac{1}{\sqrt{2}}\left[\overline\Phi_{\bar{a}}\lsup{0\,\dagger}
(q_{k1} \widehat v_a+q_{k2}\widehat w_a e^{-i\theta_{23}})+(q^*_{k1}
\widehat v^*_{\bar{a}}+q^*_{k2}\widehat w^*_{\bar{a}}e^{i\theta_{23}})
\overline\Phi_a\lsup{0}\right]\,,
\eeq
where $q_{k\ell}$ are basis-independent quantities composed of the
invariant mixing angles $\theta_{12}$ and $\theta_{13}$
given in Table~\ref{tabq}.
 \begin{table}[ht]
\centering
\caption{The U(2)-invariant quantities $q_{k\ell}$, defined in
  ref.~\cite{haberoneil}, are reproduced below. The $q_{k\ell}$
are functions of the
the invariant mixing angles $\theta_{12}$ and $\theta_{13}$, where
$c_{ij}\equiv\cos\theta_{ij}$ and $s_{ij}\equiv\sin\theta_{ij}$.
By convention, we choose $-\half\pi\leq\theta_{12}\,,\,
\theta_{13}<\half\pi$.}
\label{tabq}
\vskip 0.1in
\begin{tabular}{|c||c|c|}\hline
$\phaa k\phaa $ &\phaa $q_{k1}\phaa $ & \phaa $q_{k2} \phaa $ \\ \hline
$1$ & $c_{12} c_{13}$ & $-s_{12}-ic_{12}s_{13}$ \\
$2$ & $s_{12} c_{13}$ & $c_{12}-is_{12}s_{13}$ \\
$3$ & $s_{13}$ & $ic_{13}$ \\
$4$ & $i$ & $0$ \\ \hline
\end{tabular}
\end{table}

The charged Goldstone and Higgs bosons are immediately identified
in terms of Higgs basis fields as: $G^\pm\equiv H_1^\pm$ and
$H^\pm=H_2^\pm$.  The latter implies that $H^\pm\to ({\rm det}U)^{\pm 1}H^\pm$
under the U(2) transformation.  If necessary, one can define an
invariant charged Higgs field, $e^{\pm i\theta_{23}}H^{\pm}$.
The charged Higgs mass is given by:
\beq m^2_{H^\pm} = Y_2 +\half Z_3 v^2, \label{mc}
\eeq
Finally,
inverting \eq{hmassinv} yields:
\beq
\Phi_a=\left(\begin{array}{c}G^+\widehat v_a+H^+ \widehat w_a\\[6pt]
\displaystyle \frac{v}{\sqrt{2}}\widehat
v_a+\frac{1}{\sqrt{2}}\sum_{k=1}^4 \left(q_{k1}\widehat
v_a+q_{k2}e^{-i\theta_{23}}\widehat w_a\right)h_k
\end{array}\right)\,.
 \eeq
Inserting this result into \eq{genericpot} immediately yields the
basis-independent form of the Higgs self-couplings given in
Appendix~A.  Likewise, the invariant forms of the
Higgs boson couplings to vector bosons can be obtained by expanding
out the covariant derivatives that appear in the Higgs kinetic energy
terms; these couplings are also given in Appendix~A.

The Higgs boson couplings to the
fermions arise from the Yukawa Lagrangian, which can
be written in terms of the quark mass-eigenstate fields
as:\footnote{\Eq{yuklag} corrects an error in eq.~(75) of
ref.~\cite{haberoneil}.}
\beq \label{yuklag}
-\mathscr{L}_{\rm Y}=\anti U_L \Phi_{\abar}^{0\,*}{{\eta^U_a}} \ur -\anti
D_L K^\dagger\Phi_{\abar}^- {{\eta^U_a}}\ur
+\anti U_L K\Phi_a^+{{\eta^{D\,\dagger}_{\abar}}} \dr
+\anti D_L\Phi_a^0 {{\eta^{D\,\dagger}_{\abar}}}\dr+{\rm h.c.}\,,
\eeq
where
$K$ is the CKM mixing matrix.  The $\eta^{U,D}$
are $3\times 3$ Yukawa coupling matrices.
We can construct invariant and pseudo-invariant matrix Yukawa couplings:
\beq \label{kapparho}
\kappa^{Q}\equiv \widehat v^*_\abar\eta^{Q}_a\,,\qquad\qquad
\rho^{Q}\equiv \widehat w^*_\abar\eta^{Q}_a\,,
\eeq
where $Q=U$ or $D$.  Inverting these equations yields
$\eta^Q_a=\kappa^Q\widehat
v_a+\rho^Q\widehat w_a$.
One can rewrite \eq{yuklag} in the Higgs basis,\footnote{\Eq{yukhbasis}
corrects an error in eq.~(76) of ref.~\cite{haberoneil}.}
\beqa \label{yukhbasis}
-\mathscr{L}_{\rm Y}&=&\anti U_L (\kappa^U H_1^{0\,\dagger}
+\rho^U H_2^{0\,\dagger})\ur
-\anti D_L K^\dagger(\kappa^U H_1^{-}+\rho^U H_2^{-})\ur \nonumber \\
&& +\anti U_L K (\kappa^{D\,\dagger}H_1^++\rho^{D\,\dagger}H_2^+)\dr
+\anti D_L (\kappa^{D\,\dagger}H_1^0+\rho^{D\,\dagger}H_2^0)\dr+{\rm h.c.}
\eeqa
Note that under the U(2) transformation,
\beq \label{rhotrans}
\kappa^Q~~\hbox{is invariant and}~~\rho^Q\to (\det U)\rho^Q\,.
\eeq

By construction, $\kappa^U$ and $\kappa^D$ are proportional to the
(real non-negative) diagonal quark mass matrices $M_U$ and $M_D$,
respectively.  In particular,
\beq \label{MQ}
M_U=\frac{v}{\sqrt{2}}\kappa^U={\rm diag}(m_u\,,\,m_c\,,\,m_t)\,,\qquad
M_D=\frac{v}{\sqrt{2}}\kappa^{D\,\dagger}={\rm
diag}(m_d\,,\,m_s\,,\,m_b) \,.
\eeq
The matrices $\rho^U$ and $\rho^D$ are independent complex
$3\times 3$ matrices.
The final form for the Yukawa couplings of the mass-eigenstate Higgs
bosons and the Goldstone bosons to the quarks is:
\beqa
 && \hspace{-0.5in} -\mathscr{L}_Y = \frac{1}{v}\overline D
\biggl\{M_D (q_{k1} P_R + q^*_{k1} P_L)+\frac{v}{\sqrt{2}}
\left[q_{k2}\,[e^{i\theta_{23}}\rho^D]^\dagger P_R+
q^*_{k2}\,e^{i\theta_{23}}\rho^D P_L\right]\biggr\}Dh_k \nonumber \\
&&\quad  \hspace{-0.2in} +\frac{1}{v}\overline U \biggl\{M_U (q_{k1}
P_L + q^*_{k1} P_R)+\frac{v}{\sqrt{2}}
\left[q^*_{k2}\,e^{i\theta_{23}}\rho^U P_R+
q_{k2}\,[e^{i\theta_{23}}\rho^U]^\dagger P_L\right]\biggr\}U h_k
\nonumber \\
&&\quad \hspace{-0.3in} +\biggl\{\overline U\left[K[\rho^D]^\dagger
P_R-[\rho^U]^\dagger KP_L\right] DH^+ +\frac{\sqrt{2}}{v}\,\overline
U\left[K\mdd P_R-\mud KP_L\right] DG^+ +{\rm
h.c.}\biggr\}\,.\label{Yukawas}
\eeqa
By writing $[\rho^Q]^\dagger H^+=[\rho^Q
e^{i\theta_{23}}]^\dagger[e^{i\theta_{23}}H^+]$, we see that the
Higgs-fermion Yukawa couplings depend only on invariant quantities:
the diagonal quark mass matrices, $\rho^Q e^{i\theta_{23}},$ and the
invariant angles $\theta_{12}$ and $\theta_{13}$.
Since $\rho^Q e^{i\theta_{23}}$ is in general a complex matrix,
\eq{Yukawas} contains CP-violating neutral-Higgs--fermion interactions.
Moreover, \eq{Yukawas}
exhibits Higgs-mediated flavor-changing neutral currents
(FCNCs) at tree-level in cases where the $\rho^Q$ are
not flavor-diagonal. Thus, for a phenomenologically acceptable
theory, the off-diagonal elements of $\rho^Q$ must be small.

Note that the parameter $\tan\beta$ [where the angle $\beta$ is defined
in \eq{emvev}] does not appear in any of the Higgs couplings
[cf.~Appendix A and \eq{Yukawas}].  This is to be expected,
since $\tan\beta$ is a basis-dependent
quantity in the general 2HDM and is therefore
an unphysical parameter~\cite{haberoneil}.
Of course, $\tan\beta$ can be promoted to a physical
parameter in special situations in which a particular basis
is physical (e.g., in the presence of a discrete symmetry
or supersymmetry, which restricts the form of the scalar potential
in a particular basis).  In this paper, we do not assume that any
basis (apart from the Higgs basis and the neutral scalar mass-eigenstate
basis) has physical significance.

\section{Basis-independent conditions for CP-Conservation}


At present, all known CP-violating effects can be attributed to a
phase in the CKM matrix $K$.  The source of this CP-violation is an
unremovable complex phase in the Higgs--fermion Yukawa couplings
of the Standard Model.  When
we extend the Standard Model by adding a second Higgs doublet, new
sources of CP-violation can arise from potentially complex
Higgs self-couplings and new Higgs--fermion Yukawa couplings.  In this
section, we determine the basis-independent conditions that yield no
new sources of CP-violation (at tree-level) beyond the one non-trivial
phase of the CKM matrix, and explore some of its consequences.

The Higgs scalar potential is explicitly CP-conserving if there exists
a basis, called the \textit{real basis}, in
which all scalar potential parameters are simultaneously real~\cite{cpbasis}.
In addition, if there exists  a real basis in which the Higgs
vacuum expectation values are simultaneously real, then
CP is also preserved by the vacuum (and is not spontaneously broken).
In the latter case, it is then possible to
perform an O(2) global transformation on the fields of the Higgs basis, which
maintains the reality of the
scalar potential parameters.  Hence, the condition
for a CP-conserving Higgs potential and vacuum is the existence of
a \textit{real} Higgs basis.  The only surviving basis freedom in
defining the Higgs basis is the rephasing of $H_2$.  Thus, it follows
from \eq{hbasispot} that the Higgs scalar potential and vacuum are
CP-conserving if and only if
\footnote{No separate
condition is required for the complex parameter
$Y_3$ due to the potential minimum condition
of \eq{Y3}.}
\beq \label{z56}
\Im(Z_5^* Z_6^2)=\Im(Z_5^* Z_7^2)=\Im(Z_6^* Z_7)=0\,,
\eeq
which are equivalent to conditions first established in
ref.~\cite{cpx2}, and subsequently rederived in refs.~\cite{davidson}
and~\cite{cpbasis}.

We now add in the Higgs-fermion interactions and impose the
requirement of CP-conserving neutral Higgs boson-fermion interactions.
This requirement is satisfied if the coefficients of the neutral Higgs
boson-fermion interactions are simultaneously real in a real Higgs basis.
It then follows from \eq{yukhbasis} that\footnote{\Eq{cprho} corrects an error
in eq.~(D3) of ref.~\cite{haberoneil}, which incorrectly stated
that the matrices of \eq{cprho} must be hermitian.
To derive this result, consider the interaction Lagrangian,
$$
\mathscr{L}_{\rm int}=A_{ij}\overline Q_i P_L Q_j+{\rm h.c.},
$$
and note that $(A_{ij}\overline Q_i  P_L Q_j)^\dagger=A^*_{ij}\overline{Q}_j P_R Q_i=
(A^\dagger)_{ij} \overline{Q}_i P_L Q_j$.
Under a CP transformation,
$$
U_{\rm CP} (A_{ij}\overline Q_i P_L Q_j)U^{-1}_{\rm CP}=A_{ij} \overline{Q}_j P_R Q_i
=(A^T)_{ij}\overline{Q}_i P_R Q_j\,.
$$
Imposing CP-invariance of the interaction Lagrangian yields $A^\dagger=A^T$; i.e., $A$ is
a real matrix.}
\beq \label{cprho}
Z_5(\rho^Q)^2\,,\,Z_6\,\rho^Q\,,\,{\rm and}\,\,Z_7\,\rho^Q\,\,
\hbox{\rm are real matrices}\quad (Q=U,D\,\,{\rm and}\,\, E)\,.
\eeq
Note that if \eq{cprho} is satisfied then $Z_5^{1/2}\rho^Q$ is either a purely real
or a purely imaginary matrix.  In particular, given a basis in which $Z_5$ is real
and $Z_6$, $Z_7$ and the matrix $\rho^Q$ are purely imaginary, one can always transform
to a real Higgs basis via $H_2\to iH_2$.

It is instructive to provide the explicit basis-independent form of the CP
transformation law.  In the Higgs basis, it is convenient to
employ the invariant Higgs fields, $H_1$ and $e^{i\theta_{23}}H_2$.
Then, under a CP transformation,
\beq \label{UCPhiggs}
U_{\rm CP}H_1(\boldsymbol{\vec{x}},t)  U^{-1}_{\rm CP}
=H_1^\dagger(-\boldsymbol{\vec{x}},t) \,,\qquad \qquad
U_{\rm CP}[\eta^* e^{i\theta_{23}}H_2(\boldsymbol{\vec{x}},t)] U^{-1}_{\rm CP}
= [\eta^* e^{i\theta_{23}}H_2(-\boldsymbol{\vec{x}},t)]^\dagger
\,,
\eeq
where $U_{\rm CP}$ is a unitary operator acting on the Hilbert space
of fields, and $\eta$ is a basis-independent
complex phase factor to be determined.  Applying this transformation
to the Higgs scalar potential in the Higgs basis [\eq{hbasispot}], it
follows that the Higgs scalar potential and vacuum is CP-invariant,
i.e.~$U_{\rm CP}\,\mathcal{V}\,U^{-1}_{\rm CP}=\mathcal{V}$ and
$U_{\rm CP}|\,0\,\rangle=|\,0\,\rangle$, if
\beq \label{etaZ}
\Im(\eta^2 Z_5 e^{-2i\theta_{23}})=0\,,\,\qquad\quad
\Im(\eta Z_6 e^{-i\theta_{23}})=0\,,\qquad\quad
\Im(\eta Z_7 e^{-i\theta_{23}})=0\,.
\eeq
These results immediately yield the conditions of \eq{z56}.
Likewise, if we demand that the neutral Higgs-fermion Yukawa interaction
is CP-invariant, it follows that
\beq \label{etarho}
\Im(\eta^*\rho^Q e^{i\theta_{23}})=0\,,\qquad\quad (Q=U,D~{\rm and}~E)\,.
\eeq
Combining eqs.~(\ref{etaZ}) and (\ref{etarho}), we obtain the conditions
of \eq{cprho}.

In a generic basis, the CP transformation law is easily obtained by
applying a global U(2) transformation to the Higgs basis fields in
\eq{UCPhiggs}.  Using \cite{haberoneil}
\beq \left(
\begin{array}{c}H_1 \\ H_2 \end{array}\right) = \left(
\begin{array}{cc} \phm\widehat{w}_2 & -\widehat{w}_1\\ - \widehat{v}_2
& \phm\widehat{v}_1 \end{array}\right) \left( \begin{array}{c}\Phi_1
\\ \Phi_2 \end{array}\right).
\label{eqnten}
\eeq
it follows that
\beq \label{CPgen}
\Phi_a (\boldsymbol{\vec{x}},t) \rightarrow (\widehat{v}_a \widehat{v}_b
+\eta^{2}\thetdoub \widehat{w}_a \widehat{w}_b)
\Phi_\bbar^*(-\boldsymbol{\vec{x}},t)\,.
\eeq
One can easily check that the invariance
of the scalar potential in
the generic basis [\eq{genericpot}]
with respect to the transformation law of \eq{CPgen}
again yields
\eq{z56}, as expected.  Note that the matrix
\beq \label{Uab}
\mathcal{U}_{ab}\equiv\widehat{v}_a \widehat{v}_b
+\eta^{2}\thetdoub \widehat{w}_a \widehat{w}_b\,,
\eeq
is unitary and symmetric.  Thus, the CP-transformation law in the
generic basis takes the general form (cf. ref.~\cite{cpbasis}):
\beq \label{cptrans}
U_{\rm CP}\Phi_a (\boldsymbol{\vec{x}},t)U^{-1}_{\rm CP}=\mathcal{U}_{ab}
\Phi_\bbar^*(-\boldsymbol{\vec{x}},t)\,,
\eeq
and  invariance of the vacuum under CP requires \cite{branco}:
\beq  \langle\Phi_a\rangle= \mathcal{U}_{ab} \langle\Phi_\bbar\rangle^*\,,
\label{vac}
\eeq
where $\mathcal{U}$ is any symmetric unitary $2\times 2$ matrix.
Indeed, \eq{Uab} satisfies the above conditions.

If \eqs{z56}{cprho} are satisfied, then the neutral Higgs boson
tree-level interactions are CP-conserving, and the neutral Higgs
fields are eigenstates of CP.  We follow the standard
notation~\cite{hhg} and denote the CP-odd Higgs field by $A^0$ and the
lighter and heavier CP-even neutral Higgs fields by $h^0$ and $H^0$,
respectively.

The neutral Higgs mass eigenstates determine the mixing angles
$\theta_{ij}$.  Thus, in the CP-conserving case, the
requirement\footnote{In the case of non-degenerate neutral Higgs boson
masses, it is automatic that the neutral Higgs mass eigenstates are
simultaneously CP-eigenstates.  In the case where the CP-odd Higgs
boson is mass-degenerate with a CP-even Higgs boson, it is always
convenient (though not strictly necessary)
to choose the physical mass-degenerate states to be
CP-eigenstates.}
that the neutral Higgs bosons are CP-eigenstates determines the
phase factor $\eta$ that appears in eqs.~(\ref{UCPhiggs}),
(\ref{etaZ}), (\ref{CPgen}), (\ref{Uab}) and (\ref{cpodd}).  By examining the
Higgs interaction terms given in Appendix~\ref{app:couplings},
one can determine a consistent set of
assignments for the CP quantum numbers of the neutral Higgs bosons
such that their interactions with gauge bosons and Higgs bosons is
CP-invariant.  For example, the CP-odd Higgs boson can be identified in general as\footnote{In
the case of $Z_6=Z_7=\rho^Q=0$, one of the three neutral Higgs bosons
is CP-even and the the other two neutral Higgs bosons have opposite CP quantum numbers.
But for this special case, one cannot determine which of these latter two scalars
is CP-odd.  See Section~\ref{z6700} for further details.}
\beq
\label{cpodd}
A^0=\Im(\eta^*\thet H_2^0)\,.
\eeq
In Sections~\ref{subsecz6}--\ref{sec:herquetcpcons}, we have examined all possible cases for
the Higgs scalar parameters in which the scalar potential and vacuum is CP-conserving,
and for each case the value of the phase factor $\eta^2$ is determined.
For simplicity, we assume that
the three neutral Higgs masses are non-degenerate.  The
mass-degenerate cases are treated in Appendix~\ref{app:howie}.

\subsection{The CP-conserving 2HDM with $\boldsymbol{Z_6\neq 0}$}
\label{subsecz6}

For $Z_6 \neq 0$ (and no restrictions on the possible values of
$Z_5$ or $Z_7$), a
CP-invariant Higgs potential can arise in the 2HDM under one of the
three cases listed in Table~\ref{Z6cond}.  The derivation of
these results (given in \Ref{haberoneil})
is reviewed in Appendix~\ref{app:howie}.
Note that \eqs{etaZ}{etarho} correlate the overall phases of
$Z_6$, $Z_7$ and the $\rho^Q$.  In particular, in Case~I,
$\Im(Z_6 e^{-i\theta_{23}})=\Im(Z_7 e^{-i\theta_{23}})
=\Im(\rho^Q e^{i\theta_{23}})=0$, whereas
$\Re(Z_6 e^{-i\theta_{23}})=\Re(Z_7 e^{-i\theta_{23}})
=\Re(\rho^Q e^{i\theta_{23}})=0$ in Cases IIa and b.

The U(2)-invariant quantities $q_{k\ell}$ for each of the three cases
shown in Table~\ref{Z6cond} are exhibited in Tables \ref{c1}, \ref{c2}
and \ref{c3}.
\begin{table}[ht!]
\centering
\caption{Basis-independent conditions
for a CP-conserving scalar potential and vacuum
when $Z_6\neq 0$.
The neutral Higgs mixing angles $\theta_{ij}$ are defined with respect to
the mass-ordering $m_{h_1}\leq m_{h_2}\leq m_{h_3}$.
The phase factor $\eta^2$ governs the CP transformation law
[cf.~\eq{CPgen}].  Additional conditions in which $Z_6$ is replaced
by $Z_7$ and by $\rho^{Q\,*}$ ($Q=U,D$ or $E$),
respectively, must also hold due to
the phase correlations implicit in \eqs{etaZ}{etarho}.  In the case
where two of the neutral Higgs masses are equal, one linear
combination of neutral Higgs states
will be CP-even and the orthogonal linear combination
will be CP-odd.  The latter defines the relevant mixing angle,
$\theta_{12}$ in Case I and $\theta_{13}$ in Case~II, respectively.
\label{Z6cond}}
\vskip 0.2in
\begin{tabular}{|c||c|c|c|c|c|}\hline
\phaa Cases\phaa & \phaa conditions \phaa & \phaa $\eta^2$ \phaa
& \phaa $A^0$ \phaa & \phaa $h^0$ \phaa & \phaa $H^0$ \phaa \\ \hline
I & \phaa $s_{13}=\Im(Z_5\,e^{-2i\theta_{23}})
=\Im(Z_6\,e^{-i\theta_{23}})=0$\phaa
& $+1$ & $h_3$ & $h_1$ & $h_2$ \\
IIa & $\phaa s_{12}=\Im(Z_5\,e^{-2i\theta_{23}})=\Re(Z_6\,e^{-i\theta_{23}})=0$
\phaa  & $-1$ & $h_2$ & $h_1$ & $h_3$ \\
IIb & $\phaa c_{12}=\Im(Z_5\,e^{-2i\theta_{23}})=\Re(Z_6\,e^{-i\theta_{23}})=0$
\phaa & $-1$ & $h_1$ & $h_2$ & $h_3$ \\ \hline
\end{tabular}
\end{table}

\begin{table}[ht!]
\begin{minipage}[t]{2.0in}
\centering
\parbox[t]{1.75in}{\caption{The U(2)-invariant quantities $q_{k\ell}$
for Case I}\label{c1}} \\ \vspace{0.1in}
\begin{tabular}{|c||c|c|}\hline
$\phaa k\phaa $ &\phaa $q_{k1}\phaa $ & \phaa $q_{k2} \phaa $ \\ \hline
$1$ & $c_{12}$ & $-s_{12}$ \\
$2$ & $s_{12}$ & $\phm c_{12}$ \\
$3$ & $0$ & $\phm i$ \\ \hline
\end{tabular}
\end{minipage}
\hfill
\begin{minipage}[t]{2.0in}
\centering
\parbox[t]{1.75in}{\caption{The U(2)-invariant quantities $q_{k\ell}$
for Case IIa}\label{c2}} \\ \vspace{0.1in}
\begin{tabular}{|c||c|c|}\hline
$\phaa k\phaa $ &\phaa $q_{k1}\phaa $ & \phaa $q_{k2} \phaa $ \\ \hline
$1$ & $\phm 0$ & $1$ \\
$2$ & $-c_{13}$ & $is_{13}$ \\
$3$ & $\phm s_{13}$ & $ic_{13}$ \\ \hline
\end{tabular}
\end{minipage}
\hfill
\begin{minipage}[t]{2.0in}
\centering
\parbox[t]{1.75in}{\caption{The U(2)-invariant quantities $q_{k\ell}$
for Case IIb}\label{c3}} \\ \vspace{0.1in}
\begin{tabular}{|c||c|c|}\hline
$\phaa k\phaa $ &\phaa $q_{k1}\phaa $ & \phaa $q_{k2} \phaa $ \\ \hline
$1$ & $c_{13}$ & $-is_{13}$ \\
$2$ & $0$ & $1$ \\
$3$ & $s_{13}$ & $\phm ic_{13}$ \\ \hline
\end{tabular}
\end{minipage}
\end{table}

It is convenient to define an invariant quantity,
$\varepsilon_{56}$, by the relation
\beq  \label{zz56}
\Re(Z_5^* Z_6^2)=\varepsilon_{56} |Z_5|\,|Z_6|^2\,,\qquad
\varepsilon_{56}\equiv\pm 1\,.
\eeq
Since ${\rm Im}~(Z_5 e^{-2i\theta_{23}})=0$ is satisfied in
Cases I and II, it follows that
\beq \label{zz62}
\Re(Z_5^* Z_6^2)={\rm Re}(Z_5 e^{-2i\theta_{23}})
\left[{\rm Re}(Z_6 e^{-i\theta_{23}})^2-{\rm Im}(Z_6
  e^{-i\theta_{23}})^2\right]=\pm|Z_6|^2 {\rm Re}(Z_5
  e^{-2i\theta_{23}})\,,
\eeq
where we take the positive [negative] sign depending on whether
$\Im(Z_6\,e^{-i\theta_{23}})=0$ [$\Re(Z_6\,e^{-i\theta_{23}})=0$].
Hence, \eqs{zz56}{zz62} yield
\beq \label{rez56}
{\rm Re}(Z_5 e^{-2i\theta_{23}})=\begin{cases} \phm\epsilon_{56}|Z_5|\,,&
\quad \text{if}~~\Im(Z_6\,e^{-i\theta_{23}})=0\,, \\
-\epsilon_{56}|Z_5|\,,&
\quad \text{if}~~\Re(Z_6\,e^{-i\theta_{23}})=0\,.\end{cases}
\eeq
Note that $\varepsilon_{56}$
is the sign of $Z_5$ in the real basis.\footnote{In the real Higgs basis as defined above, $\theta_{23}=n\pi$ for
integer $n$.  Since ${\rm Im}(Z_5 e^{-2i\theta_{23}})=0$, it follows from \eq{rez56}
that $Z_5=\epsilon_{56}|Z_5|$.  That is, $\varepsilon_{56}$ is the sign of $Z_5$ in the real Higgs basis.}
\Eq{rez56} can be rewritten more compactly as:
\beq \label{rez56compact}
{\rm Re}(Z_5 e^{-2i\theta_{23}})=\eta^2\epsilon_{56}|Z_5|\,.
\eeq

One can use \eq{cpodd} to identify the CP-odd Higgs boson, $A^0$.
The identity of $A^0$ is also easily discerned from
Tables~\ref{c1}, \ref{c2} and \ref{c3}, since any neutral Higgs state $h_k$
with $q_{k1}\neq 0$ must be CP-even.  As there is one CP-odd state
in the neutral Higgs spectrum, it must correspond to the $q_{k1}=0$
entries of Tables~\ref{c1}, \ref{c2} and~\ref{c3}.

The squared-masses of the neutral bosons are given by:
\beqa
m^2_{h^0,H^0}&=&\half v^2\left[Y_2/v^2+Z_1+\half(Z_3+Z_4+\varepsilon_{56}|Z_5|)
\mp\sqrt{\left[Y_2/v^2-Z_1+\half(Z_3+Z_4+\varepsilon_{56}|Z_5|)\right]^2
+4|Z_6|^2}\right],\nonumber \\
&&\phantom{line} \label{m2hH}\\
m_{A^0}^2&=&Y_2+\half v^2(Z_3+Z_4-\varepsilon_{56}|Z_5)|)\,,\label{m2A}
\eeqa
where $\epsilon_{56}$ is defined above [cf.~\eqs{zz56}{rez56}].
In particular, Case I corresponds to the mass ordering $m_{A^0} > m_{H^0}$,
and Cases IIa and IIb correspond to $m_{A^0} < m_{H^0}$.
Moreover, the two separate parameter regimes corresponding to
Cases IIa and IIb correspond to the two possible mass orderings
$m_{A^0}<m_{h^0}$ and $m_{A^0}>m_{h^0}$, respectively, as exhibited in
Table~\ref{Z6cond}.

\subsection{The CP-conserving 2HDM with $\boldsymbol{Z_6=0}$
and $\boldsymbol{Z_7\neq 0}$}
\label{z6700}

For the case of $Z_6 = 0$ and $Z_7\neq 0$, a
CP-invariant Higgs potential can arise in the 2HDM under any one of the
following six conditions listed in Table~\ref{Z60cond}.
The U(2)-invariant quantities $q_{k\ell}$ for the cases
shown in Table~\ref{Z60cond} are exhibited in Tables \ref{d1},
\ref{d2} and \ref{d3}.  A derivation of these results is given
in Appendix~\ref{app:howie}.

\begin{table}[ht!]
\centering
\caption{Basis-independent conditions
for a CP-conserving scalar potential and vacuum
when $Z_6=0$, $Z_7\neq 0$.
The neutral Higgs mixing angles $\theta_{ij}$ are defined with
respect to
the mass-ordering $m_{h_1}\leq m_{h_2}\leq m_{h_3}$, and
$\overline{\theta}_{23}\equiv\theta_{23}-\theta_{12}$.
The phase factor $\eta^2$ governs the CP transformation law
[cf.~\eq{CPgen}].  Additional conditions in which $Z_7$ is replaced
by $\rho^{Q\,*}$ ($Q=U,D$ and $E$), respectively, must also hold due to
the phase correlations implicit in \eqs{correlation1}{correlation2}.
The two CP-even states are denoted as $h_1^0$ and
$h_2^0$, where $m_{h_1^0}^2=Z_1 v^2$ and
$m_{h_2^0}^2=Y_2+\half(Z_3+Z_4+\varepsilon_{57}|Z_5|)v^2$.
The couplings of $h_1^0$ coincide with those of the Standard Model
Higgs boson.
The squared-mass of the CP-odd Higgs boson is given by:
$\mha^2 =Y_2+\half(Z_3+Z_4-\varepsilon_{57}|Z_5|)v^2$.
If $Z_1$ is chosen such that $h_1^0$ is degenerate
in mass with either $h_2^0$ or $A^0$, then $\theta_{12}$
in Cases I$^\prime$ and II$^\prime$ or $\theta_{13}$ in Case
III$^\prime$ are fixed by the requirement that the properties of
the mass-degenerate state $h_1^0$ coincide with
those of the Standard Model Higgs boson.
Additional cases not included in this table that can arise when two
of the neutral Higgs bosons are degenerate in mass are treated
in Table~\ref{Z60cond2}.
\label{Z60cond}}
\vskip 0.2in
\begin{tabular}{|c||c|c|c|c|c|}\hline
\phaa Cases\phaa & \phaa conditions \phaa & \phaa $\eta^2$ \phaa
& \phaa $A^0$  \phaa & \phaa $h_1^0$ \phaa &
 \phaa $h_2^0$ \phaa  \\ \hline
I$^\prime$a & \phaa $s_{13}=s_{12}=\Im(Z_5\,e^{-2i\theta_{23}})
=\Im(Z_7\,e^{-i\theta_{23}})=0$\phaa
& $+1$ & $h_3$ & $h_1$ & $h_2$ \\
I$^\prime$b & \phaa $s_{13}=s_{12}=\Im(Z_5\,e^{-2i\theta_{23}})
=\Re(Z_7\,e^{-i\theta_{23}})=0$\phaa
& $-1$ & $h_2$ & $h_1$ & $h_3$ \\
II$^\prime$a & \phaa $s_{13}=c_{12}=\Im(Z_5\,e^{-2i\theta_{23}})
=\Im(Z_7\,e^{-i\theta_{23}})=0$\phaa
& $+1$ & $h_3$ & $h_2$ & $h_1$ \\
II$^\prime$b & \phaa $s_{13}=c_{12}=\Im(Z_5\,e^{-2i\theta_{23}})
=\Re(Z_7\,e^{-i\theta_{23}})=0$\phaa
& $-1$ & $h_1$& $h_2$ & $h_3$ \\
III$^\prime$a & $\phaa c_{13}=\Im(Z_5\,e^{-2i\overline\theta_{23}})
=\Im(Z_7\,e^{-i\overline\theta_{23}})=0$
\phaa & $\phm e^{2i\theta_{12}}$ & $h_1$ & $h_3$ & $h_2$\\
III$^\prime$b & $\phaa c_{13}=\Im(Z_5\,e^{-2i\overline\theta_{23}})
=\Re(Z_7\,e^{-i\overline\theta_{23}})=0$
\phaa  & $-e^{2i\theta_{12}}$ & $h_2$ & $h_3$ & $h_1$\\ \hline
\end{tabular}
\end{table}

\begin{table}[ht!]
\begin{minipage}[t]{2.0in}
\centering
\parbox[t]{1.75in}{\caption{The U(2)-invariant quantities $q_{k\ell}$
for Cases I$^\prime$a and I$^\prime$b}\label{d1}} \\ \vspace{0.1in}
\begin{tabular}{|c||c|c|}\hline
$\phaa k\phaa $ &\phaa $q_{k1}\phaa $ & \phaa $q_{k2} \phaa $ \\ \hline
$1$ & $1$ & $0$ \\
$2$ & $0$ & $1$ \\
$3$ & $0$ & $i$ \\ \hline
\end{tabular}
\end{minipage}
\hfill
\begin{minipage}[t]{2.0in}
\centering
\parbox[t]{1.75in}{\caption{The U(2)-invariant quantities $q_{k\ell}$
for Cases II$^\prime$a and II$^\prime$b}\label{d2}} \\ \vspace{0.1in}
\begin{tabular}{|c||c|c|}\hline
$\phaa k\phaa $ &\phaa $q_{k1}\phaa $ & \phaa $q_{k2} \phaa $ \\ \hline
$1$ & $\phm 0$ & $1$ \\
$2$ & $-1$ & $0$ \\
$3$ & $\phm 0$ & $i$ \\ \hline
\end{tabular}
\end{minipage}
\hfill
\begin{minipage}[t]{2.0in}
\centering
\parbox[t]{1.75in}{\caption{The U(2)-invariant quantities $q_{k\ell}$
for Cases III$^\prime$a and III$^\prime$b}\label{d3}} \\ \vspace{0.1in}
\begin{tabular}{|c||c|c|}\hline
$\phaa k\phaa $ &\phaa $q_{k1}\phaa $ & \phaa $q_{k2} \phaa $ \\ \hline
$1$ & $\phm 0$ & $ie^{i\theta_{12}}$ \\
$2$ & $\phm 0$ & $e^{i\theta_{12}}$ \\
$3$ & $-1$ & $0$ \\ \hline
\end{tabular}
\end{minipage}
\end{table}
\clearpage

Cases I$^\prime$a and I$^\prime$b  correspond to
the combination of Cases I and IIa of Table~\ref{Z6cond}.
Cases II$^\prime$a and II$^\prime$b correspond to
the combination of Cases I and IIb of Table~\ref{Z6cond}.
Finally, Cases III$^\prime$a and III$^\prime$b are new.  In these
last two cases,
\beq \label{bardef}
\overline\theta_{23}\equiv\theta_{23}-\theta_{12}\,,\qquad\quad \overline\eta^2=
\eta^2 e^{-2i\theta_{12}}=\pm 1\,,
\eeq
play the roles of $\theta_{23}$ and $\eta^2$, respectively.
Note that \eqs{etaZ}{etarho} correlate the overall phases of
$Z_7$ and the $\rho^Q$.  In particular,
\beqa
&&\hspace{-0.4in}\text{Cases I}^\prime\text{a and II}^\prime\text{a:} \quad
\Im(Z_7 e^{-i\theta_{23}})=\Im(\rho^Q e^{i\theta_{23}})=0,
\!\!\!\qquad
\text{Case III}^\prime\text{a:}\quad
\Im(Z_7 e^{-i\overline\theta_{23}})=\Im(\rho^Q
e^{i\overline\theta_{23}})=0, \label{correlation1} \\
&&\hspace{-0.4in}\text{Cases I}^\prime\text{b and II}^\prime\text{b:} \quad
\Re(Z_7 e^{-i\theta_{23}})=\Re(\rho^Q e^{i\theta_{23}})=0,\!\!\!\!\qquad
\text{Case III}^\prime\text{b:} \quad
\Re(Z_7 e^{-i\overline\theta_{23}})=\Re(\rho^Q e^{i\overline\theta_{23}})=0.
\label{correlation2}
\eeqa


The Higgs state corresponding to $q_{k1}\neq 0$ in Tables
\ref{d1}, \ref{d2} and \ref{d3} is a CP-even Higgs boson.
Moreover, as $q_{k1}=\pm 1$ and $|q_{k2}|=0$
in each case, it follows from Appendix~\ref{app:couplings} that this state
has precisely the couplings of the Standard Model Higgs boson!
Note that the $q_{k1}$ vanish for the other two neutral Higgs states,
and thus cannot be used
to fix the absolute CP quantum numbers of these two states.
In the $Z_6=0$ model, it is $Z_7$ and/or $\rho^Q$ that determine
which of these two states is CP-even and which is CP-odd.

It is convenient to define an invariant quantity,
$\varepsilon_{57}$, by the relation
\beq  \label{zz57}
\Re(Z_5^* Z_7^2)=\varepsilon_{57} |Z_5|\,|Z_7|^2\,,\qquad
\varepsilon_{57}\equiv\pm 1\,.
\eeq
Since ${\rm Im}(Z_5 e^{-2i\theta_{23}})=0$ is satisfied in
Cases I$^\prime$ and II$^\prime$, it follows that
\beq \label{rez57}
{\rm Re}(Z_5 e^{-2i\theta_{23}})=\begin{cases} \phm\epsilon_{57}|Z_5|\,,&
\quad \text{if}~~\Im(Z_7\,e^{-i\theta_{23}})=0\,, \\
-\epsilon_{57}|Z_5|\,,&
\quad \text{if}~~\Re(Z_7\,e^{-i\theta_{23}})=0\,.\end{cases}
\eeq
Note that $\varepsilon_{57}$ is the sign of $Z_5$ in the real basis.
\Eq{rez57} can be rewritten more compactly as:
\beq \label{rez57compact}
{\rm Re}(Z_5 e^{-2i\theta_{23}})=\eta^2\epsilon_{57}|Z_5|\,.
\eeq
In Case III$^\prime$, ${\rm Im}(Z_5 e^{-2i\overline\theta_{23}})=0$,
in which case, \eqs{rez57}{rez57compact} hold with $\theta_{23}$ and $\eta^2$ are replaced by
$\overline\theta_{23}$ and $\overline{\eta}^2$, respectively.

The masses of the neutral Higgs bosons are as follows.  There is
one CP-even Higgs boson whose squared-mass is given by:\footnote{In
\eqs{mH1}{mH2}, we employ the notation $h_1^0$ and $h_2^0$ for the
two CP-even Higgs bosons (rather than $h^0$ and $H^0$),
since the mass ordering of these states depends on the
the choice of the 2HDM parameters.}
\beq \label{mH1}
m^2_{h_1^0}=Z_1 v^2\,.
\eeq
As noted above, the mass and couplings of $h_1^0$
are exactly the same as those of the
Standard Model Higgs boson.\footnote{The Standard Model properties
of $h_1^0$ are independent of its mass and the masses of $h_2^0$ and
$A^0$.  In this sense, this case is not a decoupling limit, although
the properties of $h_1^0$ are identical to the corresponding
properties of the lightest CP-even Higgs boson in the decoupling limit.}
The squared-masses of the remaining two
neutral Higgs bosons (a CP-even state $h_2^0$ and a CP-odd
state~$\ha$) are given by:
\beqa
m_{h_2^0}^2&=&Y_2+\half(Z_3+Z_4+\varepsilon_{57}|Z_5|)v^2\,,\label{mH2} \\
\mha^2 &=&Y_2+\half(Z_3+Z_4-\varepsilon_{57}|Z_5|)v^2\,.\label{mH3}
\eeqa
Cases involving
mass-degenerate neutral Higgs bosons are examined in Appendix C.

The above results are valid as long as either $Z_6$ or $Z_7$ is
non-vanishing.  If both $Z_6=0$ and $Z_7=0$, the model has some extra
features, which we examine in the following section.

\subsection{The 2HDM with
$\boldsymbol{Z_6=Z_7=0}$}\label{sec:herquetcpcons}

In Sections~\ref{subsecz6} and \ref {z6700}, we established
basis-independent conditions for which the
2HDM scalar potential and vacuum were CP-conserving.  If $Z_6=Z_7=0$
(and $Y_3=0$ by virtue of the potential minimum condition), then
$Z_5$ is the the only potentially complex parameter of the scalar
potential in the Higgs basis.  Consequently, one can rephase the Higgs
field $H_2$ to obtain a real Higgs basis (where $Z_5$ is real).
Hence, if $Z_6=Z_7=0$ then the 2HDM scalar potential and vacuum
automatically preserve the CP symmetry.\footnote{One can implement
$Y_3=Z_6=Z_7=0$ by imposing a $\mathbb{Z}_2$ symmetry in the Higgs basis.
If the Higgs-fermion couplings also respect this discrete symmetry,
then the resulting 2HDM is the Inert Doublet Model~\cite{inert},
since the model contains
no interaction vertices with an odd number of $H_2$ fields.}

Starting from any real basis of a CP-invariant 2HDM scalar potential,
one can always apply an O(2) transformation to the Higgs fields
to define another generic real basis.
In general, all possible real basis choices can be reached in this way.
However, in the case of $Y_3=Z_6=Z_7=0$, there exists a particular U(2)
transformation, ${\rm diag}(1,i)$,
that is not an O(2) transformation, which has the
effect of changing the sign of $Z_5$.  This corresponds to redefining
the second Higgs field by
\beq H_2
\rightarrow i H_2 \,.\label{h2t}
\eeq
Following Appendix A of \Ref{cpbasis} (where the analogous
arguments for the time-reversal-invariant 2HDM is presented),
the CP transformation
law is unique only if all real basis choices are related by an O(2)
transformation.  If all real basis choices are related by
a larger global transformation group, O(2)$\times\mathbb{D}\subset$U(2),
then the CP transformation law (within the Higgs/gauge boson sector)
is not unique and the
number of inequivalent CP transformation laws is equal to the number
of elements of the (non-trivial) discrete group $\mathbb{D}$.
Applying this to the 2HDM with $Y_3=Z_6=Z_7=0$, we
identify $\mathbb{D}=\mathbb{Z}_2$, which is
the discrete group consisting of the identity element and
${\rm diag}(1,i)\in$U(2) [the latter
changes the sign of $Z_5$].  We conclude
that for the $Y_3=Z_6=Z_7=0$ model, there are two inequivalent
definitions of CP in the Higgs/gauge boson sector.  For example,
in Cases I$^\prime$ and II$^\prime$ of Table~\ref{Z60cond}, the two definitions
of CP correspond to $\eta^2=\pm 1$ in \eq{CPgen} [for Case III$^\prime$,
simply replace $\theta_{23}$ with $\overline\theta_{23}$ and $\eta^2$
with $\overline{\eta}^2$].

In particular, consider the U(2)-invariant couplings $q_{k\ell}$
given in Tables \ref{d1}, \ref{d2} and \ref{d3}.  The Higgs boson $h_1^0$,
defined here as the scalar $h_k$ corresponding to $|q_{k1}|=1$, is CP-even.
For either choice of the two inequivalent definitions of CP,
the couplings of $h_1^0$ precisely match those of the Standard Model
Higgs boson [as previously noted below \eq{correlation2}].  But, for the two Higgs states $h_2^0$ and $h_3^0$
with $q_{k1}=0$, the  Higgs/gauge boson
interactions are insufficient to uniquely identify the CP-odd Higgs
field as noted above.  The squared-mass of the neutral Higgs bosons
must be the same as in the previous subsection (where $Z_6=0$ and $Z_7\neq 0$),
since the neutral Higgs squared-mass matrix is independent of $Z_7$.
However, when $Z_7=0$, we cannot employ \eqs{mH2}{mH3} since
$\varepsilon_{57}$ is no longer defined.  Nevertheless, one can
directly analyze the squared-mass matrix given by \eq{mtilmatrix}, which
is diagonal.  Defining $Z_5\equiv |Z_5|e^{2i\theta_5}$, and noting
that $\Im(Z_5 e^{-2i\theta_{23}})=0$, it follows that $\theta_5-\theta_{23}=\half n\pi$
for some integer $n$.  Hence
$\Re(Z_5 e^{-2i\theta_{23}})=\pm|Z_5|$, where the $\pm$ corresponds to the two possible
choices $\theta_5-\theta_{23}=0$ or $\half\pi$.  We conclude that
the squared-masses of $h_2^0$ and $h_3^0$ are given by:
\beq \label{m2hh}
m^2_{h_2^0,h_3^0}=Y_2+\half v^2(Z_3+Z_4\mp |Z_5|)\,,
\eeq
where by convention, we choose $m_{h_2^0}<m_{h_3^0}$.

If the neutral Higgs--fermion Yukawa interactions are CP-conserving, then
the ambiguity of the CP quantum numbers of $h_2^0$ and $h_3^0$ can
be resolved.
The results of Table~\ref{Z60cond} still apply if $Z_7$ is replaced by
$\rho^{Q\,*}$ (for either $Q=U,D$ or $E$).
It is convenient to define an invariant quantity,
$\varepsilon_{5Q}$, by the relation,
\beq  \label{zz5Q}
\Re[Z_5 (\rho^Q_{ij})^2]=\varepsilon_{5Q} |Z_5|\,|\rho^Q_{ij}|^2\,,\qquad
\varepsilon_{5Q}\equiv\pm 1\,,
\eeq
where $\rho^Q_{ij}$ is any non-vanishing matrix element of $\rho^Q$.
Following the derivation
of \eqthree{zz57}{rez57}{rez57compact}, it then follows that
\beq \label{rez5Q}
{\rm Re}(Z_5 e^{-2i\theta_{23}})=\begin{cases} \phm\epsilon_{5Q}|Z_5|\,,&
\quad \text{if}~~\Im(e^{i\theta_{23}}\rho^Q)=0\,,\\
-\epsilon_{5Q}|Z_5|\,,&
\quad \text{if}~~\Re(e^{i\theta_{23}}\rho^Q)=0\,,\end{cases}
\eeq
for Cases I$^\prime$ and II$^\prime$
(for Case III$^\prime$, $\theta_{23}$ and ${\eta}^2$ are replaced by
$\overline\theta_{23}$ and $\overline{\eta}^2$, respectively).
Note that $\varepsilon_{5Q}$ is the sign of
$Z_5$ in the real Higgs basis in which the scalar potential parameters and
the Higgs-fermion Yukawa coupling matrices are simultaneously real.
In particular, $\varepsilon_{5Q}$ is independent of the choice of $i$ and~$j$
in \eq{zz5Q} [assuming $\rho^Q_{ij}\neq 0$].
Even though $Z_6=Z_7=0$, the sign of $Z_5$ in the real Higgs basis is
meaningful due to the presence of the Yukawa couplings.
\Eq{rez5Q} can be rewritten more compactly as:
\beq \label{rez5Qcompact}
{\rm Re}(Z_5 e^{-2i\theta_{23}})=\eta^2\epsilon_{5Q}|Z_5|\,.
\eeq

The two choices of $\eta^2=\pm 1$ are now distinguishable.
For example, in Case I$^\prime$,
the diagonal parts of the $QQh_k$ interactions are given by:
\beqa
\mathcal{L}_{QQh_2}&=&\frac{-1}{\sqrt{2}}\sum_{i=1}^3
\overline{Q}_i \left[\Re(e^{i\theta_{23}}\rho^Q)\pm
i\gamma\ls{5}\Im(e^{i\theta_{23}}\rho^Q)\right]_{ii}Q_ih_2\,,\\
\mathcal{L}_{QQh_3}&=&\frac{-1}{\sqrt{2}}\sum_{i=1}^3
\overline{Q}_i \left[\mp i\gamma\ls{5}\Re(e^{i\theta_{23}}\rho^Q) +
\Im(e^{i\theta_{23}}\rho^Q)\right]_{ii}Q_ih_3\,,\label{yukthree}
\eeqa
where the upper (lower) sign corresponds to $Q=U$ ($Q=D,E$).
It follows that in Case I$^\prime$, $h_3$ is CP-odd
if $\eta^2=1$, i.e. $\Im(e^{i\theta_{23}}\rho^Q)=0$ and $h_2$ is
CP-even if $\eta^2=-1$, i.e. $\Re(e^{i\theta_{23}}\rho^Q)=0$.
That is, the neutral-Higgs-fermion Yukawa interaction selects one of
the two inequivalent definitions of CP.  Cases II$^\prime$ and III$^\prime$
can be similarly treated.  In a real Higgs basis,
the unique CP transformation law depends on whether $\rho^Q$ is a
purely real or purely imaginary matrix.
If the  neutral Higgs--fermion Yukawa interactions are CP-violating,
then neither $h_2^0$ nor $h_3^0$ can be assigned a definite CP quantum
number.

\subsection{CP symmetries in the 2HDM}
\label{cp1}

Generalized CP-transformations (GCPs) in the 2HDM have been examined in
\refs{Ferreira:2009wh}{Ferreira:2010yh}.  In a generic basis,
a GCP transformation is of the form given in \eq{cptrans}, where
$\mathcal{U}$ is an arbitrary
$2\times 2$ unitary matrix.  Three classes of GCPs were
identified in \refs{Ferreira:2009wh}{Ferreira:2010yh} according to the value
of $\mathcal{U}\mathcal{U}^*$:
\beqa
\rm{(i)}~~CP1:&&\,\,\, \mathcal{U}\mathcal{U}^*=\mathds{1}_{2\times 2}\,,
\quad \phm\text{i.e.,
$\mathcal{U}$ is a unitary symmetric matrix}\,,\\
\rm{(ii)}~~CP2:&&\,\,\, \mathcal{U}\mathcal{U}^*=-\mathds{1}_{2\times 2}\,,\quad \text{i.e.,
$\mathcal{U}$ is a
unitary antisymmetric matrix}\,,\\
\rm{(iii)}~~CP3:&&\,\,\, \mathcal{U}\mathcal{U}^*\neq \pm\mathds{1}_{2\times 2}\,,
\eeqa
where $\mathds{1}_{2\times 2}$ is the $2\times 2$ identity matrix.  The
CP1 transformation corresponds to \eq{Uab}.  Imposing CP1 on the 2HDM scalar potential implies
that there exists a basis in which all the scalar potential parameters are real.
Imposing CP2 and CP3 yields additional constraints on the scalar potential,
which are not especially relevant to the matters addressed in this paper.
In \Ref{Ferreira:2010yh}, the possibility of imposing symmetries in a specific basis
is discussed.  This can lead to additional conditions on the scalar potential parameters,
which may or may not correspond to a higher symmetry of the 2HDM.

In \refs{herquet}{herquet2}, the CP1 transformation is applied directly in the Higgs
basis.  In particular, these authors examine:
\beq
\begin{pmatrix} H_1 \\ H_2\end{pmatrix} = \begin{pmatrix} e^{i\xi} & \,\,\, 0 \\
0 & \,\,\, e^{-i\xi}\end{pmatrix} \begin{pmatrix} H_1^* \\ H_2^*\end{pmatrix}\,,
\eeq
where $0\leq\xi\leq\pi$.
Imposing this CP1 transformation on the Higgs basis constrains the
potentially complex scalar potential
parameters as follows:
\beqa
&&1.~~\text{If}~\xi=0\phantom{\,,\pi}\quad\Longrightarrow\quad Y_3, Z_5, Z_6, Z_7\in\mathbb{R}\,,\\
&&2.~~\text{If}~\xi=\pi\phantom{\,,\pi}\quad\Longrightarrow\quad Z_5\in\mathbb{R},\quad Y_3=Z_6=Z_7=0\,,\\
&&3.~~\text{If}~\xi\neq 0\,,\,\pi\quad\Longrightarrow\quad Y_3=Z_5=Z_6=Z_7=0\,.
\eeqa
This analysis singles out the importance of the $Z_6=Z_7=0$ model discussed in
Section~\ref{sec:herquetcpcons}, which is designated as a
``twisted'' CP-conserving model in \Ref{herquet}.\footnote{In \Ref{herquet},
the twisted model
is associated with the $Z_6=Z_7=0$ model with custodial symmetry.
Here, we see that
custodial symmetry has nothing to do with the existence of this class
of models, but is an additional constraint that can be imposed on the
CP-conserving scalar potential.
See Section~\ref{twisted} for further discussions of this point.}
  The case of $Z_5=Z_6=Z_7=0$ possesses
similar properties to the former model, with the added feature that
the two Higgs scalars of indefinite CP quantum number are mass-degenerate.

\section{Basis-independent conditions for custodial symmetry in the 2HDM}   

In the Standard Model,
the tree-level relation, $m_W^2=m_Z^2\cos\theta_W$, is a consequence
of an accidental global symmetry of the Higgs potential.  In
particular, the SM Higgs Lagrangian possesses an
SO(4)$\iso$SU(2)$_L\times$SU(2)$_R/\mathbb{Z}_2$ global symmetry, whereas
the full electroweak Lagrangian is invariant under
SU(2)$\times$U(1)$_Y$, which is a subgroup of the larger global
symmetry group.  The global \textit{custodial} SU(2)$_V$ symmetry, which
is the diagonal (vector) subgroup of SU(2)$_L\times$SU(2)$_R$
(where $V=L+R$),
is responsible for the gauge boson tree-level mass relation.

The U(1)$_Y$ hypercharge gauge interactions and the
Higgs--fermion Yukawa couplings break the custodial symmetry.  This
leads to finite one-loop radiative corrections to the
gauge boson tree-level mass relation.  The dominant part of these
corrections can be parameterized by a single quantity called $T$,
introduced by Peskin and Takeuchi~\cite{peskin}.  It is
convenient to define $T$ relative to a ``reference Standard Model,''
in which the Higgs mass is fixed.  A convenient choice is to define
$T=0$ for a Standard Model Higgs mass of 117~GeV.\footnote{The
choice of Higgs mass is dictated by the global Standard Model fit
to precision electroweak data~\cite{lepewwg,erlerPDG,gfitter},
which suggests that the Higgs mass
must lie above but not too far away from the lower Higgs mass bound
(at 95\% CL) of 114.4 GeV established at LEP~\cite{lepbound}.
In Ref.~\cite{erlerPDG}, a Higgs mass of 117~GeV is chosen for
the reference Standard Model in the analysis of new physics
contributions to the Peskin-Takeuchi $S$, $T$ and $U$ parameters.}
Deviations from
$T=0$ can be accommodated by either changing the value of the Higgs
mass or adding new custodial-violating interactions to the theory.
Experimentally, $T$ is observed to be quite small, which suggests that
the custodial-breaking effects of the electroweak Lagrangian
due to new physics beyond the Standard Model (or a Standard Model
Higgs mass that differs significantly from 117~GeV) must be
quite small.

In the 2HDM with a generic scalar potential, the Higgs Lagrangian
does not possess a global custodial symmetry.  One can therefore
write the Higgs Lagrangian in the form
\beq \label{separate}
\mathcal{L}_{\rm Higgs}=\mathcal{L}_{\rm CSC}+\mathcal{L}_{\rm CSV}\,,
\eeq
where $\mathcal{L}_{\rm CSC}$ and $\mathcal{L}_{\rm CSV}$ are the custodial
symmetry conserving and violating pieces, respectively.
The terms that contribute to $\mathcal{L}_{\rm CSV}$ reside in the scalar
potential, and do not effect the gauge boson mass relation at tree-level.
Hence, these terms only contribute a finite correction
at one-loop to the $T$ parameter.  Nevertheless, the experimental
determination of $T$ can place significant constraints on the
parameters of the 2HDM scalar potential.
In this section, we formulate a basis-independent characterization of
custodial symmetry.  This will permit a clean basis-independent
separation of the  custodial
symmetry conserving and violating pieces of the Higgs Lagrangian as in
\eq{separate}.



\subsection{Custodial symmetry of the Higgs Sector}

\subsubsection{Basis-dependent conditions for custodial symmetry}
\label{pv}

Conditions for custodial symmetry of the Higgs sector in the 2HDM
doublet model has been previous addressed by Pomarol and Vega~\cite{pomarol}.
Consider the 2HDM scalar potential in a generic basis,
\beqa
\mathcal{V}&=& m_{11}^2\Phi_1^\dagger\Phi_1+m_{22}^2\Phi_2^\dagger\Phi_2
-[m_{12}^2\Phi_1^\dagger\Phi_2+{\rm h.c.}]
+\half\lambda_1(\Phi_1^\dagger\Phi_1)^2
+\half\lambda_2(\Phi_2^\dagger\Phi_2)^2
+\lambda_3(\Phi_1^\dagger\Phi_1)(\Phi_2^\dagger\Phi_2)
\nonumber\\[6pt]
&&\quad
+\lambda_4(\Phi_1^\dagger\Phi_2)(\Phi_2^\dagger\Phi_1)
+\left\{\half\lambda_5(\Phi_1^\dagger\Phi_2)^2
+\big[\lambda_6(\Phi_1^\dagger\Phi_1)
+\lambda_7(\Phi_2^\dagger\Phi_2)\big]
\Phi_1^\dagger\Phi_2+{\rm h.c.}\right\}\,,\label{genpot}
\eeqa
where $m_{11}^2$, $m_{22}^2$, $\lambda_1,\lambda_2,\lambda_3,
\lambda_4\in\mathbb{R}$ are
real parameters, and $m_{12}^2$, $\lambda_5$,
$\lambda_6$, $\lambda_7\in\mathbb{C}$ are potentially complex.
The vacuum expectation values of the neutral Higgs fields, denoted
by
\beq
\langle\Phi^0_a\rangle\equiv \frac{v_a}{\sqrt{2}}\in\mathbb{C}\,,\qquad
a=1,2\,,
\eeq
are also generically complex.
Pomarol and Vega asserted that the imposition of custodial
symmetry on the 2HDM scalar potential
yields two independent cases [in the notation of \eq{genpot}]:
\beqa
{\rm Case}~1:&\quad& v_1,v_2\in\mathbb{R}\,,\quad \lambda_4=\lambda_5\,,\quad
{\rm and}\quad m^2_{12},\lambda_5,\lambda_6,\lambda_7\in\mathbb{R}
\,,\label{cas1} \\[6pt]
{\rm Case}~2:&\quad &v_1=v_2^*\in\mathbb{C}\,,\quad
m_{11}^2=m_{22}^2\,,\quad \lambda_1=\lambda_2=\lambda_3\,,\quad \lambda_6=\lambda_7\,,\quad
{\rm and}\quad m_{12}^2,\lambda_5,\lambda_6,\lambda_7\in\mathbb{C}\,.\label{cas2}
\eeqa
These conditions are derived as follows.

In Case~1, one constructs two
$2\times 2$ matrix fields,\footnote{In the notation of \eq{m1m2},
$\widetilde\Phi_a$ is the first column and
$\Phi_a$ is the second column of the matrix $M_a$ (for $a=1,2$).}
\beq \label{m1m2}
M_1 \equiv (\widetilde{\Phi}_1\,, \Phi_1), \hspace{5mm}M_2 \equiv
(\widetilde{\Phi}_2\,,  \Phi_2),
\eeq
where $\widetilde{\Phi}\equiv i \sigma_2 \Phi^*$.
The matrix fields transform under SU(2)$_L\times$SU(2)$_R$ as
\beq \label{cust1}
M_a \rightarrow L~M_a~R^\dagger\,,\qquad a=1,2\,.
\eeq
The SU(2)$_{L}\times$SU(2)${}_{R}$ scalar potential is constructed by
employing the manifestly invariant combinations, $\Tr[M_1^\dagger M_1]$,
$\Tr[M_2^\dagger M_2]$
and $\Tr[M_1^\dagger M_2]$.\footnote{One can check that
$\Tr[M_1^\dagger M_2]=\Tr[M_2^\dagger M_1]$ and $\det M_a=\half\Tr[M_a^\dagger M_a]$ (for $a=1,2$), so
only three independent invariant quadratic forms are relevant.}  Explicitly,
\beqa
\mathcal{V} &=& \half m_{11}^2 \Tr[M_1^\dagger M_1]
+\half m_{22}^2 \Tr[M_2^\dagger M_2]-
m_{12}^2 \Tr[M_1^\dagger M_2]+
\tfrac{1}{8} \lambda_1 \left(\Tr[M_1^\dagger M_1]\right)^2
+\tfrac{1}{8} \lambda_2 \left(\Tr[M_2^\dagger M_2]\right)^2\nonumber\\
&& + \tfrac{1}{4} \lambda_3 \Tr[M_1^\dagger M_1] \Tr[M_2^\dagger M_2]
+\half \lambda \left(\Tr[M_1^\dagger M_2]\right)^2
+\half \left( \lambda_6 \Tr[M_1^\dagger M_1]
+\lambda_7 \Tr[M_2^\dagger M_2] \right)\Tr[M_1^\dagger M_2]\,,
\label{pomvegpot1}
\eeqa
where hermiticity implies that all the coefficients above are real.
Thus, imposing the SU(2)$_L\times$SU(2)$_R$ symmetry on the scalar potential
and comparing with \eq{genpot} then yields $\lambda=\lambda_4=\lambda_5$ and
$m^2_{12},\lambda_5,\lambda_6,\lambda_7\in\mathbb{R}$.
If the scalar field vacuum expectation values satisfy:
\beq  \label{vevs1}
\vev{M_a} = \frac{1}{\sqrt{2}}
\left(\begin{array}{cc}v^*_a&0\\0&v_a\end{array}\right)
=\frac{v_a}{\sqrt{2}}\,\mathbb{1}_{2\times 2}\,,
\eeq
then $\vev{M_a}$ is invariant
under the SU(2)$_V$ custodial symmetry group, since
$\vev{M_a} \rightarrow L \vev{M_a} R^\dagger
= \vev{M_a}$ when $L=R$.  \Eq{vevs1} imposes
the condition $v_a\in\mathbb{R}$,
and \eq{cas1} is thus established.

In Case 2, one constructs the $2\times 2$ matrix field,
\beq
M_{12} \equiv (\widetilde{\Phi}_1 \,,\Phi_2), \label{matdeftwo}\eeq
which transforms under SU(2)$_L\times$SU(2)$_R$ as
\beq \label{cust2}
M_{12}\rightarrow L~M_{12}~R^\dagger\,.
\eeq
The SU(2)$_{L}\times$SU(2)${}_{R}$ scalar potential is constructed by
employing the manifestly invariant combinations $\Tr[M_{12}^\dagger
M_{12}]$, $\det M_{12}$, $\det (M_{12})^2$
and $\det[M_{12}^\dagger M_{12}]$.  Explicitly,
\beqa
\mathcal{V}&=& m^2\Tr[M_{12}^\dagger M_{12}]-\left(m_{12}^2\det
M_{12}+{\rm h.c.}\right)+\half\lambda \left(\Tr[M_{12}^\dagger
M_{12}]\right)^2 \nonumber \\
&& +\lambda_4 \det[M_{12}^\dagger M_{12}]
+\half\left(\lambda_5\det (M_{12})^2+{\rm h.c.}\right)
+\left(\lambda^\prime \det M_{12} \Tr[M_{12}^\dagger
M_{12}]+{\rm h.c.}\right)\,.
\eeqa
Thus, imposing the SU(2)$_L\times$SU(2)$_R$ symmetry on the scalar potential
and comparing with \eq{genpot}
then yields $m^2=m_{11}^2=m_{22}^2$, $\lambda=\lambda_1=\lambda_2=\lambda_3$,
$\lambda^\prime=\lambda_6=\lambda_7$
and $m_{12}^2,\lambda_5,\lambda_6,\lambda_7\in\mathbb{C}$.
If the scalar field vacuum expectation values satisfy:
\beq  \label{vevs2}
\vev{M_{12}} = \frac{1}{\sqrt{2}}
\left(\begin{array}{cc}v^*_1&0\\0&v_2\end{array}\right)
=\frac{v_2}{\sqrt{2}}\,\mathbb{1}_{2\times 2}\,,
\eeq
then $\vev{M_{21}}$ is invariant
under the SU(2)$_V$ custodial symmetry group.
\Eq{vevs2} imposes
the condition $v_1^*=v_2\in\mathbb{C}$,
and \eq{cas2} is thus established.

Although the two cases of Pomarol and Vega appear to be distinct, a
more careful analysis shows that the two cases are in fact equivalent,
and correspond to the formulation of the 2HDM in two different choices
of the scalar field basis.
This can be established as follows.
First, we note that in both Cases 1 and 2, the scalar potential
depends on three independent squared-mass parameters and six independent
scalar self-coupling parameters.  Now, suppose one begins with
a 2HDM subject to the constraints of Case 2 [\eq{cas2}].  It is
convenient to define:
\beq
\widehat v_1=\widehat v_2^*=\frac{1}{\sqrt{2}}e^{i\theta}\,,\qquad m^2\equiv
m_{11}^2=m_{22}^2\,,\qquad
\lambda\equiv\lambda_1=\lambda_2=\lambda_3\,,\qquad
\lambda^\prime\equiv \lambda_6=\lambda_7\,, \eeq where $\widehat v_a$ is
defined in \eq{emvev}.  By performing a basis transformation,
$\Phi_a\to U_{a\bar{b}}\Phi_b$, with
\beq \label{umatrix}
U=\frac{1}{\sqrt{2}}\begin{pmatrix} \phm e^{-i\theta} & \,\,\,
  e^{i\theta} \\ -ie^{-i\theta} & \,\,\,
  ie^{i\theta}\end{pmatrix}\,,
  \eeq
  the coefficients of the scalar
potential [cf.~\eq{genpot}] are transformed to [in the notation of
\eq{hbasispot}]: \beqa
Y_1 &=& m^2-{\rm Re}(m_{12}^2 e^{-2i\theta})\,,\label{yone}\\[5pt]
Y_2 &=& m^2+{\rm Re}(m_{12}^2 e^{-2i\theta})\,,\\[5pt]
Y_3 &=& -{\rm Im}(m_{12}^2 e^{-2i\theta})\,,\\[5pt]
Z_1 &=& \lambda+\half\lambda_4+\half {\rm Re}(\lambda_5 e^{-4i\theta})
+2{\rm Re}(\lambda^\prime e^{-2i\theta}) \,,\\[5pt]
Z_2 &=& \lambda+\half\lambda_4+\half {\rm Re}(\lambda_5 e^{-4i\theta})
-2{\rm Re}(\lambda^\prime e^{-2i\theta}) \,,\\[5pt]
Z_3 &=& \lambda-\half\lambda_4-\half {\rm Re}(\lambda_5 e^{-4i\theta})
\,,\\[5pt]
Z_4&=&Z_5 = \half\lambda_4-\half {\rm Re}(\lambda_5 e^{-4i\theta})
\,,\label{z45eq}\\[5pt]
Z_6&=& \half {\rm Im}(\lambda_5 e^{-4i\theta})
+{\rm Im}(\lambda^\prime e^{-2i\theta})\,,\\[5pt]
Z_7 &=& -\half {\rm Im}(\lambda_5 e^{-4i\theta}) +{\rm
  Im}(\lambda^\prime e^{-2i\theta})\,,\label{zseven}
\eeqa
while the normalized scalar field
vacuum expectation values are transformed to:
\beq \label{hbasis}
\widehat{v}_a\longrightarrow U_{a\bar b}\widehat v_b=\delta_{a1}\,.
\eeq
\Eq{hbasis} defines the Higgs
basis $\{H_1\,,\,H_2\}$, up to a phase redefinition of $H_2$.  We
immediately note that $Z_4=Z_5$ and the vacuum expectation values and
all the scalar potential parameters are real.  Thus, the Higgs basis
satisfies all the conditions of Case 1 of Pomarol and Vega
[cf.~\eq{cas1}].  Moreover, it is easy to check that any additional
O(2) basis transformation preserves $\lambda_4=\lambda_5$ and
the reality of the scalar potential parameters.  Thus, we have
confirmed that Cases 1 and 2 of Pomarol and Vega are equivalent and
simply represent different choices of the scalar field
basis.\footnote{In ref.~\cite{haberpomarol}, it was shown that in
Type-I and Type-II 2HDMs, the corresponding Higgs-fermion Yukawa couplings
(defined in the standard basis where the discrete symmetry $\Phi_2\to -\Phi_2$
is manifest) are custodial symmetric if and only if the scalar potential
parameters satisfy \eqs{cas1}{cas2}, respectively.   The
two ways to implement custodial symmetry given by \eqs{cust1}{cust2}, respectively,
can be distinguished based on the presence or absence of the $A^0 G G$
effective interaction.  This is possible, as the special forms of
the Type I and II Higgs-Yukawa interactions effectively select a ``preferred'' basis.}

Of course, one can also perform arbitrary U(2) transformations of the Higgs
fields.  The resulting scalar potential parameters and vacuum
expectation values will in general satisfy neither case 1 nor case 2
of Pomarol and Vega.  Yet, all these parameterizations are physically
equivalent and maintain the custodial symmetry.  Clearly, it is desirable
to formulate a basis-independent description of custodial symmetry.
We shall provide such a formulation in the next subsection.

\subsubsection{The Basis-Independent Condition for Custodial Symmetry in the Scalar Sector}
\label{invcust}

It is possible generalize the two implementations of custodial
symmetry presented in the previous subsection by constructing an
SU(2)$_{L}\times$SU(2)${}_{R}$ invariant scalar potential using the Higgs basis fields, $H_1$ and $H_2$.  The advantage of this basis choice is that no supplementary conditions on the vevs are required.  In particular,
we define $2\times 2$ matrix fields:
\beq
\mathbb{M}_1 \equiv (\widetilde{H}_1\,,H_1), \hspace{5mm}\mathbb{M}_2
\equiv (\widetilde{H_2}\,,H_2),  \hspace{5mm}
\mathbb{M}_{12} \equiv (\widetilde{H}_1\,,H_2), \hspace{5mm}\mathbb{M}_{21} \equiv (\widetilde{H_2}\,,H_1)\,.
\label{matdefgen}
\eeq
Since $\langle{H_1^0}\rangle=v/\sqrt{2}$ (where $v=246$~GeV) and $\langle{H_2^0}\rangle=0$, it follows that
\beq
\langle{\mathbb{M}_1}\rangle= \frac{v}{\sqrt{2}}\,\mathbb{1}_{2\times 2}\,,\qquad\quad
 \langle{\mathbb{M}_2}\rangle =0\,,
 \eeq
 whereas neither $\langle\mathbb{M}_{12}\rangle$ nor $\langle\mathbb{M}_{21}\rangle$
 are proportional to the identity matrix.
 Consequently, if we wish to preserve a custodial SU(2)$_V$ after electroweak symmetry
 breaking, the scalar potential in the Higgs basis \textit{must} be solely a function
of $\mathbb{M}_1$ and $ \mathbb{M}_2$.

In the Higgs basis, the field $H_1$ is basis-invariant, as it is
defined such that $\langle{H_1^0}\rangle \equiv v/\sqrt{2}$ is real
and positive.  On the other hand, since $\langle{H_2^0}\rangle=0$ it
follows that $H_2$ is only defined up to an overall rephasing.  That
is, $H_1$ is an invariant field with respect to basis transformations,
whereas $H_2$ is a pseudo-invariant field.  As a result, the
SU(2)$_{L}\times$SU(2)${}_{R}$ transformation laws for $\mathbb{M}_1$
and $\mathbb{M}_{2}$ are given by:
\beq \label{LRRtrans}
\mathbb{M}_1\to L\,\mathbb{M}_1 R^\dagger\,,\qquad\quad
\mathbb{M}_2\to L\,\mathbb{M}_2 R^{\prime\,\dagger}\,, \qquad L\in
{\rm SU}(2)_L\,,\quad R\,,\,R^\prime \in {\rm SU}(2)_R\,.
\eeq
Since
$H_i$ and $\widetilde H_i$ ($i=1,2$) are doublets under the weak
SU(2)$_L$ gauge transformation, the transformation matrices $L$
appearing in \eq{LRRtrans} must be the same in the
SU(2)$_L\times$SU(2)$_R$ transformation laws of $\mathbb{M}_1$ and
$\mathbb{M}_2$.  As noted by \cite{herquet}, the same requirement does
not hold for the SU(2)$_R$ transformation; hence in general
$R^\prime\neq R$.  However, $R$ and $R^\prime$ are related by the
fact that the gauged U(1) hypercharge operator,
$Y\equiv {\rm diag}(-1\,,\,+1)$, is a diagonal generator of SU(2)$_R$.
In particular, if we write $R\equiv\exp(i\theta n^a T_R^a)$
where $(n^1,n^2,n^3)$ is a unit vector, then
$T_R^3$ is proportional to $Y$.  Since the $H_i$ are
hypercharge $+1$ fields and the $\widetilde H_i$ are hypercharge $-1$
fields, the relation between $R$ and $R^\prime$ is fixed by
$R^\prime\equiv PRP^{-1}$, where $P$ is an SU(2) matrix and
$P\exp{(i\theta Y)P^{-1}}=\exp{(i\theta Y)}$ for all
$\theta$~\cite{herquet}.  By expanding in $\theta$, it follows that
$PY=YP$, which constrains $P$ to be of the form
$P={\rm diag}(e^{-i\chi}\,,\, e^{i\chi})$,
where $0\leq\chi<2\pi$.  We conclude that the most general form for
the SU(2)$_L\times$SU(2)$_R$ transformation laws for $\mathbb{M}_1$
and $\mathbb{M}_2$ is given by
\beq \label{LRtrans}
\mathbb{M}_1\to L\,\mathbb{M}_1 R^\dagger\,,\qquad\quad \mathbb{M}_2\to
L\,\mathbb{M}_2 PR^\dagger P^{-1}\,,
\eeq
where
\beq \label{Pdef}
L\in {\rm
  SU}(2)_L\,,\quad R\in {\rm SU}(2)_R\,,\qquad P\equiv\begin{pmatrix}
  e^{-i\chi} & \quad 0 \\ 0 & \quad e^{i\chi}\end{pmatrix}\,,\quad
(0\leq\chi<2\pi)\,.
\eeq
The phase angle $\chi$ can be interpreted as
representing the freedom to rephase the field $H_2$.  In particular,
if one defines $\mathbb{M}^\prime_1\equiv \mathbb{M}_1$ and
$\mathbb{M}^\prime_2\equiv \mathbb{M}_2 P$, then the
transformation laws for $\mathbb{M}_1^\prime$ and $\mathbb{M}_2^\prime$ are the
same, i.e. $\mathbb{M}^\prime_a\to L\,\mathbb{M}^\prime_a R^\dagger$
for $a=1,2$.

The SU(2)$_{L}\times$SU(2)${}_{R}$ scalar potential is constructed by employing the manifestly
invariant combinations, $\Tr[\mathbb{M}_1^\dagger \mathbb{M}_1]$, $\Tr[\mathbb{M}_2^\dagger \mathbb{M}_2]$
and $\Tr[\mathbb{M}_1^\dagger \mathbb{M}_2P]$.\footnote{Note that
$\Tr[\mathbb{M}_1^\dagger \mathbb{M}_2P]=\Tr[\mathbb{M}_2^\dagger \mathbb{M}_1P^{-1}]$, so
only three independent invariant quadratic forms are possible.}  Explicitly,
\beqa
\mathcal{V} &=& \half Y_1 \Tr[\mathbb{M}_1^\dagger \mathbb{M}_1]
+\half Y_2 \Tr[\mathbb{M}_2^\dagger \mathbb{M}_2]+
\widetilde{Y}_3  \Tr[\mathbb{M}_1^\dagger \mathbb{M}_2P]+
\tfrac{1}{8} Z_1 \left(\Tr[\mathbb{M}_1^\dagger \mathbb{M}_1]\right)^2
+\tfrac{1}{8} Z_2 \left(\Tr[\mathbb{M}_2^\dagger \mathbb{M}_2]\right)^2\nonumber\\
&& + \tfrac{1}{4} Z_3 \Tr[\mathbb{M}_1^\dagger \mathbb{M}_1] \Tr[\mathbb{M}_2^\dagger \mathbb{M}_2]
+\half \lambda \left(\Tr[\mathbb{M}_1^\dagger \mathbb{M}_2 P]\right)^2
+\half \left( \widetilde{Z}_6 \Tr[\mathbb{M}_1^\dagger \mathbb{M}_1]
+\widetilde{Z}_7 \Tr[\mathbb{M}_2^\dagger \mathbb{M}_2] \right)\Tr[\mathbb{M}_1^\dagger \mathbb{M}_2 P]\,,
\label{pompotgen}
\eeqa
where hermiticity implies that all coefficients above are real.
\Eq{pompotgen} is equivalent to
\beqa \label{custpotgen}
\mathcal{V}&=& Y_1 H_1^\dagger H_1+ Y_2 {H_2}^{\dagger} H_2
+[Y_3  H_1^\dagger {H_2}+{\rm h.c.}]\nonumber \\[5pt]
&&\quad+\half Z_1(H_1^\dagger H_1)^2 +\half Z_2({H_2}^\dagger {H_2})^2
+Z_3(H_1^\dagger H_1)({H_2}^\dagger {H_2})
+Z_4( H_1^\dagger {H_2})({H_2}^\dagger H_1) \nonumber \\[5pt]
&&\quad +\left\{\half Z_5  (H_1^\dagger {H_2})^2
+\big[Z_6 (H_1^\dagger H_1)+Z_7 ({H_2}^\dagger {H_2})\big]
H_1^\dagger {H_2}+{\rm h.c.}\right\}\,,
\eeqa
where
\beqa
\widetilde{Y_3}&=&Y_3 e^{-i\chi}=Y_3^* e^{i\chi}\in\mathbb{R}\,,\qquad\quad
\lambda=Z_4=Z_5 e^{-2i\chi}=Z_5^* e^{2i\chi}\in\mathbb{R}\,, \nonumber \\
\widetilde{Z_6}&=&Z_6 e^{-i\chi}=Z_6^* e^{i\chi}\in\mathbb{R}\,,\qquad\quad
\widetilde{Z_7}=Z_7 e^{-i\chi}=Z_7^* e^{i\chi}\in\mathbb{R}\,.
\label{custconds}
\eeqa
\Eq{custconds} implies that:
\beq \label{yzzz}
\Im(Y_3 e^{-i\chi})=\Im(Z_5 e^{-2i\chi})=\Im(Z_6 e^{-i\chi})=\Im(Z_7 e^{-i\chi})=0\,,
\eeq
which immediately implies that the scalar potential is CP-conserving.
Hence according to \eq{custconds}, the conditions for custodial
symmetry are given by
\beq \label{z45cond}
Z_4=Z_5 e^{-2i\chi}\in\mathbb{R}\,,\qquad Z_6e^{-i\chi}\,,\,Z_7e^{-i\chi}\in\mathbb{R}\,.
\eeq
Note that the conditions of \eq{z45cond} are basis-independent.  In particular, under
a basis transformation $\Phi_a\to U_{a\bar{b}}\Phi_b$,
\beq \label{chitrans}
e^{i\chi}\to (\det U)^{-1} e^{i\chi}\,.
\eeq
In the case of $Z_6\neq 0$ and/or $Z_7\neq 0$, one can relate the angle $\chi$ to
$\theta_{23}$.  In particular, by comparing \eqs{etaZ}{yzzz} it follows that
$e^{-i\chi}=\pm\eta e^{-i\theta_{23}}$.  The $\pm$ ambiguity is removed by
squaring this result, which yields
\beq \label{relate}
e^{-2i\chi}=\eta^2 e^{-2i\theta_{23}}\,.
\eeq
The phase $\eta^2$ is specified in Tables~\ref{Z6cond} and
\ref{Z60cond} for the various cases under which CP conservation holds.
In general, the basis-independent condition for custodial symmetry is:
\beq \label{z45eta}
Z_4=\eta^2 \Re(Z_5 e^{-2i\theta_{23}})\,,
\eeq
where we have used the fact that $\Im(Z_5 e^{-2i\theta_{23}})=0$ for
a CP-conserving 2HDM scalar potential.\footnote{For
Case~III$^\prime$ of Table~\ref{Z60cond},
one must replace $\theta_{23}$ and $\eta^2$ with
$\overline{\theta}_{23}$ and $\overline{\eta}^2$, respectively
[cf.~\eq{bardef}].
\label{replace}}

One can also eliminate the phase angle $\chi$ using \eq{custconds}:
\beq
e^{-2i\chi}=\frac{Z_6^*}{Z_6}=\frac{Z_7^*}{Z_7}\,.
\eeq
Consequently, if $Z_6\neq 0$ then \eq{z45cond} is equivalent to\footnote{CP
conservation requires that $\Im(Z_5 Z_6^{*2})=\Im(Z_5 Z_7^{*2})=0$.  Hence the numerators
of \eqs{invcondit}{altcond} are manifestly real, as required since $Z_4$ is a real
parameter.}
\beq
Z_4 = \frac{Z_5 Z_6^{*2}}{|Z_6|^2}=\varepsilon_{56}|Z_5|\,,
\label{invcondit}
\eeq
where the invariant quantity $\varepsilon_{56}$ was introduced
initially in \eq{zz56}.
The condition for custodial symmetry given by \eq{invcondit} is manifestly
basis-independent.\footnote{\Eq{invcondit} can also be obtained
from \eqs{rez56compact}{z45eta}, after noting that $\eta^4=1$.}
Similarly, if $Z_7\neq 0$, the basis-independent
condition for custodial symmetry can be written in the following form:
\beq \label{altcond}
Z_4 = \frac{Z_5 Z_7^{*2}}{|Z_7|^2}=\varepsilon_{57}|Z_5|\,,
\eeq
where the invariant quantity $\varepsilon_{57}$ was introduced
initially in \eq{zz57}.  Finally, if $\rho^Q\neq 0$, then under the
assumption that $Z_6\neq 0$ and/or $Z_7\neq 0$, one can use
\eqs{rez5Qcompact}{z45eta} to obtain:\footnote{In deriving \eq{z4q}
we used the fact that $\eta^4=1$ in Cases I$^\prime$
and II$^\prime$, and $\overline{\eta}^4=1$ in Case III$^\prime$}
\beq \label{z4q}
Z_4=\varepsilon_{5Q}|Z_5|\,.
\eeq
In the real Higgs basis, defined as the basis in which the scalar
potential parameters and the Yukawa coupling matrices $\rho^Q$ are
simultaneously real,  $\varepsilon_{56}$, $\varepsilon_{57}$
and $\varepsilon_{5Q}$ coincide with the sign of $Z_5$.  Thus
\eqthree{invcondit}{altcond}{z4q} reduce to the simple
relation,  $Z_4=Z_5$, in the real Higgs basis.  This result is consistent with \eq{z45eq}
obtained previously.  Note that the condition $Z_4=Z_5$ is invariant
with respect to $H_2\to -H_2$, which is the only remaining basis
freedom within the real basis.


The special case of $Z_6=Z_7=0$ must be treated separately.
In this case, $Y_3=0$ by virtue of \eq{Y3} and $Z_5$ is the only potentially
complex parameter of the scalar potential in the Higgs basis.  The condition
for a custodial symmetric scalar potential is now given by the single condition,
$Z_4=Z_5 e^{-2i\chi}\in\mathbb{R}$ [cf.~\eq{z45cond}].  Writing $Z_5=|Z_5| e^{2i\theta_5}$,
it follows that $\theta_5+\chi=n\pi/2$ for some integer $n$.  That is,
the basis-independent condition for custodial symmetry is given simply by:
\beq \label{modz5}
Z_4=\pm |Z_5|\,.
\eeq
In contrast to \eqthree{invcondit}{altcond}{z4q}, where $Z_4$ is uniquely determined
(and is equal to $Z_5$ in the real Higgs basis), in the special case of
$Z_6=Z_7=0$, there are two solutions for $Z_4$ that are consistent with a custodial symmetric
scalar potential.  This result can also be deduced by noting that
when $Z_6=Z_7=0$, \eqs{etaZ}{yzzz} yield
$e^{-2i\chi}=\pm\eta e^{-2i\theta_{23}}$, where the $\pm$ ambiguity cannot be removed
[in contrast to \eq{relate}].
Hence, when $Z_6=Z_7=0$, the basis-independent condition for custodial symmetry
is [cf.~footnote~\ref{replace}]:
\beq \label{z45etaalt}
Z_4=\pm\eta^2 \Re(Z_5 e^{-2i\theta_{23}})\,,
\eeq
which again exhibits two possible solutions.  Since $\eta^2=\pm 1$ and
$\Im(Z_5 e^{-2i\theta_{23}})=0$, \eq{z45etaalt} is equivalent to \eq{modz5} as expected.

The above results do not depend on the Yukawa coupling matrix $\rho^Q$.
If $\rho^Q=0$, then one
is free to redefine $H_2\to iH_2$, which has the effect of transforming $Z_5\to -Z_5$.
In this case, the sign of $Z_5$ in the real Higgs basis is basis-dependent,
and the two conditions $Z_4=\pm Z_5$ are equivalent.  Nevertheless,
there are still two solutions for $Z_4$ since a custodial symmetric scalar potential is possible with
either sign for $Z_4$.  If $\rho^Q\neq 0$, then the transformation
$H_2\to iH_2$ has the effect of transforming $\rho^Q$ to $i\rho^Q$.
If the neutral Higgs--fermion interactions are CP-conserving, then a real
Higgs basis exists in which $Z_5$ and $\rho^Q$ are simultaneously real.
In this case, the sign of $Z_5$ in the real Higgs basis is meaningful.
In contrast to \eq{z4q}, the condition for a custodial symmetric scalar
potential, which can be obtained directly from \eqs{rez5Qcompact}{z45etaalt},
is given by:
\beq \label{z4qalt}
Z_4=\pm\varepsilon_{5Q}|Z_5|\,.
\eeq
The existence of these two possible solutions when $Z_6=Z_7=0$ has a critical
impact on the nature of the Higgs mass degeneracy in the custodial limit,
as shown in the next subsection.
If the neutral Higgs--fermion interactions are CP-violating, then $\varepsilon_{5Q}$
has no meaning and \eq{z4qalt} must be discarded.  Nevertheless, the conclusion
that $Z_4=\pm |Z_5|$ for a custodial symmetric scalar potential with $Z_6=Z_7=0$ still applies.

In summary,
the basis-independent condition for a custodial-symmetric scalar potential is given by:
\beq \label{basindcust}
\hspace{-0.65in} Z_4=\begin{cases}  \varepsilon_{56}|Z_5|\,, & \quad \text{for}~~Z_6\neq 0\,, \\[5pt]
 \varepsilon_{57}|Z_5|\,, & \quad \text{for}~~Z_7\neq 0\,, \\[5pt]
\pm |Z_5|\,, & \quad \text{for}~~Z_6=Z_7=0\,.
\end{cases}
 \eeq
The above conditions do not depend on the form of the neutral Higgs--fermion interactions.
However, if the neutral Higgs--fermion interactions are CP-conserving, then
there exists an invariant quantity $\varepsilon_{5Q}$, defined in \eq{zz5Q}, which is equal
to the sign of $Z_5$ in the real Higgs basis (where all scalar potential parameters
and $\rho^Q$ are real).  In this case, we also have
\beq \label{basindcust2}
Z_4=\begin{cases}
\varepsilon_{5Q} |Z_5|\,, & \quad \text{for}~Z_6\neq 0~\text{and/or}~Z_7\neq 0\,,\\[5pt]
\pm \varepsilon_{5Q}|Z_5|\,, & \quad \text{for}~Z_6=Z_7=0\,.\end{cases}
\eeq
In a real Higgs basis, the general condition for a custodial symmetric scalar potential is
$Z_4=Z_5$.  In the special case of $Z_6=Z_7=0$, the condition $Z_4=-Z_5$ also yields
a custodial symmetric scalar potential.  These two conditions
are physically inequivalent when $\rho^Q\neq 0$.

\subsubsection{Higgs mass degeneracy in the custodial limit}\label{twisted}

The squared-mass of the charged Higgs boson is given by \eq{mc}.
If $Z_6\neq 0$ and/or $Z_7\neq 0$ and if CP is conserved in
the neutral Higgs sector, then the
squared-mass of the CP-odd Higgs boson is given by \eqs{m2A}{mH3}, which we
can rewrite as:
\beq
m_{A^0}^2=\begin{cases} m_{H^\pm}^2+\half v^2(Z_4-\varepsilon_{56}|Z_5|)\,,
\quad \text{if}~Z_6\neq 0\,,\\
 m_{H^\pm}^2+\half v^2(Z_4-\varepsilon_{57}|Z_5|)\,,
\quad \text{if}~Z_7\neq 0\,.\end{cases}
\eeq
In the custodial limit \eq{basindcust} applies, and it follows that
\beq \label{equalmasses}
m^2_{H^\pm}=m^2_{A^0}= Y_2 +\half Z_3 v^2\,,
\eeq
in agreement with the results of \cite{pomarol}.
That is, the charged Higgs boson and the CP-odd Higgs boson are
mass-degenerate in the custodial limit.

The case of $Z_6=Z_7=0$ is special, as discussed in
Section~\ref{sec:herquetcpcons}.  In this case, there is one neutral
CP-even Higgs boson, denoted by $h_1^0$,
with squared-mass $m_{h_1^0}^2=Z_1 v^2$ and two
neutral Higgs states of indeterminate CP quantum number, denoted
by $h_2^0$ and $h_3^0$, with squared-masses given by \eq{m2hh},
which yields:
\beq \label{m2hhcust}
m^2_{h_2^0,h_3^0}=m_{H^\pm}^2+\half v^2 (Z_4\mp |Z_5|)\,.
\eeq
According to \eq{basindcust}, $Z_4=\pm |Z_5|$ in the custodial limit.
We conclude that
either one of the states $h_2^0$ or $h_3^0$ can be degenerate in
mass with the charged Higgs boson.  However, the CP-quantum number of
$h_2^0$ and $h_3^0$ are indeterminate (if the Higgs-fermion interactions are
neglected), since there are two inequivalent definitions of CP when $Z_6=Z_7=0$.
This ambiguity can be resolved if the neutral Higgs-fermion
interactions are CP-conserving.\footnote{If the neutral Higgs--fermion
interactions are CP-violating, then the neutral Higgs state that is
degenerate in mass with the charged Higgs boson does not possess a
well-defined CP quantum number.}  In this case, the two neutral states can
be identified as a CP-even state $h^0$ or $H^0$ and a CP-odd state $A^0$.
Using \eqs{basindcust2}{m2hh}, it follows that:
\beq \label{equalmasses2}
m_{H^\pm}^2=\begin{cases} m_{A^0}^2 & \quad \text{if}~Z_4
=\varepsilon_{5Q} |Z_5|\quad\text{and}\quad Z_6=Z_7=0\,,\\
m_{h^0}^2~ &
\quad \text{if}~Z_4=-\varepsilon_{5Q}|Z_5|\,,\,\quad m_{H^\pm}^2< Z_1 v^2
\quad\text{and}\quad Z_6=Z_7=0\,,\\
m_{H^0}^2~ &
\quad \text{if}~Z_4=-\varepsilon_{5Q} |Z_5|\,,\,\quad m_{H^\pm}^2> Z_1 v^2
\quad\text{and}\quad Z_6=Z_7=0\,.\end{cases}
\eeq
where $m_{H^0}>m_{h^0}$ by convention.
In particular, $m_{H^\pm}^2=m_{A^0}^2$ if $Z_4=Z_5$ in the real Higgs basis,
whereas $m_{H^\pm}^2=m_{H^0}^2$ if $Z_4=-Z_5$ in the real Higgs basis.
This result is easy to understand.  If $Z_4=-Z_5$, we can perform a basis
transformation $H_2\to iH_2$, which yields $Z_4=Z_5$ and $\rho^Q\to i\rho^Q$.
The effect of the latter is to transform the pseudoscalar Yukawa coupling of the neutral
Higgs boson into a scalar Yukawa coupling.
The case in which the charged Higgs boson is
mass-degenerate with the CP-even neutral Higgs boson
corresponds to the case of ``twisted custodial symmetry''
introduced in \cite{herquet}.

Although this final conclusion is the same, we disagree with the
interpretation of ``twisted custodial symmetry'' given in
\Ref{herquet}.  As employed in \Ref{herquet}, the term ``twisted'' is
associated with a particular choice of the angle $\chi$ in
the SU(2)$_L\times$SU(2)$_R$ transformation law of $\mathbb{M}_2$
given in \eq{LRtrans}.  However, we have shown above that this angle
is basis-dependent and thus has no physical significance.  It is also argued
in \Ref{herquet} that custodial symmetry plays a critical role in
formulating the ``twisted'' scenario.  We have shown above
that the ``twisted'' scenario is a consequence of the two-fold ambiguity in
the definition of CP in the special case of $Z_6=Z_7=0$ (in the
absence of the Higgs--fermion Yukawa couplings).  This
ambiguity exists whether or not the custodial symmetry is present,
as shown in Section~\ref{sec:herquetcpcons}.  The custodial symmetry
is relevant in the following sense.  The possibility that
$m_{H^\pm}^2=m_{h^0}^2$ or $m_{H^\pm}^2=m_{H^0}^2$ arises precisely because
the custodial symmetry condition $Z_4=\pm \varepsilon_{5Q}|Z_5|$
allows for a negative sign in this relation if and only if $Z_6=Z_7=0$.

\subsection{Custodial symmetry in the Higgs-fermion sector for
the general 2HDM}

We now examine the Higgs-fermion Yukawa interactions in more detail,
and discuss the implications of custodial symmetry for this sector.

Custodial symmetry in the Yukawa Lagrangian was analyzed for the Type
I and II 2HDM in \cite{haberpomarol}.  Here we will shall examine the
general 2HDM without assuming additional conditions to restrict the
terms of the Higgs--fermion Yukawa Lagrangian.  In a generic basis,
the Higgs-fermion Lagrangian is given by \eq{yuklag}.
It is convenient to rewrite this Lagrangian in the following compact
form:
\beq \label{ymodeliiicute}
-\mathscr{L}_{\rm Y}=\overline{\qcal}_L \widetilde\Phi_{\bar{a}}\eta^U_a \mathcal{U}_R
+\overline{\qcal}_L\Phi_a \eta^{D\,\dagger}_{\bar{a}}D_R +{\rm h.c.}\,,
\eeq
where $\mathcal{U}\equiv K^\dagger U$, $\overline{\mathcal{Q}}_L \equiv
(\overline{\mathcal{U}}\,\,\,\overline{D})_L$, and
$\Phi_a\equiv\left(\begin{array}{c} \Phi^+_a \\ \Phi^0_a\end{array}\right)$.
In the Higgs basis, the corresponding Lagrangian given in \eq{yukhbasis}
can likewise be expressed compactly as:
\beq \label{hbasisymodeliii}
-\mathscr{L}_{\rm Y}=\overline{\qcal}_L (\widetilde H_1\kappa^U+
\widetilde H_2\rho^U) \mathcal{U}_R+\overline{\qcal}_L (H_1\kappa^{D\,\dagger}
+ H_2\rho^{D\,\dagger}) D_R +{\rm h.c.}\,,
\eeq
where the basis-invariant coupling matrices $\kappa^Q$ and $\rho^Q$ are defined
in \eq{kapparho}.

\subsubsection{Basis-dependent formulation of custodial symmetry in the Higgs--fermion sector}

We first examine the conditions for custodial symmetry of the
Higgs-fermion Yukawa interactions in the two basis choices of
Pomarol and Vega following the results of Section~\ref{pv}.
In Case 1, one writes the Yukawa interactions in terms of the $2\times 2$
matrix fields $M_1$ and $M_2$ defined in \eq{m1m2}.
The form of the Yukawa interactions invariant under
SU(2)$_{L}\times$SU(2)${}_{R}$ is then given by:
\beq \label{custinvy}
-\mathscr{L}_{\rm Y}=\eta_1 \overline{\qcal}_L~ M_1
~\left(\begin{array}{c}\mathcal{U}_R\\D_R\end{array}\right)+\eta_2
\overline{\qcal}_L~ M_2
~\left(\begin{array}{c}\mathcal{U}_R\\D_R\end{array}\right)+{\rm
  h.c.}\,,
\eeq
One can easily check that \eq{custinvy} is manifestly invariant
under the $SU(2)_L \times SU(2)_R$  transformations
\beq \label{transf}
{M}_i \rightarrow LM_i R^\dagger\,,\qquad
\overline{\qcal}_L  \rightarrow \overline{\qcal}_L L^\dagger\,,\qquad
\left(\begin{array}{c}\mathcal{U}_R\\D_R\end{array}\right)
\rightarrow R \left(\begin{array}{c}\mathcal{U}_R\\D_R\end{array}\right)\,.
\eeq
Comparing with \eq{ymodeliiicute} then yields the custodial symmetry
conditions,
\beq \label{caseonecon}
\eta_1=\eta^U_1= \eta^{D\dagger}_1\,,\qquad\qquad
\eta_2=\eta^U_2= \eta^{D\dagger}_2\,.
\eeq

In Case 2, one writes the Yukawa interactions in terms of the $2\times 2$
matrix fields
\beq
M_{12} \equiv (\widetilde{\Phi}_1 \,,\Phi_2)\,,\qquad\quad
M_{21} \equiv (\widetilde{\Phi}_2 \,,\Phi_1)\,,
\label{mat12def}
\eeq
which transforms under SU(2)$_L\times$SU(2)$_R$ as
\beq \label{cust12}
M_{12}\rightarrow L~M_{12}~R^\dagger\,,\qquad\quad
M_{21}\rightarrow L~M_{21}~R^\dagger\,.
\eeq
The form of the Yukawa interactions invariant under
SU(2)$_{L}\times$SU(2)${}_{R}$ is then given by:
\beq \label{custinvy2}
-\mathscr{L}_{\rm Y}=\eta_{12}\overline{\qcal}_L~ {M}_{12}
~\left(\begin{array}{c}\mathcal{U}_R\\D_R\end{array}\right)
+\eta_{21} \overline{\qcal}_L~ {M}_{21}
~\left(\begin{array}{c}\mathcal{U}_R\\D_R\end{array}\right)+{\rm h.c.}\,.
\eeq
Comparing with \eq{ymodeliiicute} then yields the custodial symmetry
conditions,
\beq \label{casetwocon}
\eta_{12}=\eta^U_1= \eta^{D\dagger}_2\,,\qquad\qquad
\eta_{21}=\eta^U_2= \eta^{D\dagger}_1\,.
\eeq

As in Section~~\ref{pv}, we can demonstrate that Cases 1 and 2 are
equivalent and simply represent different choices of the scalar field
basis.  To prove this assertion, we start from the basis of Case 2
and perform the basis transformation to the Higgs basis as specified
by the unitary matrix given by \eq{umatrix}.  Then, $\kappa^Q$,
$\rho^Q$ are related to the Yukawa coupling matrices $\eta_1^Q$, $\eta_2^Q$
via
\beq
\begin{pmatrix} \kappa^Q \\ \rho^Q \end{pmatrix}=
\frac{1}{\sqrt{2}}\begin{pmatrix} \phm e^{-i\theta} & \,\,\,
  e^{i\theta} \\ -ie^{-i\theta} & \,\,\,
  ie^{i\theta}\end{pmatrix}
\begin{pmatrix} \eta_1^Q \\ \eta_2^Q \end{pmatrix}\,.
\eeq
Using the Case 2 custodial symmetry conditions given in \eq{casetwocon},
it follows that:\footnote{Using \eq{kapparho} with $\widehat v_1=\widehat v_2^*=\frac{1}{\sqrt{2}}\,e^{i\theta}$,
one immediately reproduces \eq{kapQ}.  The corresponding result for $\rho^Q$ differs by an overall factor of
$i$.  But, we are free to redefine the Higgs-basis field $H_2\to iH_2$, which yields
$\rho^Q\to i\rho^Q$ in agreement with \eq{rhoQ}.}
\beqa
\kappa^U&=&\frac{1}{\sqrt{2}}\left(e^{i\theta}\eta_2^U+e^{-i\theta}\eta_1^U\right)=\kappa^{D\,\dagger}\,,\label{kapQ}\\
\rho^U &=&\frac{i}{\sqrt{2}}\left(e^{i\theta}\eta_2^U-e^{-i\theta}\eta_1^U\right)=\rho^{D\,\dagger}\,.\label{rhoQ}
\eeqa
That is, in the Higgs basis, the Case 1 custodial symmetry conditions given by \eq{caseonecon}
are satisfied.  Moreover, these conditions are preserved under any
additional O(2) basis transformation.
Thus, we have verified that Cases 1 and 2 of Pomarol and Vega, including the
SU(2)$_L\times$SU(2)$_R$ Higgs--fermion Yukawa interactions specified above, are equivalent and simply
represent different choices of the scalar field basis.

Using \eq{MQ}, the condition $\kappa^U=\kappa^{D\,\dagger}$ is equivalent to the equality
of the up and down-type fermion mass matrices,
\beq \label{mud}
M_U=M_D\,,
\eeq
which is clearly a basis-independent condition.  However, the condition $\rho^U=\rho^{D\,\dagger}$ is
not quite basis-independent, as $\rho^Q$ is a pseudo-invariant quantity.   At this stage,
\eq{rhoQ} has been obtained in a real Higgs basis.   In the next subsection, we obtain the
basis-independent conditions for custodial symmetry of the
Higgs--fermion Yukawa interactions.

\subsubsection{Basis-independent formulation of custodial symmetry in the Higgs--fermion sector}

Following Section~\ref{invcust}, we introduce the $2\times 2$ matrix fields in the Higgs
basis, denoted by $\mathbb{M}_1$ and $\mathbb{M}_2$ [cf.~\eq{matdefgen}], whose transformation
properties under SU(2)$_L\times$SU(2)$_R$ are given by \eqs{LRtrans}{Pdef}.  Note that the
transformation law for $\mathbb{M}_2$ includes a phase angle degree of freedom $\chi$
that reflects the freedom to rephase the Higgs-basis field $H_2$.
The form of the Yukawa interactions invariant under
SU(2)$_{L}\times$SU(2)${}_{R}$ is then given by:
\beq \label{custinvy3}
-\mathscr{L}_{\rm Y}=\kappa\, \overline{\qcal}_L~ \mathbb{M}_1
~\left(\begin{array}{c}\mathcal{U}_R\\D_R\end{array}\right)+\rho\,
\overline{\qcal}_L~ \mathbb{M}_2 P
~\left(\begin{array}{c}\mathcal{U}_R\\D_R\end{array}\right)+{\rm
  h.c.}\,,
\eeq
where $P\equiv{\rm diag}(e^{-i\chi}\,,\,e^{i\chi})$.    Comparing with \eq{hbasisymodeliii}
yields,
\beq \label{kr}
\kappa=\kappa^U=\kappa^{D\,\dagger}\,,\qquad\quad
\rho=e^{i\chi}\rho^U=e^{-i\chi}\rho^{D\,\dagger}\,.
\eeq
The first condition above implies $M_U=M_D$, which reproduces the result of \eq{mud}.
The second condition is basis independent in light of \eqs{rhotrans}{chitrans}.

For a generic custodial-symmetric Higgs--fermion Yukawa interaction, the matrices $\rho^U$
and $\rho^D$ are correlated according to \eq{kr}, but they can be non-diagonal and complex.
Thus, the custodial symmetry does \text{not} imply CP-conserving neutral Higgs--fermion couplings.
However, we can impose CP-conservation of the neutral Higgs-fermion interactions if
the conditions listed in \eq{cprho} are respected.   An equivalent set of conditions
(which are more useful as they do not rely on $Z_5$, $Z_6$ and $Z_7$) is given by \eq{etarho}.
In this case, it is convenient to use \eq{relate} to rewrite the
second condition of \eq{kr} as follows,\footnote{As usual, in Case III$^\prime$
of Table~\ref{Z60cond}, one must replace $\theta_{23}$ and $\eta^2$ with
$\overline{\theta}_{23}$ and $\overline{\eta}^2$, respectively
[cf.~\eq{bardef}].}
\beq \label{rhocondition}
e^{i\theta_{23}}\rho^U=\eta^2[e^{i\theta_{23}}\rho^D]^\dagger\,,
\eeq
which is manifestly basis-independent.  The sign factor $\eta^2$ is given in
Tables~\ref{Z6cond} and \ref{Z60cond}.
If $Z_6=Z_7=0$, then Table~\ref{Z60cond}
applies with $Z_7 e^{-i\theta_{23}}$ replaced by $\rho^Q e^{i\theta_{23}}$ ($Q=U,D$).
In particular, note that for a CP-conserving Higgs--fermion interaction,
$\Im(\rho^U e^{i\theta_{23}})=\Im(\rho^D e^{i\theta_{23}})=0$
if $\eta=+1$ and $\Re(\rho^U e^{i\theta_{23}})=\Re(\rho^D e^{i\theta_{23}})=0$ if $\eta=-1$.

\section{The oblique parameters $\boldsymbol{S}$, $\boldsymbol{T}$ and $\boldsymbol{U}$}
The $S$, $T$, and $U$ parameters, introduced by Peskin and
Takeuchi~\cite{peskin}, are independent ultraviolet-finite combinations of
radiative corrections to gauge boson two-point functions (the
so-called ``oblique'' corrections). The parameter $T$ is related to
the well known $\rho$-parameter of electroweak physics~\cite{veltman} by
$\rho - 1 = \alpha T$.  The oblique parameters can be expressed in
terms of the transverse part of the gauge boson two-point
functions~\cite{kennedy,erlerPDG}:\footnote{In the definition of $U$,
we differ slightly from that of \Ref{erlerPDG} by evaluating
$\Pi^{\rm new}_{Z\gamma}$ and $\Pi^{\rm new}_{\gamma\gamma}$
at $m_W^2$ (instead of $m_Z^2$).  This choice was advocated in
\Ref{kennedy}.}
\beqa
\label{obliqueSdef}
\overline\alpha\, T &\equiv& \frac{\Pi^{\rm new}_{WW}(0)}{m_W^2}
-\frac{\Pi^{\rm new}_{ZZ}(0)}{m_Z^2}\,, \\
\frac{\overline\alpha}{4\overline s^2_Z\overline c^2_Z}\,S &\equiv &
\frac{\Pi^{\rm new}_{ZZ}(m_Z^2)-\Pi^{\rm new}_{ZZ}(0)}{m_Z^2}
-\left(\frac{\overline c^2_W-\overline s^2_W}{\overline c_W\overline s_W}
\right)\frac{\Pi^{\rm new}_{Z\gamma}(m_Z^2)}{m_Z^2}
-\frac{\Pi^{\rm new}_{\gamma\gamma}(m_Z^2)}{m_Z^2}\,,\\
\frac{\overline\alpha}{4\overline s^2_Z\overline c^2_Z}\,(S+U) &\equiv &
\frac{\Pi^{\rm new}_{WW}(m_W^2)-\Pi^{\rm new}_{WW}(0)}{m_W^2}
-\frac{\bar{c}_W}{\bar{s}_W}\,\frac{\Pi^{\rm new}_{Z\gamma}(m_W^2)}{m_W^2}
-\frac{\Pi^{\rm new}_{\gamma\gamma}(m_W^2)}{m_W^2}\,,
\eeqa
where $\bar{s}_W\equiv \sin\theta_W(m_Z)$,
$\bar{c}_W\equiv \cos\theta_W(m_Z)$, and $\overline\alpha\equiv
\bar{g}^2\bar{s}^2\ls{Z}/(4\pi )$
are defined in the $\overline{\rm MS}$ scheme evaluated
at $m_Z$.  The $\Pi_{V_a V_b}^{\rm new}$ are the new physics contributions
to the one-loop $V_a$---$V_b$ vacuum polarization functions.
New physics contributions are defined as those that enter relative
to the Standard Model with a particular choice of
the Standard Model Higgs mass (denoted in what follows by $m_\phi$).
In \Ref{erlerPDG}, the value of $m_\phi=117$~GeV is chosen.

In the linear approximation~\cite{peskin}, which is a good
approximation if the energy scale new physics that contributes to
the oblique parameters is significantly larger than $m_Z$, we may
approximate:
\beq
\label{piA}
\Pi^{\rm new}_{ij}(q^2) \simeq A_{ij}(0) + q^2 F_{ij}(q^2)\,.
 \eeq
Electromagnetic gauge invariance implies that:\footnote{Although
$\Pi_{Z\gamma}(0)\neq 0$ (when all Standard Model
contributions are included),  the new physics contributions
to $\Pi_{Z\gamma}(0)$ considered in this paper can be shown to vanish
as a consequence of electromagnetic gauge invariance.}
\beq
A_{\gamma\gamma}(0)=A_{Z\gamma}(0)=0\,.
\eeq
In the linear approximation, the oblique parameters take the following
form~\cite{habertasi}
\beqa
\label{Tdef}\alpha T &\equiv& \frac{A_{WW}(0)}{m_W^2}- \frac{A_{ZZ}(0)}{m_Z^2}\\
\label{Sdef}
\frac{g^2}{16\pi c_W^2} S &\equiv& F_{ZZ}(m_Z^2)-F_{\gamma\gamma}(m_Z^2)
-\left(\frac{c_W^2-s_W^2}{s_W c_W}\right)F_{Z\gamma}(m_Z^2)\\
\frac{g^2}{16\pi}(S + U)&\equiv& F_{WW}(m_W^2)-F_{\gamma\gamma}(m_W^2)
-\frac{c_W}{s_W}F_{Z\gamma}(m_W^2)\,,
\eeqa
where we have dropped the bars for ease of
notation.

The $S$, $T$ and $U$ parameters are defined relative to the
Standard Model, so that $S=T=U=0$ corresponds to the Standard Model
with a particular ``reference'' choice of the Higgs mass $m_\phi$.
The 2HDM yields new contributions to $S$, $T$ and $U$ that in general shift
their values away from zero.  To compute the 2HDM contributions to
$S$, $T$ and $U$, we evaluate the relevant one-loop gauge boson polarization
functions in which the Higgs bosons appear as intermediate states,
and then subtract out the corresponding contributions due to the
Standard Model Higgs boson of mass $m_\phi$.  In our computations, we
initially leave $m_\phi$ as a free parameter.

\subsection{2HDM contributions to  $\boldsymbol{S}$,
$\boldsymbol{T}$ and $\boldsymbol{U}$}

The derivations of $S$, $T$ and $U$ are provided in
Appendix~\ref{app:one}.
The 2HDM contributions to $S$ are given by:
\beqa \label{seqn}\nonumber
 S&=& \frac{1}{\pi m_Z^2} \Biggl\{\sum_{k=1}^3 q_{k1}^2\biggl[
\mathcal{B}_{22}(m_Z^2;m_Z^2,m_k^2)- m_Z^2
\mathcal{B}_{0}(m_Z^2;m_Z^2,m_k^2)\biggr]\\[6pt] \nonumber
& &+q_{11}^2 \mathcal{B}_{22}(m_Z^2;m_2^2,m_3^2)
+q_{21}^2\mathcal{B}_{22}(m_Z^2;m_1^2,m_3^2)
+q_{31}^2\mathcal{B}_{22}(m_Z^2;m_1^2,m_2^2)\\[6pt]
& & - \mathcal{B}_{22}(m_Z^2;{m^2_{H^\pm}},{m^2_{H^\pm}})
-\mathcal{B}_{22}(m_Z^2;m_Z^2,m_\phi^2)
+ m_Z^2\mathcal{B}_{0}(m_Z^2;m_Z^2,m_\phi^2)\Biggr\}\,,
\eeqa
where
\beqa
\mathcal{B}_{22}(q^2;m_1^2,m_2^2) &\equiv&
B_{22}(q^2;m_1^2,m_2^2)-B_{22}(0;m_1^2,m_2^2)\,,\label{b22} \\[6pt]
\mathcal{B}_{0}(q^2;m_1^2,m_2^2) &\equiv& B_{0}(q^2;m_1^2,m_2^2)-B_{0}(0;m_1^2,m_2^2)\,,\label{bzero}
\eeqa
and the $m_k$ are the masses of the neutral Higgs $h_k$ ($k =1,2,3$).
The functions $B_{22}$ and $B_{0}$ appearing in \eqs{b22}{bzero},
defined in ref.~\cite{passario}, arise in the evaluation of
the two-point loop integrals.  They can be evaluated in dimensional
regularization using the
following formulae of ref.~\cite{habertasi}:
\beqa
\label{eqnb}
B_{22}(q^2;m_1^2,m_2^2) &=& \tfrac{1}{4}(\Delta+1)[m_1^2+m_2^2-\tfrac{1}{3}q^2]-\half\int^1_0\,dx X \ln(X-i\epsilon), \\
B_{0}(q^2;m_1^2,m_2^2) &=& \Delta-\int^1_0\,dx \ln(X-i\epsilon),
\eeqa
where
\beq
X \equiv m_1^2 x + m_2^2(1-x) -q^2x(1-x)\,,\qquad\quad
\Delta \equiv \frac{2}{4-d}+\ln 4\pi-\gamma\,,
\eeq
in $d$ space-time dimensions.  Note that
\beq
B_{22}(q^2;m_1^2,m_2^2)=B_{22}(q^2;m_2^2,m_1^2)\,,\qquad\qquad
B_{0}(q^2;m_1^2,m_2^2)=B_{0}(q^2;m_2^2,m_1^2)\,.
\eeq

The 2HDM contributions to $T$ and $U+S$ are given by:
\beqa
T &=& \frac{1}{16\pi m_W^2s_W^2} \Biggl\{\,\sum_{k=1}^3|q_{k2}|^2
\mathcal{F}({m^2_{H^\pm}},m_k^2)-q_{11}^2 \mathcal{F}(m_2^2,m_3^2)
-q_{21}^2 \mathcal{F}(m_1^2,m_3^2)-q_{31}^2 \mathcal{F}(m_1^2,m_2^2)\nonumber\\
& &+\sum_{k=1}^3 q_{k1}^2\biggl[\mathcal{F}(m_W^2,m_k^2)-\mathcal{F}(m_Z^2,m_k^2)
-4m_W^2 B_0(0;m_W^2,m_k^2)+4m_Z^2 B_0(0;m_Z^2,m_k^2)\bigr]\biggr]\nonumber\\
&&+\mathcal{F}(m_Z^2,m_\phi^2) -\mathcal{F}(m_W^2,m_\phi^2)
+4m_W^2 B_0(0;m_W^2,m_\phi^2)-4m_Z^2 B_0(0;m_Z^2,m_\phi^2)\Biggr\},
\label{thdmt}
\eeqa
\beqa
\hspace{-0.5in}
S+U &=&\frac{1}{\pi m_W^2} \Biggl\{-\sum_{k=1}^3 q_{k1}^2 m_W^2
\mathcal{B}_{0}(m_W^2;m_W^2,m_k^2)
+ m_W^2 \mathcal{B}_{0}(m_W^2;m_W^2,m_\phi^2)
- \mathcal{B}_{22}(m_W^2;m_W^2,m_\phi^2) \nonumber\\
& & + \sum_{k=1}^3\biggl[q_{k1}^2\mathcal{B}_{22}(m_W^2;m_W^2,m_k^2)
+|q_{k2}|^2 \mathcal{B}_{22}(m_W^2;{m^2_{H^\pm}},m_k^2)\biggr]
-2 \mathcal{B}_{22}(m_W^2;{m^2_{H^\pm}},{m^2_{H^\pm}})\Biggr\}\,,\label{tsu} \eeqa
where the function $\mathcal{F}$ is defined by
\beq
\mathcal{F}(m_1^2,m_2^2) \equiv \half(m_1^2+m_2^2)-\frac{m_1^2m_2^2}{m_1^2-m_2^2}
\ln\left(\frac{m_1^2}{m_2^2}\right)\,.
\eeq
Note that
\beq
\mathcal{F}(m_1^2,m_2^2)=\mathcal{F}(m_2^2,m^2_1)\,,\qquad\qquad
\mathcal{F}(m^2,m^2)=0\,.
\eeq
One can simplify the expression for $T$ by making use of the identity:
\beq \label{abid}
m_1^2 B_0(0;m_1^2,m_3^2)-m_2^2 B_0(0;m_2^2,m_3^2)=
\mathcal{F}(m_1^2,m_3^2)-\mathcal{F}(m_2^2,m_3^2)
+A_0(m_1^2)-A_0(m_2^2)-\half(m_1^2-m_2^2)\,,
\eeq
where
\beq
A_0(m^2)\equiv m^2(\Delta+1-\ln m^2)\,.
\eeq
Applying the identity of \eq{abid} in the expression for $T$ then yields:
\beqa
T &=& \frac{1}{16\pi m_W^2s_W^2} \Biggl\{\,\sum_{k=1}^3|q_{k2}|^2
\mathcal{F}({m^2_{H^\pm}},m_k^2)-q_{11}^2 \mathcal{F}(m_2^2,m_3^2)
-q_{21}^2 \mathcal{F}(m_1^2,m_3^2)
-q_{31}^2 \mathcal{F}(m_1^2,m_2^2)\nonumber\\
& &+3\sum_{k=1}^3 q_{k1}^2\bigl[\mathcal{F}(m_Z^2,m_k^2)-\mathcal{F}(m_W^2,m_k^2)\bigr]
-3\bigl[\mathcal{F}(m_Z^2,m_\phi^2) -\mathcal{F}(m_W^2,m_\phi^2)\bigr]\Biggr\}\,, \label{thdmt2}
\eeqa
which reproduces the result first obtained in \Ref{pomarol}.
In particular, note that terms in \eq{abid} of the form
$A_0(m_1^2)-A_0(m_2^2)-\half(m_1^2-m_2^2)$
are independent of $m_3^2$ and hence cancel out exactly in \eq{thdmt2}.

Using \eqs{seqn}{tsu}, we can isolate the $U$-parameter,
\beqa
U&=&\mathcal{G}(m_W^2)-\mathcal{G}(m_Z^2)+\frac{1}{\pi m_W^2}\left\{\sum_{k=1}^3 [|q_{k2}|^2
\mathcal{B}_{22}(m_W^2;m^2_{H^\pm},m_k^2)- 2\mathcal{B}_{22}(m_W^2;{m^2_{H^\pm}},{m^2_{H^\pm}})\right\}
\nonumber \\
&&
-\frac{1}{\pi m_Z}\Biggl\{q_{11}^2 \mathcal{B}_{22}(m_Z^2;m_2^2,m_3^2)
+q_{21}^2\mathcal{B}_{22}(m_Z^2;m_1^2,m_3^2)
+q_{31}^2\mathcal{B}_{22}(m_Z^2;m_1^2,m_2^2)
- \mathcal{B}_{22}(m_Z^2;{m^2_{H^\pm}},{m^2_{H^\pm}})\Biggr\}\,,\nonumber \\
\phantom{line}\label{uparm}
\eeqa
where
\beq \label{gcaldef}
\mathcal{G}(m_V^2)\equiv \frac{1}{\pi m_V^2}\left\{\sum_{k=1}^3 q_{k1}^2\left[%
\mathcal{B}_{22}(m_V^2; m_V^2,m_k^2)
-m_V^2\mathcal{B}_0(m_V^2;m_V^2,m_k^2)\right]-\mathcal{B}_{22}(m_V^2; m_V^2,m_\phi^2)
+m_V^2\mathcal{B}_0(m_V^2;m_V^2,m_\phi^2)\right\}\,.
\eeq

\subsection{$\boldsymbol{S}$, $\boldsymbol{T}$, and $\boldsymbol{U}$
in the CP-conserving limit}

To obtain $S$, $T$ and $U$ in the CP-conserving limit,
we must identify the values of $q_{k1}$ and
$q_{k2}$ and the corresponding neutral Higgs masses $m_k$ in the
CP-conserving limit.  For example, in \Ref{haberoneil}, the
values of the $q_{k\ell}$ and $m_k$ were obtained
for Cases I, IIa and IIb defined in Section~\ref{subsecz6}.  For the
reader's convenience, we reproduce those results here in
Tables~X--XII.   These three cases correspond to
three different mass orderings of the neutral Higgs bosons
(by assumption, we assume here that $m_{h_1}<m_{h_2}<m_{h_3}$).

%
%
\vspace{-0.1in}
%
\begin{table}[t!]
\begin{minipage}[t]{2.0in}
\centering
\parbox[t]{1.75in}\caption{\phantom{Case} Case~I: $s_{13}=0$.
In a real basis,
$e^{-i\theta_{23}}= {\rm sgn}~Z_6\equiv\varepsilon_6$.
The neutral Higgs fields are $h_1=\hl$, $h_2=-\varepsilon_6\hh$
and $h_3=\varepsilon_6\ha$. The angular factors are
$c_{12}=\sbma$ and $s_{12}=-\varepsilon_6\cbma$.\label{cpcons1}} \vskip 0.08in
\begin{tabular}{|c||c|c|}\hline
$\phaa k\phaa $ &\phaa $q_{k1}\phaa $ & \phaa $\phm q_{k2} \phaa $ \\ \hline
$1$ & $c_{12}$ & $-s_{12}$ \\
$2$ & $s_{12}$ & $\phm c_{12}$ \\
$3$ & $0$ & $\phm i$ \\ \hline
\end{tabular}
\end{minipage}
\hfill
\begin{minipage}[t]{2.0in}
\centering
\parbox[t]{1.75in}\caption{\phantom{Case} Case~IIa: $s_{12}=0$.
In a real basis, $e^{-i\theta_{23}}=i\,{\rm sgn}~Z_6\equiv i\varepsilon_6$.
The neutral Higgs fields are $h_1=\hl$,
$h_2=\varepsilon_6\ha$ and $h_3=\varepsilon_6\hh$.
The angular factors are
$c_{13}=\sbma$ and $s_{13}=\varepsilon_6\cbma$.\label{cpcons2}}
\vskip 0.08in
\begin{tabular}{|c||c|c|}\hline
$\phaa k\phaa $ &\phaa $q_{k1}\phaa $ & \phaa $\phm q_{k2} \phaa $ \\ \hline
$1$ & $c_{13}$ & $ -is_{13} $\\
$2$ & $0$ & $\phm 1$ \\
$3$ & $s_{13}$ & $\phm i c_{13}$ \\ \hline
\end{tabular}
\end{minipage}
\hfill
\begin{minipage}[t]{2.0in}
\centering
\parbox[t]{1.75in}\caption{\phantom{Case} Case~IIb: $c_{12}=0$.
In a real basis, $e^{-i\theta_{23}}=i\,{\rm sgn}~Z_6\equiv i\varepsilon_6$.
The neutral Higgs fields are $h_1=\varepsilon_6\ha$,
$h_2=-\hl$ and $h_3=\varepsilon_6\hh$. The angular factors are
$c_{13}=\sbma$ and $s_{13}=\varepsilon_6\cbma$.\label{cpcons3}}
\vskip 0.08in
\begin{tabular}{|c||c|c|}\hline
$\phaa k\phaa $ &\phaa $\phm q_{k1}\phaa $ & \phaa $q_{k2} \phaa $ \\ \hline
$1$ & $\phm 0$ & $1$ \\
$2$ & $-c_{13}$ & $is_{13}$ \\
$3$ & $\phm s_{13}$ & $i c_{13}$ \\ \hline
\end{tabular}
\end{minipage}
\end{table}
%
\vskip 0.1in

In the CP-conserving limit, it is traditional to employ the factors
$\cos(\beta-\alpha)$ and $\sin(\beta-\alpha)$, where $\alpha$ is the
mixing angle obtained from diagonalizing the $2\times 2$ CP-even
Higgs squared-mass matrix in a generic real basis~\cite{hhg}.
These angle factors are
related to the $q_{k\ell}$ as indicated in the captions to Tables X--XII.
The results for $S$, $T$ and $U$ do not depend on which Case is
employed to compute the $q_{k\ell}$, since the different cases simply
correspond to different mass-orderings of the neutral Higgs bosons.
Plugging in the values of the $q_{k\ell}$ parameters from
any of the Cases exhibited in Tables
X--XII into \eqst{seqn}{tsu}, and choosing the reference Higgs
mass $m_\phi=m_{h^0}$ (where $h^0$ is the lightest CP-even
neutral Higgs boson), we obtain:
\beqa \nonumber
S&=& \frac{1}{\pi m_Z^2} \biggl\{
-\mathcal{B}_{22}(m_Z^2;{m^2_{H^\pm}},{m^2_{H^\pm}})+\sin^2(\beta-\alpha)
\mathcal{B}_{22}(m_Z^2;m_{H^0}^2,m_{A^0}^2)
\nonumber \\
& & \qquad\qquad+\cos^2(\beta-\alpha)
\bigl[\mathcal{B}_{22}(m_Z^2;m_{h^0}^2,m_{A^0}^2)
 + \mathcal{B}_{22}(m_Z^2;m_Z^2,m_{H^0}^2)-
\mathcal{B}_{22}(m_Z^2;m_Z^2,m_{h^0}^2)\nonumber\\
& &\qquad\qquad\qquad\qquad\qquad - m_Z^2\mathcal{B}_{0}(m_Z^2;m_Z^2,m_{H^0}^2)
+m_Z^2\mathcal{B}_{0}(m_Z^2;m_Z^2,m_{h^0}^2)\bigl]
\biggr\}\,,\\
T &=& \frac{1}{16\pi s_W^2 m_W^2} \biggl\{\mathcal{F}({m^2_{H^\pm}},m_{A^0}^2)
+\sin^2(\beta-\alpha)\bigl[\mathcal{F}({m^2_{H^\pm}},m_{H^0}^2)
-\mathcal{F}(m_{A^0}^2,m_{H^0}^2)\bigr]\nonumber \\[6pt]
&
&\qquad\qquad+\cos^2(\beta-\alpha)\bigl[\mathcal{F}({m^2_{H^\pm}},m_{h^0}^2)
-\mathcal{F}(m_{A^0}^2,m_{h^0}^2)
+\mathcal{F}(m_W^2,m_{H^0}^2)-\mathcal{F}(m_W^2,m_{h^0}^2)\nonumber\\[6pt]
& & \qquad\qquad\qquad\qquad\qquad
-\mathcal{F}(m_Z^2,m_{H^0}^2)+\mathcal{F}(m_Z^2,m_{h^0}^2)+4m_Z^2
B_0(0;m_Z^2,m_{H^0}^2)-4m_Z^2 B_0(0;m_Z^2,m_{h^0}^2)
\nonumber\\[6pt]& &\qquad\qquad\qquad\qquad\qquad
-4m_W^2 B_0(0;m_W^2,m_{H^0}^2)+4m_W^2 B_0(0;m_W^2,m_{h^0}^2)\bigr]
\biggr\}\,,\\[6pt]
S+ U &=& \frac{1}{\pi m_W^2} \biggl\{
\mathcal{B}_{22}(m_W^2;{m^2_{H^\pm}},m_{A^0}^2)
-2\mathcal{B}_{22}(m_W^2;{m^2_{H^\pm}},{m^2_{H^\pm}})
+\sin^2(\beta-\alpha)\mathcal{B}_{22}(m_W^2;{m^2_{H^\pm}},m_{H^0}^2)
\biggr.\nonumber \\[6pt]
&&\qquad\qquad
+\cos^2(\beta-\alpha)\bigl[\mathcal{B}_{22}(m_W^2;m_{h^0}^2,{m^2_{H^\pm}})
+\mathcal{B}_{22}(m_W^2;m_W^2,m_{H^0}^2)
-\mathcal{B}_{22}(m_W^2;m_W^2,m_{h^0}^2)\nonumber\\[6pt]
& &\qquad\qquad\qquad\qquad\qquad
+m_W^2\mathcal{B}_{0}(m_W^2;m_W^2,m_{H^0}^2)
-m_W^2\mathcal{B}_{0}(m_W^2;m_W^2,m_{h^0}^2)\bigr]\biggr\}\,.
\eeqa
The above results agree with
results previously obtained in \refs{habertasi}{Froggatt}.\footnote{In
eq.~(3.47) of \Ref{habertasi}, there is a typographical error in the
expression for $T$.  The right bracket at the end of the third line of
eq.~(3.47) is misplaced and should appear at the end of the fifth line.}

\subsection{$\boldsymbol{T}$ and $\boldsymbol{U}$ and the custodial limit}

In the custodial symmetric limit, both the $T$ and $U$-parameters must vanish~\cite{peskin}.
Using \eq{thdmt2}, we can verify this behavior.  In the 2HDM,
custodial symmetry-breaking arises from two sources.  The first source
is the gauged U(1)-hypercharge interactions that are always present.
The second source is the custodial symmetry-breaking terms of the
scalar potential.  Let us look at both sources in turn.

We can formally restore custodial symmetry in the gauge sector by
taking the limit of $g^\prime \rightarrow 0$ (in which case, $m_Z = m_W$).
If we set $m_W=m_Z$ in \eq{thdmt2}, we see that
the second line of this equation vanishes.  That is, the second line
of \eq{thdmt2} is a consequence of the gauged U(1)-hypercharge interactions.
Formally, this term must be proportional to $g^\prime$.  Noting that
\beq \label{alpha}
\frac{\alpha}{s_W^2 m_W^2}=\frac{g^2}{4\pi^2 m_W^2}=\frac{g^{\prime\,2}}{4\pi(m_Z^2-m_W^2)}\,,
\eeq
it follows that
\beqa
\alpha T &=& \frac{3g^{\prime\,2}}{64\pi^2(m_Z^2-m_W^2)}\Biggl\{
\sum_{k=1}^3 q_{k1}^2\bigl[\mathcal{F}(m_Z^2,m_k^2)
-\mathcal{F}(m_W^2,m_k^2)\bigr]
-\mathcal{F}(m_Z^2,m_\phi^2) +\mathcal{F}(m_W^2,m_\phi^2)\Biggr\}
\nonumber\\
&&+\frac{g^2}{64\pi m_W^2} \Biggl\{\,\sum_{k=1}^3|q_{k2}|^2
\mathcal{F}({m^2_{H^\pm}},m_k^2)-q_{11}^2 \mathcal{F}(m_2^2,m_3^2)
-q_{21}^2 \mathcal{F}(m_1^2,m_3^2)-q_{31}^2 \mathcal{F}(m_1^2,m_2^2)\Biggr\}
\,. \label{thdmt3}
\eeqa
In this form, one can explicitly identify the term in $T$ proportional
to $g^\prime$ as the piece that arises from the
gauged U(1)-hypercharge interactions.\footnote{Note that
the expression in \eq{thdmt3} that multiplies $g^{\prime\,2}$
approaches a finite limit as $m_Z\to m_W$.  Hence, the entire term
does indeed vanish in the custodial symmetry limit as expected.}

The term proportional to $g$ in \eq{thdmt3} arises as a consequence of
custodial symmetry breaking in the scalar potential.  Thus, we should
verify that this term vanishes in the limit of a custodial symmetric
scalar potential.  In this limit, CP is conserved, so we may use
the results of Table X--XII to evaluate \eq{thdmt3} [any one of the three Cases can
be used as noted in the previous subsection].  For convenience,
we again choose $m_\phi=m_{h^0}$, in which case,
\beqa
\alpha T &=& \frac{3g^{\prime\,2}\cos^2(\beta-\alpha)}
{64\pi^2(m_Z^2-m_W^2)}\Biggl\{
\mathcal{F}(m_Z^2,m_{H^0}^2)
-\mathcal{F}(m_W^2,m_{H^0}^2)
-\mathcal{F}(m_Z^2,m_{h^0}^2) +\mathcal{F}(m_W^2,m_{m_h^0})\Biggr\}
\nonumber\\[6pt]
&&\qquad\quad +\frac{g^2}{64\pi m_W^2} \Biggl\{
\mathcal{F}({m^2_{H^\pm}},m_{A^0}^2)+\sin^2(\beta-\alpha)\left[
\mathcal{F}({m^2_{H^\pm}},m_{H^0}^2)-\mathcal{F}({m^2_{A^0}},m_{H^0}^2)\right]
\nonumber \\
&&\qquad\qquad\qquad\qquad\qquad\quad
+\cos^2(\beta-\alpha)\left[
\mathcal{F}({m^2_{H^\pm}},m_{h^0}^2)-\mathcal{F}({m^2_{A^0}},m_{h^0}^2)\right]
\Biggr\}\,. \label{thdmtcp}
\eeqa
For a custodial symmetric scalar potential, the term proportional to
$g^2$ in \eq{thdmtcp} must vanish, i.e.
\beq \label{vanish}
\mathcal{F}({m^2_{H^\pm}},m_{A^0}^2)+\sin^2(\beta-\alpha)\left[
\mathcal{F}({m^2_{H^\pm}},m_{H^0}^2)-\mathcal{F}({m^2_{A^0}},m_{H^0}^2)\right]
+\cos^2(\beta-\alpha)\left[
\mathcal{F}({m^2_{H^\pm}},m_{h^0}^2)-\mathcal{F}({m^2_{A^0}},m_{h^0}^2)\right]
=0\,.
\eeq

In Section~\ref{twisted}, we demonstrated that for a custodial symmetric scalar potential,
$m_{H^\pm}^2=m_{A^0}^2$ [cf.~\eq{equalmasses}], in nearly all cases.
Indeed, for $\sin(\beta-\alpha)\cos(\beta-\alpha)\neq 0$, the only
solution to \eq{vanish} is $m_{H^\pm}^2=m_{A^0}^2$.  However, we
identified the special case of $Z_6=Z_7=0$ in which the
custodial symmetric scalar potential could also yield
$m_{H^\pm}^2=m_{h^0}^2$ or $m_{H^\pm}^2=m_{H^0}^2$[cf.~\eq{equalmasses2}].
For example, by comparing Table X with Tables~\ref{d1} and \ref{d2},
we see that the special case of $Z_6=Z_7=0$ corresponds to
$\cos(\beta-\alpha)=0$ and $\sin(\beta-\alpha)=0$, respectively.  In these
two cases, \eq{vanish} reduces to the following two equations:
\beqa
&& \mathcal{F}({m^2_{H^\pm}},m_{A^0}^2)+
\mathcal{F}({m^2_{H^\pm}},m_{H^0}^2)-\mathcal{F}({m^2_{A^0}},m_{H^0}^2)=0\,,
\qquad\quad \text{if}~\cos(\beta-\alpha)=0\,,\label{vanishI} \\[6pt]
&& \mathcal{F}({m^2_{H^\pm}},m_{A^0}^2)+
\mathcal{F}({m^2_{H^\pm}},m_{h^0}^2)-\mathcal{F}({m^2_{A^0}},m_{h^0}^2)
=0\qquad\qquad\, \text{if}~\sin(\beta-\alpha)=0\,.\label{vanishII}
\eeqa
Of course, $m_{H^\pm}^2=m_{A^0}^2$  remains as a possible solution to
both of the above equations.  But, for each equation above, a second
solution exists, namely  $m_{H^\pm}^2=m_{H^0}^2$ for \eq{vanishI} and
$m_{H^\pm}^2=m_{h^0}^2$ for \eq{vanishII}.  Thus, we confirm that
in the case of $Z_6=Z_7=0$, the custodial symmetric mass relations
identified in \eq{equalmasses2} are consistent with the vanishing of
the $T$ parameter (in the limit of $g^\prime=0$ and $m_W=m_Z$).

So far, we have focused on the contributions of the
bosonic sector of the 2HDM to the $T$ parameter.  In addition,
there are also fermion loop contributions since the Higgs--fermion
Yukawa interactions can also violate the custodial symmetry.
However, at one-loop, the only custodial-violating contribution to the
$T$-parameter arises due to the non-degeneracy of the up and down
fermion mass matrices.  But, this effect also is present in the
Standard Model with one Higgs doublet, as first noted in \Ref{veltman}.
New custodial symmetry breaking effects in the Higgs--fermion
Yukawa interactions that are present due to the second Higgs doublet
must involve $\rho^Q$. Since the dependence of the gauge boson
polarization functions on $\rho^Q$ only enters at two loops
in the perturbative expansion, we shall not include them in the present analysis.
It would be an interesting exercise to verify that the corresponding
two-loop contributions to the $T$ parameter vanish exactly in the
custodial symmetric limit specified in \eq{rhocondition}.

The analysis of the $U$-parameter is similar.  Using \eqs{uparm}{gcaldef},
we see that when $m_W=m_Z$, the general expression for $U$ reduces to:
\beqa
U&=&\frac{1}{\pi m_W^2}\Biggl\{\sum_{k=1}^3 [|q_{k2}|^2
\mathcal{B}_{22}(m_W^2;m^2_{H^\pm},m_k^2)-q_{11}^2 \mathcal{B}_{22}(m_W^2;m_2^2,m_3^2)
-q_{21}^2\mathcal{B}_{22}(m_W^2;m_1^2,m_3^2) \nonumber \\
&&\qquad\qquad\qquad
-q_{31}^2\mathcal{B}_{22}(m_W^2;m_1^2,m_2^2)
- \mathcal{B}_{22}(m_W^2;{m^2_{H^\pm}},{m^2_{H^\pm}})\Biggr\}\,.
\eeqa
In the CP-conserving limit (with $m_W=m_Z$),
\beqa
U&=&\frac{1}{\pi m_W^2}\Biggl\{\mathcal{B}_{22}(m_W^2;m_{H^\pm}^2,m_A^2)-
\mathcal{B}_{22}(m_W^2;m_{H^\pm}^2,m_{H^\pm}^2)\nonumber \\
&& \qquad
+\sin^2(\beta-\alpha)\left[\mathcal{B}_{22}(m_W^2;m_{H^\pm}^2,m_{H^0}^2)-
\mathcal{B}_{22}(m_W^2;m_{H^0}^2,m_{A^0}^2)\right] \nonumber \\[5pt]
&& \qquad
+\cos^2(\beta-\alpha)\left[\mathcal{B}_{22}(m_W^2;m_{H^\pm}^2,m_{h^0}^2)-
\mathcal{B}_{22}(m_W^2;m_{h^0}^2,m_{A^0}^2)\right]\Biggr\}\,.\label{UCP}
\eeqa
If $\sin(\beta-\alpha)\cos(\beta-\alpha)\neq 0$, then $U=0$ if and only
if $m_{H^\pm}^2=m_{A^0}^2$.  In the special case of $Z_6=Z_7=0$,
it follows that either $\sin(\beta-\alpha)=0$ or $\cos(\beta-\alpha)=0$,
in which case $U=0$ when
\beqa
&&\mathcal{B}_{22}(m_W^2;m_{H^\pm}^2,m_A^2)-
\mathcal{B}_{22}(m_W^2;m_{H^\pm}^2,m_{H^\pm}^2)+\mathcal{B}_{22}(m_W^2;m_{H^\pm}^2,m_{H^0}^2)-
\mathcal{B}_{22}(m_W^2;m_{H^0}^2,m_{A^0}^2)=0\,,\nonumber \\
&&\hspace{4in} \text{if}~\cos(\beta-\alpha)=0\,,\\[6pt]
\label{UvanishI} &&
\mathcal{B}_{22}(m_W^2;m_{H^\pm}^2,m_A^2)-
\mathcal{B}_{22}(m_W^2;m_{H^\pm}^2,m_{H^\pm}^2)+\mathcal{B}_{22}(m_W^2;m_{H^\pm}^2,m_{h^0}^2)-
\mathcal{B}_{22}(m_W^2;m_{h^0}^2,m_{A^0}^2)=0\,,\nonumber \\
&&\hspace{4in} \text{if}~\sin(\beta-\alpha)=0\,.
\label{UvanishII}
\eeqa
Of course, $m_{H^\pm}^2=m_{A^0}^2$  remains as a possible solution to
both of the above equations.  But, for each equation above, a second
solution exists, namely  $m_{H^\pm}^2=m_{H^0}^2$ for \eq{UvanishI} and
$m_{H^\pm}^2=m_{h^0}^2$ for \eq{UvanishII}.  Thus, we confirm that
in the case of $Z_6=Z_7=0$, the custodial symmetric mass relations
identified in \eq{equalmasses2} are consistent with the vanishing of
the $U$ parameter.

\subsection{$\boldsymbol{S}$, $\boldsymbol{T}$ and
$\boldsymbol{U}$ in the decoupling limit}

In the decoupling limit of the 2HDM~\cite{decoupling}, one
neutral Higgs boson, conventionally denoted by $h_1$, is kept
light, with mass $m_1\lsim\mathcal{O}(m_Z)$, and the other neutral
Higgs bosons $h_2$ and $h_3$ and the charged Higgs boson $H^\pm$ have
masses of order $\Lambda\gg m_Z$.  In Appendix~\ref{app:two},
the Higgs masses and invariant mixing angles are evaluated in the decoupling limit.
The resulting masses are given in \eqst{massrelations1}{secondorder} and
the invariant mixing angles are given in \eq{sandc}.
The explicit forms for the $q_{ki}$ given in Table~\ref{tabq} are
given by
\beq \label{qdecoupling}
q_{11}^2\simeq |q_{22}|^2\simeq |q_{32}|^2
\simeq 1-\mathcal{O}\left(\frac{v^4}{\Lambda^4}\right)\,,\qquad\quad
q_{21}^2\simeq q_{31}^2\simeq |q_{12}|^2\simeq
\mathcal{O}\left(\frac{v^4}{\Lambda^4}\right)\,.
\eeq

We now turn to the computation of the oblique parameters in the
decoupling limit.  As a first step, we eliminate $q_{11}$ in favor of
$q_{21}$ and $q_{31}$ using the identity,
\beq \label{qid}
\sum_{i=1}^3 q_{k1}^2=1\,.
\eeq
It is also convenient to set the reference Higgs mass $m_\phi=m_1$.
Then, one can write the general expression for $S$ as follows:
\beqa
S&=& \frac{1}{\pi m_Z^2} \biggl\{\mathcal{B}_{22}(m_Z^2;m_2^2,m_3^2)-
\mathcal{B}_{22}(m_Z^2;{m^2_{H^\pm}},{m^2_{H^\pm}})\nonumber\\
& &\qquad\,\,\, +q_{21}^2\left[\mathcal{B}_{22}(m_Z^2;m_Z^2,m_2^2)+
\mathcal{B}_{22}(m_Z^2;m_1^2,m_3^2) - m_Z^2 \mathcal{B}_{0}(m_Z^2;m_Z^2,m_2^2) \right]\nonumber\\
&&\qquad\,\,\,  +q_{31}^2\left[\mathcal{B}_{22}(m_Z^2;m_Z^2,m_3^2)+
\mathcal{B}_{22}(m_Z^2;m_1^2,m_2^2) - m_Z^2 \mathcal{B}_{0}(m_Z^2;m_Z^2,m_3^2) \right] \nonumber\\
& &\qquad\,\,\, -(q_{21}^2+q_{31}^2)\left[\mathcal{B}_{22}(m_Z^2;m_2^2,m_3^2)+
\mathcal{B}_{22}(m_Z^2;m_Z^2,m_1^2)- m_Z^2\mathcal{B}_{0}(m_Z^2;m_Z^2,m_1^2)\right]\biggr\}\,.\eeqa
Employing the decoupling limit conditions of \eq{qdecoupling},
\beq \label{sdec}
S= \frac{1}{\pi m_Z^2} \biggl[\mathcal{B}_{22}(m_Z^2;m_2^2,m_3^2)
- \mathcal{B}_{22}(m_Z^2;{m^2_{H^\pm}},{m^2_{H^\pm}})
+\mathcal{O}\left(\frac{v^4}{\Lambda^4}\right)\biggr]\,.
\eeq
Using \eqst{m1approx}{m3approx} and noting the expansion,
\beq \label{b22expand}
\mathcal{B}_{22}(m_Z^2;\LL + a v^2,\LL+bv^2) = -\tfrac{1}{12}m_Z^2
\left[\Delta - \ln\LL - \frac{(a+b)v^2}{2\Lambda^2} + \frac{m_Z^2}{10\Lambda^2} +
\mathcal{O}\left(\frac{v^4}{\Lambda^4}\right)\right]\,,
\eeq
it then follows that:
\beq
\label{decS}
S \simeq \frac{m_2^2+m_3^2-2 {m^2_{H^\pm}}}{24\pi m_3^2}=\frac{Z_4
  v^2}{24\pi m_3^2}\,,
\eeq
where terms of $\mathcal{O}(v^2/\Lambda^4)$ have been neglected.

One can evaluate $T$ in a similar manner.  Setting $m_\phi=m_1$ in
\eq{thdmt}, and employing the decoupling limit conditions of \eq{qdecoupling},
we obtain:
\beq
T = \frac{1}{16\pi s_W^2 m_W^2}\left[\mathcal{F}({m^2_{H^\pm}},m_2^2)+
\mathcal{F}({m^2_{H^\pm}},m_3^2)-\mathcal{F}(m_2^2,m_3^2)+
\mathcal{O}\left(\frac{v^4}{\Lambda^4}\right)\right]\,.
\eeq
Using \eqst{m1approx}{m3approx} and noting the expansion,
\beq
\mathcal{F}(\LL + a v^2, \LL+ b v^2) \simeq\tfrac{1}{6} v^2
\left[\frac{(a-b)^2}{\Lambda^2}
+\mathcal{O}\left(\frac{v^4}{\Lambda^4}\right)\right]\,,
\label{expF}
\eeq
it then follows that:
\beq T \label{decT} \simeq
\frac{({m^2_{H^\pm}} - m_3^2)({m^2_{H^\pm}} - m_2^2)}
{48\pi s_W^2 m_W^2 m_3^2}=\frac{(Z_4^2-|Z_5|^2)v^2}{48\pi e^2 m_3^2}\,,
\eeq
after using $e=gs_W=2s_W m_W/v$.
In the custodial limit, $H^\pm$ is degenerate in mass with either $h_2$ or $h_3$,\footnote{In
general $H^\pm$ is degenerate in mass with $A^0$ whose identity (either $h_2$ or $h_3$)
is determined from \eqs{identity1}{identity2}.  If $Z_6=Z_7=0$, $H^\pm$ may instead be
degenerate in mass with $H^0$ in the custodial limit
as noted in \eq{equalmasses2}.  Of course, in the decoupling limit
$H^\pm$ can never be degenerate in mass with $h^0$ since $m_{h^0}\ll m_{H^\pm}$.}
in which case $T$ must vanish.  Indeed, \eq{decT} satisfies this requirement.
This result is not surprising since to leading order in the decoupling limit,
we may set $\cos(\beta-\alpha)=0$ in which case \eq{vanishI} applies.

As a check of the above computations, one can use
eqs.~(\ref{decS}) and~(\ref{decT}) to calculate the contributions of
the Higgs sector of the minimal supersymmetric Standard Model (MSSM)
to $S$ and $T$ in the decoupling limit.
The quartic couplings of the
MSSM Higgs potential, defined in a supersymmetric
basis~\cite{susybasis}, satisfy:
\beq
\lambda_1=\lambda_2=\tfrac{1}{4}(g^2+g^{\prime\,2})\,,\qquad
 \lambda_3=\tfrac{1}{4}(g^2-g^{\prime\,2})\,,\qquad
\lambda_4=-\half g^2\,,\qquad \lambda_5=\lambda_6=\lambda_7=0\,,
\eeq
where the $\lambda_i$ are defined in \eq{genpot}.
In the supersymmetric basis, the ratio of
vacuum expectation values is $\tan\beta\equiv v_2/v_1$.
Since the MSSM Higgs sector is CP-conserving, one can
transform to the \textit{real} Higgs basis, which yields:
\beqa
Z_1=Z_2&=&\tfrac{1}{4}(g^2+g^{\prime\,2})\cos^2 2\beta\,,\qquad
Z_5=\tfrac{1}{4}(g^2+g^{\prime\,2})\sin^2 2\beta
\,,\qquad Z_7=-Z_6=\tfrac{1}{4}(g^2+g^{\prime\,2})
\sin 2\beta\cos 2\beta\,,\nonumber \\
\phantom{line}\label{zz1} \\
Z_3&=&\tfrac{1}{4}\left[
(g^2+g^{\prime\,2})\sin^2 2\beta+g^2-g^{\prime\,2}\right]\,,\qquad\qquad
Z_4=\tfrac{1}{4}\left[
(g^2+g^{\prime\,2})\sin^2 2\beta-2g^2\right]\,.\label{zz2}
\eeqa
Note that $Z_5>0$ which means that $\varepsilon_{56}=\varepsilon_{57}=+1$.
Using \eqs{mc}{m2A} yields the exact (tree-level) mass relation,
\beq {m^2_{H^\pm}} = m_{A^0}^2 + m_W^2\,.\label{mssm1}
\eeq
Moreover, in the decoupling limit, \eq{m2approx} yields
$m_{H^0}^2 = m_{A^0}^2 +Z_5 v^2+ \mathcal{O}(v^2/m_{A^0}^2)$, which can
be rewritten using \eq{zz1},
\beq
m_{H^0}^2 = m_{A^0}^2 + m_Z^2 \sin^22\beta
+ \mathcal{O}(v^2/m_{A^0}^2). \label{mssm2}
\eeq
Substituting  \eqs{zz1}{zz2}  [or equivalently, \eqs{mssm1}{mssm2}]
into eqs.~(\ref{decS}) and~(\ref{decT}) yields
\beqa S_{\rm MSSM} & \simeq &\frac{m_Z^2\sin^22\beta - 2 m_W^2}{24\pi m_{A^0}^2},\nonumber\\
 T_{\rm MSSM} &\simeq& \frac{m_W^2 - m_Z^2\sin^22\beta}{48\pi s_W^2  m_{A^0}^2},\eeqa
which reproduce the results previously obtained in \Ref{habertasi}.

Finally, we examine $U$ in the decoupling limit.  Setting $m_\phi=m_1$
in \eq{tsu} and employing the decoupling limit conditions of
\eq{qdecoupling}, we obtain:
\beq
S+U=\frac{1}{\pi m_W^2}\left[\mathcal{B}_{22}(m_Z^2;m_{H^\pm}^2,m_2^2)
+\mathcal{B}_{22}(m_Z^2;m_{H^\pm}^2,m_2^3)-
2\mathcal{B}_{22}(m_Z^2;m_{H^\pm}^2,m_{H^\pm}^2)+
\mathcal{O}\left(\frac{v^4}{\Lambda^4}\right)\right]\,.
\eeq
Using \eqs{massrelations2}{massrelations3} and \eq{b22expand}, we obtain:
\beq
S+U\simeq \frac{m_2^2+m_3^2-2m^2_{H^\pm}}{24\pi c_W^2 m_3^2}=\frac{Z_4 v^2}{24\pi c_W^2 m_3^2}=\frac{S}{c^2_W}\,.
\eeq
Finally, we use \eq{decS} to isolate $U$:
\beq
U\simeq S\tan^2\theta_W\,.
\eeq
In the custodial limit where $g^\prime=0$, it follows that $\tan\theta_W=0$, in which
case $U=0$.  Remarkably, we find that $U=0$ in this limit independently of the values of the neutral
Higgs masses.  Thus, custodial symmetry breaking effects arising from the scalar
potential do \textit{not} generate a non-zero value for $U$ at
$\mathcal{O}({v^2}/{\Lambda^2})$ in the approach to the
decoupling limit.  However, \eqst{UCP}{UvanishII} imply that a non-zero value for $U$
would be generated at order $\mathcal{O}({v^4}/{\Lambda^4})$.  This observation suggests
that $U\ll T$ over a significant range of the 2HDM parameter space, a fact that can be
verified numerically.

\section{Numerical Analysis}

The parameters $S$, $T$ and $U$ obtained from an analysis of precision electroweak
data are found to be~\cite{erlerPDG}:
\beqa \label{stu}
S &=& 0.01 \pm 0.10, \nonumber\\
T &=& 0.03 \pm 0.11, \nonumber\\
U &=& 0.06 \pm 0.10, \eeqa
relative to the Standard Model, with a reference Higgs mass of
$m_\phi = 117$ GeV.  Similar results have been obtained by
the \texttt{GFITTER} collaboration~\cite{gfitter}.
Alternatively, if one assumes that $U=0$ (typically, one
expects $U\ll S$ in many models of new physics beyond the Standard Model),
then the corresponding analysis of $S$ and $T$ yields~\cite{erlerPDG}:
\beq \label{STdec}
S=0.03\pm 0.09\,,\qquad\qquad T=0.07\pm 0.08\,.
\eeq
These limits indicate that new physics contributions to the oblique parameters are tightly constrained.
In particular, if one assumes that the new physics contributions to $S$, $T$ and $U$ arise solely
from the 2HDM sector, then \eqs{stu}{STdec} would constrain the parameters of the 2HDM scalar
potential.  Such studies have appeared in the literature in a less general framework.
For example, in \Ref{kaffas}, $\rho\equiv\alpha T$ was used to constrain a
modified version of the 2HDM in which certain scalar couplings were
set equal to zero, and $\tan\beta$ was assumed to be a physical
observable.  In our approach, only basis-independent quantities are employed.
A full numerical study of the constraints of precision electroweak
data on the general 2HDM contributions to the oblique parameters will be presented elsewhere (for a
preliminary study, see \Ref{deva}).  In this section, we shall outline our analysis methods
and describe some of the key results and features of our study.

\begin{figure}[t!]
\begin{center}
$\begin{array}{c@{\hspace{.21in}}c}
\includegraphics[scale=1.02]{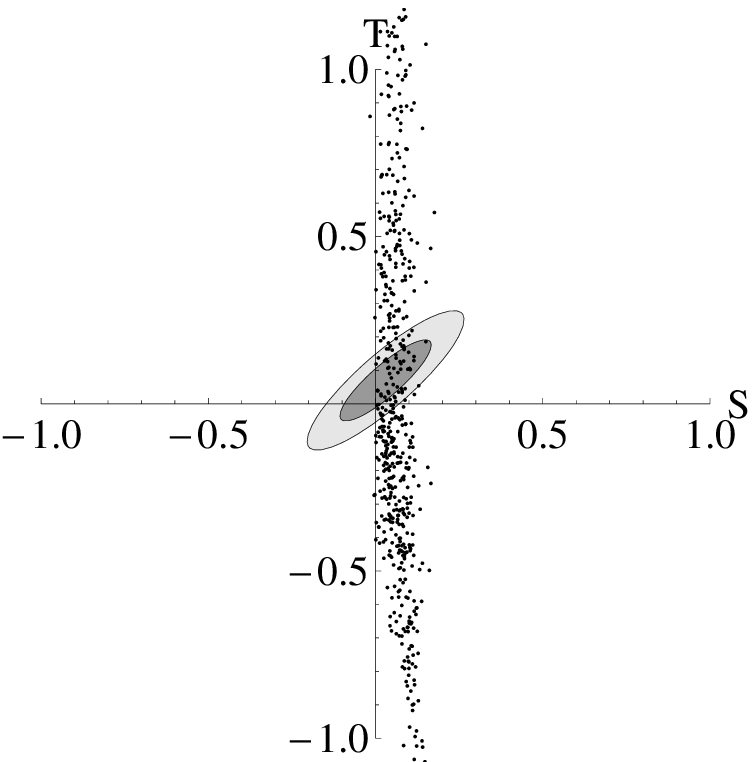} &
\includegraphics[scale=1.02]{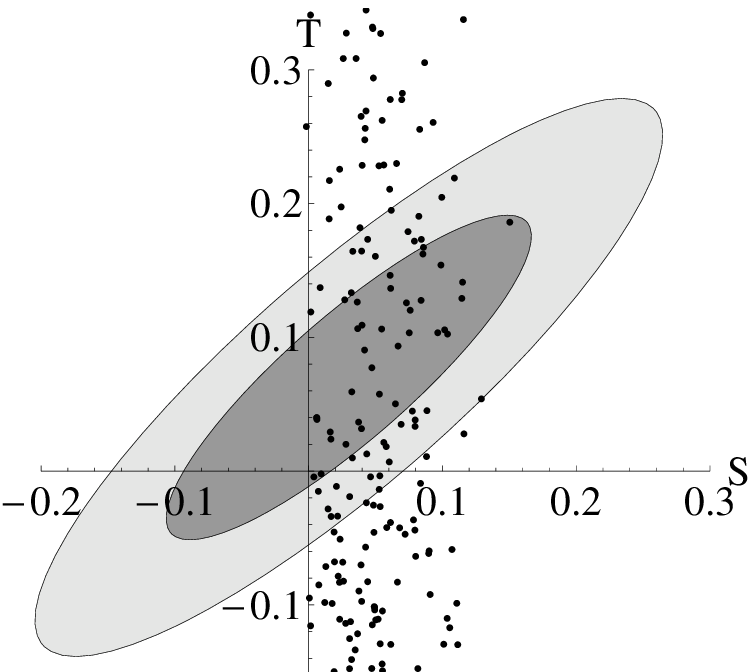}\\ [0.4cm]
\mbox{\bf (a)} & \mbox{\bf (b)}
\end{array}$
\end{center}
\caption{Scatter plots for $T$ as a function of $S$, with $m_1 = 117 \pm 10$~GeV.
The ellipses, representing the $1~\sigma$ and $2~\sigma$ contours
allowed by precision electroweak data, are based on \Ref{erlerPDG},
with the parameter $U$ fixed to zero.  Plot (a) shows the expanded view
in the $S$--$T$ plane, and
plot (b) shows a close-up view of the allowed region.}
\label{scatter}
\end{figure}

The parameters of the 2HDM that are constrained by $S$, $T$, and $U$ can be taken to be
$Z_1$, $Z_3$, $Z_3 +Z_4$, $Z_5\,e^{-2i\theta_{23}}$, $Z_6\, e^{-i\theta_{23}}$ and $Y_2$,
since these 6 quantities determine the physical Higgs masses
[cf.~eqs.~(\ref{mtilmatrix}) and~(\ref{mc})] and the invariant quantities $q_{k\ell}$
[specified in Table~\ref{tabq}].  We impose one theoretical constraint
by demanding that the $|Z_i|$ do not exceed upper bounds corresponding
to the requirement that all bosonic scattering amplitudes satisfy tree-level unitarity
(the relevant upper bounds are derived in Appendix~\ref{app:three}).
In order to compare with the determination of of $S$, $T$ and $U$
in \Ref{erlerPDG}, we fix the reference Higgs mass at $m_\phi=117$~GeV.
The procedure used here to study the effect of the 2HDM on the
oblique parameters was to choose random values of the six parameters
in the space allowed by the tree-level unitarity bounds.
Then the Higgs masses and $q_{k\ell}$ are calculated numerically and
inserted into \eqst{seqn}{tsu}
to obtain $S$, $T$, and $U$ for each point in the parameter space.

In our first study, we imposed an additional requirement that the mass
of the lightest neutral Higgs boson, $m_1$, fall within $10$~GeV of
the reference Higgs mass.  It was found that the 2HDM consistently
produces values of $U$ within $0.02$ of zero.  Thus, in order to
derive constraints on the 2HDM parameters, one can reliably set $U=0$
and compare the computed $S$ and $T$ values of the 2HDM with the
results given in \eq{STdec}.  Scanning the 2HDM parameter space and
comparing with the allowed 2~$\sigma$ contour ellipse
in $S$--$T$ plane produces the results shown in Fig.~\ref{scatter}.

From the scatter plots shown in Fig.~\ref{scatter},
it is evident that the values of $S$ produced are all consistent with the experimental constraints of \eq{STdec}.
In contrast, there are many points that lie outside the allowed range for $T$.
These points correspond to 2HDM parameters that significantly violate the custodial
symmetry of the scalar potential.  In particular, one must have a significant
splitting between the masses of the $H^\pm$ and one of the heavy neutral Higgs bosons
(identified in the generic CP-conserving 2HDM as $A^0$).
This region of parameter space is very far away from the decoupling region in which the
2HDM contributions to $T$ are quite small.  When $T$ is large, the large values
of the corresponding heavy Higgs masses are driven primarily by large values of
the $Z_iv^2$ that compete with (and in some regions dominate) the contribution of $Y_2$.
Even though the maximal values of the $Z_i$ are constrained by tree-level unitarity,
there is still a robust region of the 2HDM parameter space in which $|T|$ lies
significantly outside of the interval allowed by \eq{STdec}.
It is also interesting to note that both positive and negative signs for $T$ are allowed,
with roughly equal probability over the 2HDM parameter space.

\begin{figure}[!ht]
\begin{center}
 \includegraphics[scale=0.9]{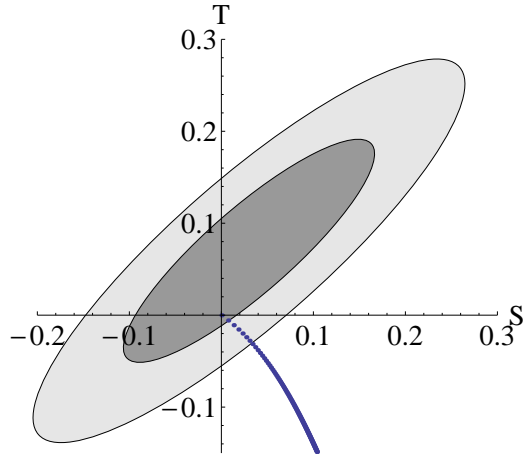} \end{center}
\caption{The effect on $S$ and $T$ when $m_1$ is increased from $117$ GeV to $500$ GeV.
Both $S$ and $T$ are zero at $m_1 = m_\phi = 117$ GeV.  When $m_1$ reaches $175$ GeV,
$S$ and $T$ exceed the boundary of the $2~\sigma$ contour ellipse.}\label{fig2added}
\end{figure}

In the analysis above, we have fixed the value of the lightest neutral Higgs mass, $m_1$,  to
be close to 117~GeV.  One can now investigate the consequence of relaxing this assumption.
First, consider the decoupling limit of the 2HDM
where $m_1\ll m_2$, $m_3$.   As $m_1$ increases (in a mass regime
in which $h_1$ is still significantly lighter than $h_2$ and $h_3$),
we should simply reproduce the known constraints of precision electroweak
observables on the mass of $m_1$.  As a concrete example, consider the
following input parameters:
\beq m_\phi = 117~{\rm GeV}, ~~Z_1 = 0.227,~~Y_2 =( 1 ~\rm TeV)^2, \eeq
with all other invariant $Z$ parameters equal to $0.01$.
The mass spectrum corresponding to these values is $m_1 = 117$ GeV,
$m_2 = m_3 = {m_{H^\pm}} = 1$~TeV.  As expected, in this limit one finds that $S \simeq T \simeq 0$.
As $Z_1$ is increased from $0.227$ to $0.505$, $m_1$ increases from $117$ GeV to $175$ GeV,
at which point $T$ and $S$ exceed the boundary of the $2~\sigma$ ellipse. In
Fig.~\ref{fig2added}, the resulting $S$ and $T$ are shown as $m_1$ is varied from $117$~GeV to $500$~GeV.

\begin{figure}[!ht]
\begin{center}
 \includegraphics[scale=0.9]{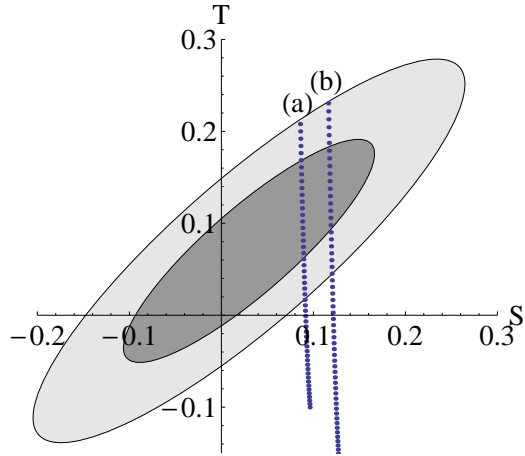} \end{center}
\caption{The effect on $S$ and $T$ when ${m_{H^\pm}}$ is varied by increasing $Z_3$, for $m_1 = 350$~GeV
and $m_\phi = 117$ GeV. In column (a), $Z_1 = 4$; in column (b), $Z_1 = 4~\pi$.
When ${m_{H^\pm}}$ is in the range $443$--$489$ GeV [$470$--$505$~GeV],
the points in column (a) [column (b)] fall within the $2~\sigma$ contour ellipse.
}\label{evade}
\end{figure}

If we depart significantly from the decoupling limit, then the
custodial-symmetry breaking mass splitting
between the $H^\pm$ and one of the heavy neutral Higgs states can
contribute positively to $T$ and offset the negative $T$ values shown in
Fig.~\ref{fig2added}.  In this way, values of the lightest Higgs boson mass above
150~GeV may still be consistent with precision electroweak data.  This possibility is
illustrated by the following example.
With $m_\phi$ fixed at $117$ GeV, let us choose $Y_2 = (50$ GeV$)^2$ and $Z_3 = 5.2$. This produces
${m_{H^\pm}} = 400$ GeV.  With $Z_1 = 4$, $Z_4$ is adjusted such that $m_1 = 350$ GeV, with all
other $Z$ parameters set equal to $0.01$. One can then dial up $Z_3$ (keeping $m_1$ fixed at
$350$ GeV by simultaneously adjusting $Z_4$) until $S$ and $T$ lie within the $2~\sigma$ contour ellipse.
For the above choice of $Z_1$, the allowed range for the charged Higgs mass is
$443~{\rm GeV}< {m_{H^\pm}} < 489$~GeV, as shown in column (a) of Fig.~\ref{evade}.
We can repeat this exercise by fixing $Z_1$ at its tree-level unitarity limit.  In this case the
allowed range is $470$ GeV $<{m_{H^\pm}} < 505$~GeV, as shown in column (b) of Fig.~\ref{evade}.
With the charged Higgs boson mass in its prescribed range, a ``light'' neutral Higgs mass of
350~GeV is consistent with precision electroweak data!

One can increase the value for $m_1$ arbitrarily high and still find values of
${m_{H^\pm}}$ that are consistent with $S$ and $T$ in the allowed range.  However, one eventually
violates the unitarity of $Z_3 +Z_4$ (if $m_1$ is too high) or $Z_3$ alone
(if ${m_{H^\pm}}$ is too high.)  As an example, by choosing $Y_2 = (50~{\rm GeV})^2$,
$Z_1 = 4 \pi$, $\zfiver = \zfivei = \zsixr = \zsixi = 0.01$, one can adjust
$Z_{3}+Z_{4}$ to get $m_1$ as high as 873~GeV before violating unitarity,
which gives a mass spectrum of
\beq
 m_1 = 873 ~\text{GeV},  \,\,\, \,\,\, m_2 = 874 ~\text{GeV},  \,\,\,\,\,\, m_3 = 875 ~\text{GeV}.
 \eeq
With this value of $m_1$, choosing $Z_3$ so that  $716~{\rm GeV} < {m_{H^\pm}} < 750$~GeV will put
$S$ and $T$ in the upper right hand corner of the allowed $2~\sigma$
contour ellipse.\footnote{Note that for this choice of parameters,
${m_{H^\pm}} < m_1$.  In fact, there are higher values of ${m_{H^\pm}}$ which
are within the allowed ellipse, but they correspond to values of $Z_3$ that exceed its unitarity bound.}

We conclude that the regions of $S$ and $T$ allowed by precision electroweak data place
significant constraints on the possible regions of the 2HDM parameter space.
In the decoupling limit of the 2HDM, the only surviving constraint is on the
mass of the lightest Higgs boson, which coincides with the corresponding Standard Model Higgs mass
upper bound deduced from precision electroweak data.  In regions of the 2HDM parameter space far
from the decoupling regime, a large chunk of the 2HDM parameter space is ruled out on the
basis of the $T$ parameter.  Nevertheless, there are regions of parameter space in the
non-decoupling regime, consistent with precision electroweak data,
in which the lightest Higgs mass is significantly larger than
the Standard Model Higgs mass upper bound.\footnote{This possibility has been considered
previously in \Ref{gunetal}.}
This possibility is realized when large negative
corrections to $T$ from $h_1$ are compensated by large positive corrections to $T$ from the other
Higgs bosons of the 2HDM.

\section{Conclusions}

In this paper, we have employed basis-independent methods in
examining the properties of the most general (CP-violating)
2HDM.  Our primary aim to provide a basis-invariant characterization
of a custodial-symmetric 2HDM scalar potential.  Since
custodial symmetry in the scalar sector of the 2HDM implies
CP-conservation, we first examined in detail the basis-independent
description of the most general CP-conserving 2HDM.  All possible
generic and special cases were examined, which depend on the values of the potentially
complex quartic Higgs self-couplings in the Higgs basis.
One special case where $Z_6=Z_7=0$ is noteworthy, due to the fact
that the CP quantum numbers of two of the three neutral Higgs states
cannot be determined by the bosonic couplings of the model.  This
behavior can be traced to the existence of two inequivalent definitions
of CP which give opposite signs for the CP-quantum numbers of each of the
two neutral states.  However, the ambiguity is resolved by the Higgs-fermion
Yukawa interactions that uniquely selects one of the two definitions for CP and thus
determines the CP quantum numbers of the two neutral states (assuming
that the Yukawa interactions are CP-conserving).  In fact, the Yukawa
interactions could be CP-violating, even if the scalar potential
\textit{and} the Yukawa interactions respect the custodial symmetry, in which
case it does not make sense to assign definite CP quantum numbers to the neutral Higgs states.

After providing a catalog of possible cases that define the
CP-conserving 2HDM, we imposed custodial symmetry and determined
the basis-independent condition that guarantees the presence of this
symmetry.  We have clarified the results of a previous analysis given
in \Ref{pomarol}, where it was asserted that there were two distinct
cases for the custodial-symmetric scalar potential.  We have demonstrated in
this paper that the two cases of \Ref{pomarol} are in fact equivalent
and simply correspond to two different basis choices for the
scalar potential.
We also showed that generically the charged Higgs boson
and the CP-odd Higgs boson are mass-degenerate in the limit of a
custodial symmetric scalar potential.  However, in the special case
of $Z_6=Z_7=0$, it is possible that the charged Higgs boson and one
of the CP-even Higgs bosons are mass-degenerate in the limit of a
custodial symmetric scalar potential, depending on the structure
of the Higgs-fermion interactions.

We have also provided a basis-independent computation of the
2HDM contributions to the oblique  parameters $S$, $T$, and $U$.
Since $T=U=0$ in the custodial-symmetry limit, our computation
provides an important check on the implications of the various
mass-degeneracies noted above.
The oblique parameters of the CP-violating 2HDM were
analyzed numerically and found to be inconsistent with the experimental
electroweak constraints over a nontrivial region of the 2HDM parameter
space.  Of course, there is still a significant region of the
parameter space in which the oblique parameters lie within the
allowed $2\sigma$ error ellipse in the $S$--$T$ plane. ($U$ is quite
small over nearly the entire 2HDM parameter space, and one can set it
to zero to good approximation.)   In the decoupling limit,
the only constraints on the 2HDM parameters are associated with
the requirement that the lightest neutral Higgs boson, which is
Standard Model-like in its properties, must have a mass below
about 150~GeV (equivalent to the constraints of the Standard Model
global fits).  In the region of the 2HDM parameter space far from
the decoupling regime, it is possible that the lightest neutral
Higgs boson mass is significantly heavier than 150~GeV.  In this
case, the large negative value of $T$ generated by the lightest
neutral Higgs boson is compensated by positive corrections to $T$ from
the other physical Higgs bosons of the 2HDM.

\bigskip

\textbf{Note Added}
\medskip

After this work was completed, a paper by 
B.~Grzadkowski, M.~Maniatis and J.~Wudka~\cite{gmw}
appeared that employs the formalism of gauge-invariant scalar
field bilinears (cf.~footnote~\ref{bilinears}) in the analysis
of custodial symmetry in the 2HDM.  They obtain 
conditions for a custodial symmetric 2HDM scalar potential that are
consistent with the results obtained in this paper.

\acknowledgments

H.E.H. greatly appreciates stimulating discussions with Jean-Marc Gerard, Pedro Ferreira,
Jack Gunion, Michel Herquet and Jo\~{a}o Silva.  D.O. is grateful to John Mason for his assistance in
some aspects of this research.
The work of H.E.H. is supported
in part by the U.S. Department of Energy, under grant
number DE-FG02-04ER41268 and in part by a Humboldt Research Award sponsored by
the Alexander von Humboldt Foundation.
The work of D.O. is supported in part by the U.S. Department of Energy, under grant
number DE-FG02-04ER41268 and in part by
a Graduate Assistance in Areas of National Need
fellowship from the U.S. Department of Education.

\appendix   
\section{Cubic and quartic bosonic couplings in the 2HDM}\label{app:couplings}

The Higgs boson interactions of the 2HDM can be expressed in terms of the basis-independent
$q_{k\ell}$ defined in Table \ref{tabq}. The cubic and quartic vector-scalar
couplings were obtained in \Ref{haberoneil} and are reproduced below:
\beqa
\mathscr{L}_{VVH}&=&\left(gm_W W_\mu^+W^{\mu\,-}+\frac{g}{2c_W}
m_Z Z_\mu Z^\mu\right)q_{k1} h_k \nonumber \\[5pt]
&&
+em_WA^\mu(W_\mu^+G^-+W_\mu^-G^+)
-gm_Zs_W^2 Z^\mu(W_\mu^+G^-+W_\mu^-G^+)
\,, \label{VVH}
\\[5pt]
\mathscr{L}_{VVHH}&=&\left[\quarter g^2  W_\mu^+W^{\mu\,-}
+\frac{g^2}{8c_W^2}Z_\mu Z^\mu\right]h_k h_k
\nonumber \\[5pt]
&&+\biggl[\half g^2 W_\mu^+ W^{\mu\,-}+e^2A_\mu A^\mu+\frac{g^2}{c_W^2}\left(\half -s_W^2\right)^2Z_\mu Z^\mu
+\frac{2ge}{c_W}\left(\half -s_W^2\right)A_\mu Z^\mu\biggr](G^+G^-+H^+H^-)
\nonumber \\[5pt]
&&+\biggl\{\left(\half eg A^\mu W_\mu^+ -\frac{g^2s_W^2}{2c_W}Z^\mu W_\mu^+\right)
(q_{k1}G^-+q_{k2}\,e^{-i\theta_{23}}H^-)h_k +{\rm h.c.}\biggr\}
\,,\label{VVHH}\\[5pt]
\mathscr{L}_{VHH}&=&\frac{g}{4c_W}\,\epsilon_{jk\ell}q_{\ell 1}
Z^\mu h_j\ddel_\mu h_k
-\half g\biggl\{iW_\mu^+\left[q_{k1} G^-\ddel\lsup{\,\mu} h_k+
q_{k2}e^{-i\theta_{23}}H^-\ddel\lsup{\,\mu} h_k\right]
+{\rm h.c.}\biggr\}\nonumber \\[5pt]
&&+\left[ieA^\mu+\frac{ig}{c_W}\left(\half -s_W^2\right)
Z^\mu\right](G^+\ddel_\mu G^-+H^+\ddel_\mu H^-)\,,\label{VHH}
\eeqa
\beqa
\mathscr{L}_{VG}&=&\left[\frac{g^2}{4} W_\mu^+W^{\mu\,-}
+\frac{g^2}{8c_W^2}Z_\mu Z^\mu\right]G^0 G^0+\half g\left(W_\mu^+G^-\ddel\lsup{\,\mu}G^0+W_\mu^-G^+\ddel\lsup{\,\mu}G^0
\right)\,\nonumber\\[5pt]
&&+\biggl\{\frac{ieg}{2} A^\mu W_\mu^+ G^- G^0
-\frac{ig^2s_W^2}{2c_W}Z^\mu W_\mu^+
G^-G^0 +{\rm h.c.}\!\biggr\}+\frac{g}{2c_W} q_{k1} Z^\mu G^0\ddel_\mu h_k\,,\label{VG}
\eeqa
where repeated indices $j,k=1,2,3$ are summed over.
In obtaining the above
interactions from \Ref{haberoneil}, we have made two simplifications.  In the $W^+W^-h_j h_k$ and
$ZZh_j h_k$ interactions, we have employed
\beq
q_{j1} q_{k1}+\Re(q_{j2}^* q_{k2})=\delta_{jk}\,,\qquad \text{for}~~j,k=1,2,3.
\eeq
In the $Zh_j h_k$ interactions ($j,k=1,2,3$), we have made use of the identity
\beq
\Im(q_{j2}^*q_{k2})=\sum_{\ell=1}^3\epsilon_{jk\ell}q_{\ell 1}\,.
\eeq

Likewise, a basis-independent form for the cubic and quartic scalar self-interactions
has been obtained in \Ref{haberoneil} and are reproduced below.  In listing
the scalar self-interactions, it is convenient to include terms involving the Goldstone
field by denoting $h_4\equiv G^0$.
\beqa \label{hcubic}
\mathcal{V}_3&=&\half v\, h_j h_k h_\ell
\biggl[q_{j1}q^*_{k1}\Re(q_{\ell 1}) Z_1
+q_{j2}q^*_{k2}\,\Re(q_{\ell 1})(Z_3+Z_4) +
\Re(q^*_{j1} q_{k2}q_{\ell 2}Z_5\,
e^{-2i\theta_{23}}) \nonumber \\
&&\qquad\qquad\qquad\quad
+\Re\left([2q_{j1}+q^*_{j1}]q^*_{k1}q_{\ell 2}Z_6\,e^{-i\theta_{23}}\right)
+\Re (q_{j2}^*q_{k2}q_{\ell 2}Z_7\,e^{-i\theta_{23}})
\biggr]\nonumber \\
&& \hspace{-0.2in} +v\,h_k
G^+G^-\biggl[\Re(q_{k1})Z_1+\Re(q_{k2}\,e^{-i\theta_{23}}Z_6)\biggr]
+v\,h_k H^+H^-\biggl[\Re(q_{k1})Z_3+\Re(q_{k2}\,e^{-i\theta_{23}}Z_7)\biggr]
\nonumber \\
&& \hspace{-0.2in} +\half v \,h_k\biggl\{G^-H^+\,e^{i\theta_{23}}
\left[q^*_{k2} Z_4
+q_{k2}\,e^{-2i\theta_{23}}Z_5+2\Re(q_{k1})Z_6
\,e^{-i\theta_{23}}\right]+{\rm h.c.}\biggr\}\,,
\\[5pt]
&& \hspace{-0.18in}
\mathcal{V}_4=\eighth h_j h_k h_l h_m
\biggl[q_{j1}q_{k1}q^*_{\ell 1}q^*_{m1}Z_1
+q_{j2}q_{k2}q^*_{\ell 2}q^*_{m2}Z_2
+2q_{j1}q^*_{k1}q_{\ell 2}q^*_{m2}(Z_3+Z_4)\nonumber \\[5pt]
&&\quad\,\,
+2\Re(q^*_{j1}q^*_{k1}q_{\ell 2}q_{m2}Z_5\,e^{-2i\theta_{23}})
+4\Re(q_{j1}q^*_{k1}q^*_{\ell 1}q_{m2}Z_6\,e^{-i\theta_{23}})
+4\Re(q^*_{j1}q_{k2}q_{\ell
  2}q^*_{m2}Z_7\,e^{-i\theta_{23}})\biggr]\nonumber \\[5pt]
&&  +\half h_j h_k G^+ G^-\biggl[q_{j1}q^*_{k1} Z_1 + q_{j2}q^*_{k2}Z_3
+2\Re(q_{j1}q_{k2}Z_6\,e^{-i\theta_{23}})\biggr]
 \nonumber \\[5pt]
&&  +\half h_j h_k H^+ H^-\biggl[q_{j2}q^*_{k2} Z_2 + q_{j1}q^*_{k1}Z_3
+2\Re(q_{j1}q_{k2}Z_7\,e^{-i\theta_{23}})\biggr] \nonumber \\
&& +\half h_j h_k\biggl\{G^- H^+\,e^{i\theta_{23}} \left[q_{j1}q^*_{k2}Z_4
+ q^*_{j1}q_{k2}Z_5\,e^{-2i\theta_{23}}+q_{j1}q^*_{k1}Z_6
\,e^{-i\theta_{23}}+q_{j2}q^*_{k2}Z_7\,e^{-i\theta_{23}}\right]+{\rm h.c.}
\biggr\} \nonumber \\[5pt]
&&
 +\half Z_1 G^+ G^- G^+ G^- +\half Z_2 H^+H^- H^+ H^-
+ (Z_3+Z_4)G^+ G^- H^+ H^- +\half Z_5 H^+H^+G^-G^-
\nonumber \\[5pt]
&&
+\half Z_5^* H^-H^- G^+G^+
+ G^+G^-(Z_6 H^+ G^-\! + Z_6^* H^- G^+) + H^+H^-(Z_7 H^+ G^-\! + Z_7^* H^- G^+)
\,,\label{scalpot}
\eeqa
summing over $j$, $k$, $\ell$, $m=1,2,3,4$.  Note that $\Re(q_{k1})=q_{k1}$ for $k=1,2,3$, whereas
$\Re(q_{41})=0$.

One can easily verify that if $q_{k1}=\pm 1$ and $q_{k2}=0$ for a fixed value of $k=1,2$ or 3.
then it follows that the couplings of the neutral Higgs field, $\pm h_k$,
are precisely those of the Standard Model Higgs boson.

\section{Neutral Higgs masses and invariant mixing angles}\label{app:invangles}

The neutral Higgs mass eigenstates are denoted by $h_k$ ($k=1,2,3$).
The corresponding squared-masses are obtained by solving
the characteristic equation of the neutral Higgs squared-mass matrix,
$\mathcal{M}^2$ [see \eq{mtilmatrix}],
\beq \label{charpoly}
\det(\mathcal{M}-x\,\mathbb{1}_{3\times 3})=-x^3+\Tr(\mathcal{M})\,x^2
-\half\left[(\Tr\mathcal{M})^2-\Tr(\mathcal{M}^2)\right]x+
\det(\mathcal{M})=0\,,
\eeq
where $\mathbb{1}_{3\times 3}$ is the $3\times 3$ identity matrix and
\beqa
\Tr(\mathcal{M}) &=& 2Y_2+(Z_1+Z_3+Z_4)v^2\,,\nonumber\\
\Tr(\mathcal{M}^2) &=& Z_1^2 v^4
+\half v^4\left[(Z_3+Z_4)^2+|Z_5|^2+4|Z_6|^2\right]
+2Y_2[Y_2+(Z_3+Z_4)v^2]\,,\nonumber \\
\det(\mathcal{M}) &=& \quarter\biggl\{Z_1 v^6[(Z_3+Z_4)^2-|Z_5|^2]
-2v^4[2Y_2+(Z_3+Z_4)v^2]|Z_6|^2 \nonumber \\
&&\qquad\qquad\qquad  +4Y_2 Z_1 v^2[Y_2+(Z_3+Z_4)v^2]
+2v^6\Re(Z_5^* Z_6^2)\biggr\}\,. \label{traces}
\eeqa
In the general (CP-violating) case, the analytic expressions for the squared-masses
are quite cumbersome, when expressed solely in terms of the scalar potential
parameters.  In ref.~\cite{haberoneil}, a more convenient expression
for the neutral Higgs squared-masses was derived in terms of the
$Z_i$ and invariant mixing angles,
\beq \label{hmassesinv}
m_k^2 = |q_{k2}|^2 A^2+ v^2\left[q_{k1}^2 Z_1
+\Re(q_{k2})\,\Re(q_{k2}Z_5\,e^{-2i\theta_{23}})
+2q_{k1}\Re(q_{k2}Z_6\, e^{-i\theta_{23}})\right]\,,
\eeq
where $m^2_k\equiv m^2_{h_k}$ (for $k=1,2,3$) and
the basis-invariant $q_{ki}$ are given in Table~I.

In \Ref{haberoneil}, we also obtained a set of
equations that determine the neutral Higgs mixing angles:\footnote{Denoting the quadratic terms in the scalar potential
by $m_{H^\pm}^2H^+H^- +\half v^2\sum_{j,k} C_{jk}h_j h_k$,
it follows that $C_{jk}=0$
for $j\neq k$.  This provides three conditions, which yield \eqst{tan13}{tan212}.}
\beqa
s_{13}\Re(Z_6\,e^{-i\theta_{23}})&=&\half c_{13}\Im(Z_5\,e^{-2i\theta_{23}})
\,, \label{tan13} \\[8pt]
(Z_1-A^2/v^2)s_{13} c_{13}&=&
(c_{13}^2-s_{13}^2)\Im(Z_6\,e^{-i\theta_{23}})\,,\label{tan213}\\[8pt]
(c_{12}^2-s_{12}^2)\bigl[c_{13}\Re(Z_6\,e^{-i\theta_{23}})
&+&\half s_{13}\Im(Z_5\,e^{-2i\theta_{23}})\bigr] =
s_{12} c_{12}\bigl[\Re(Z_5\,e^{-2i\theta_{23}}) -(Z_1-A^2/v^2) c_{13}^2
 \nonumber \\
&&\qquad\qquad\qquad\qquad\qquad
-2s_{13}c_{13}\Im(Z_6\,e^{-i\theta_{23}})\bigr] \,,\label{tan212}
\eeqa
where
\beq \label{defA2}
A^2\equiv
Y_2+\half\left[Z_3+Z_4-\Re(Z_5\,e^{-2i\theta_{23}})\right]v^2\,.
\eeq

\Eqst{hmassesinv}{tan212} can be used to derive the following results:
\beqa
\Re(Z_6\,e^{-i\theta_{23}})\,v^2 &=& c_{13}s_{12}c_{12}(m_2^2-m_1^2)\,,
\label{z6rv} \\[8pt]
\Im(Z_6\,e^{-i\theta_{23}})\,v^2 &=& s_{13}c_{13}(c_{12}^2 m_1^2+s_{12}^2
m_2^2-m_3^2) \,,\label{z6iv} \\[8pt]
\Re(Z_5\,e^{-2i\theta_{23}})\,v^2 &=& (s_{12}^2-s_{13}^2c_{12}^2)m_1^2
+(c_{12}^2-s_{12}^2 s_{13}^2)m_2^2-c_{13}^2 m_3^2\,,\label{z5rv} \\[8pt]
\Im(Z_5\,e^{-2i\theta_{23}})\,v^2 &=&  2s_{13}s_{12}c_{12}(m_2^2-m_1^2)\,.
\label{z5iv}
\eeqa
The following identity will also prove useful,
\beqa
\Im(Z_5^* Z_6^2)&=& 2\,\Re(Z_5 e^{-2i\theta_{23}})\,
\Re(Z_6\,e^{-i\theta_{23}})\,\Im(Z_6\,e^{-i\theta_{23}}) \nonumber \\
&& \qquad -\,\Im(Z_5 e^{-2i\theta_{23}})\left\{
[\Re(Z_6\,e^{-i\theta_{23}})]^2-[\Im(Z_6\,e^{-i\theta_{23}})]^2\right\}\,.
\label{z5z6e}
\eeqa
Using the results of \eqst{z6rv}{z5z6e} it then follows that:
\beq
v^6\Im(Z_5^* Z_6^2)=
s_{13}c_{13}^2 \sin{2\theta_{12}}\,(m_2^2-m_1^2)(m_3^2-m_1^2)
(m_3^2-m_2^2)\,. \label{imz56}
\eeq

\section{Basis-Independent treatment of the CP-conserving 2HDM}\label{app:howie}

In the CP-conserving Higgs sector, two of the neutral Higgs bosons,
$\hl$ and $\hh$ (with $\mhl < \mhh$) are CP-even and one neutral Higgs
boson, $\ha$, is CP-odd.
Basis-independent conditions for a CP-conserving bosonic sector have been
given in refs.~\cite{cpx,cpx2,davidson,cpbasis}.
In \Ref{davidson}, these
conditions were recast into the form given by \eq{z56}.
Since the Higgs masses and mixing angles do not depend on $Z_7$, we
focus on the implications of the condition $\Im[Z_5^*Z_6^2]=0$ for the
structure of the neutral Higgs squared-mass matrix and the invariant
mixing angles.

\subsection{The CP-conserving limit: $\boldsymbol{Z_6\neq 0}$}

If $Z_6\neq 0$, then \eqs{z6rv}{z6iv} imply that $c_{13}\neq 0$.
Suppose that the three neutral Higgs masses are non-degenerate.
Under the latter assumption, if
the CP-conserving condition $\Im(Z_5^* Z_6^2)=0$
holds, then \eq{imz56}
implies that $s_{13}s_{12}c_{12}=0$.  We examine the two
resulting cases in turn.\footnote{Note that setting \eq{z5z6e} to zero
determines $\Re(Z_5\,e^{-2i\theta_{23}})$ in terms of
$\Im(Z_5\,e^{-2i\theta_{23}})$, $\Re(Z_6\,e^{-i\theta_{23}})$
and $\Im(Z_6\,e^{-i\theta_{23}})$.  However, this is not
sufficient to impose
the conditions given in Cases I and II.  This is because the
diagonalization of the neutral Higgs squared-mass matrix
yields an extra (basis-dependent)
condition that fixes the value of $\theta_{23}$.}
\vskip 0.1in

\textbf{Case~I:} $\boldsymbol{s_{13}=0}$.  Then, \eqst{tan13}{z6iv} imply
that:
\beq \label{caseI}
\Im(Z_5\,e^{-2i\theta_{23}})=\Im(Z_6\,e^{-i\theta_{23}})=0\,.
\eeq

\textbf{Case~II:} $\boldsymbol{s_{12}c_{12}=0}$.  Then,
\eqs{z6rv}{imz56} imply that:
\beq \label{caseII}
\Im(Z_5\,e^{-2i\theta_{23}})=\Re(Z_6\,e^{-i\theta_{23}})=0\,.
\eeq
In \sec{subsecz6}, Case~II is further broken down into two subcases
(a) and (b) corresponding to $s_{12}=0$ and
$c_{12}=0$, respectively.
The three cases I, IIa and IIb simply correspond
to three possible mass orderings of the neutral Higgs bosons---the
CP-even  $h^0$ and $H^0$ (where $\mhl<\mhh$ by definition)
and the CP-odd $A^0$.

It is convenient to define the invariant angle
$\phi\equiv\theta_6-\theta_{23}$, where $\theta_6\equiv\arg Z_6$.  That is,
\beq
\Re(Z_6\,e^{-i\theta_{23}})\equiv |Z_6|\cos\phi\,,\qquad\quad
\Im(Z_6\,e^{-i\theta_{23}})\equiv |Z_6|\sin\phi\,.
\eeq
Then, Cases I and II correspond to $\sin\phi=0$ and $\cos\phi=0$,
respectively.  That is, if CP is conserved then $\sin 2\phi=0$.
Note that the converse is not necessarily valid.  In particular,
\beq
\Im(Z_5\,e^{-2i\theta_{23}})=\frac{1}{|Z_6|^2}\left[
\Re(Z_5^* Z_6^2)\sin 2\phi-\Im(Z_5^* Z_6^2)\cos 2\phi\right]\,.
\eeq
Thus, if $\sin 2\phi=0$ and $\Im(Z_5\,e^{-2i\theta_{23}})\neq 0$,
then the Higgs sector violates CP.

The quantum numbers of the neutral Higgs bosons can be determined from
the form of the Higgs self-couplings.
For example, noting that a charged Goldstone
boson pair is necessarily CP-even, the couplings of $G^+G^-$ to the
neutral Higgs bosons can be used to identify the CP-even scalars.
In Appendix~\ref{app:couplings}, the following couplings are given:
\beqa
G^+ G^- h_1: &\qquad\quad & c_{12} c_{13} Z_1 -s_{12}\Re(Z_6\,e^{-i\theta_{23}})
+c_{12} s_{13}\Im(Z_6\,e^{-i\theta_{23}})\,,\label{ggh1}\\
G^+ G^- h_2: &\qquad\quad & s_{12} c_{13} Z_1 +c_{12}\Re(Z_6\,e^{-i\theta_{23}})
+s_{12} s_{13}\Im(Z_6\,e^{-i\theta_{23}})\,,\label{ggh2}\\
G^+ G^- h_3: &\qquad\quad & s_{13} Z_1 -c_{13}\Im(Z_6\,e^{-i\theta_{23}})\,,
\label{ggh3}
\eeqa
where the mixing angles are defined such that $m_{h_1}\leq m_{h_2}\leq
m_{h_3}$.  Since one of the three neutral states is CP-odd, its
coupling to $G^+ G^-$ must vanish.  Taking $Z_1$ and $Z_6$ as
independent and non-vanishing,\footnote{Similar
conclusions can be obtained by considering the $ZZh_i$, the $Z
h_i h_j$, and the $ZG^0 h_i$ couplings.}
\beqa
{\rm if}~h_1~{\rm is~CP~odd}\,,\qquad
c_{12}&=&\Re(Z_6\,e^{-i\theta_{23}})=\cos\phi=0\,,\label{gg1}\\
{\rm if}~h_2~{\rm is~CP~odd}\,,\qquad
s_{12}&=&\Re(Z_6\,e^{-i\theta_{23}})=\cos\phi=0\,,\label{gg2}\\
{\rm if}~h_3~{\rm is~CP~odd}\,,\qquad
s_{13}&=&\Im(Z_6\,e^{-i\theta_{23}})=\sin\phi=0\,,\label{gg3}
\eeqa
which reproduces Cases I and II [\eqs{caseI}{caseII}] for
non-degenerate neutral Higgs masses.

The masses of the three neutral Higgs bosons can be evaluated
explicitly using \eqs{charpoly}{traces}.  In evaluating
$\det(\mathcal{M})$, we employ
the condition $\Im(Z_5^* Z_6^2)=0$, which implies
that $[\Re(Z_5^* Z_6^2)]^2=|Z_5|^2\,|Z_6|^4$, and leads to
two possible cases:
\beq \label{plusminus}
\Re(Z_5^* Z_6^2)=\varepsilon_{56} |Z_5|\,|Z_6|^2\,,\qquad
\varepsilon_{56}\equiv\pm 1\,.
\eeq
In both cases, \eq{charpoly} factors into a product of a linear
and a quadratic polynomial.  Solving for the roots,
the resulting neutral Higgs squared-masses are
given by \eqs{m2hH}{m2A},
where the CP-odd state is identified according to the results of
\eq{hmassesinv} and \eqst{gg1}{gg3}.
\subsection{Degenerate masses in the CP-conserving limit with $\boldsymbol{Z_6\neq 0}$}
So far, we have working under the assumption that the three Higgs
masses are unequal.  However, it is also possible that two of the
Higgs bosons are mass-degenerate.\footnote{Under the assumption
$Z_6\neq 0$, it is not possible to have
three mass-degenerate neutral Higgs bosons.}
In this case, it follows from \eq{imz56} that
$\Im(Z_5^* Z_6^2)=0$, independently of the mixing angles
(some of which may not be well-defined in the mass-degenerate limit).
If $Z_5=0$ and $Z_6\neq 0$, then \eqs{m2hH}{m2A} imply that the three three
neutral Higgs masses are distinct.  Hence, in what follows we assume
that $Z_5\neq 0$, in which case $\ha$ and one of the
CP-even scalars are degenerate in mass if:\footnote{One can also
verify the condition for degenerate roots directly from \eq{charpoly}.
The cubic equation $z^3+a_2 z^2 +a_1 z+a_0$
has (at least) two degenerate roots if and only if~\cite{abramowitz}
$$
\left[\nicefrac{1}{3} a_1-\nicefrac{1}{9}a_2^2\right]^3
+\left[\nicefrac{1}{6}(a_1
  a_2-3a_0)-\nicefrac{1}{27}a_2^3\right]^2=0\,.
$$
With a little help from Mathematica, one can show that by imposing
the above equation, \eq{z1special} is satisfied.}
\beq \label{z1special}
Z_1=Y_2/v^2+\half(Z_3+Z_4-\varepsilon_{56}|Z_5|)
+\frac{|Z_6|^2}{\varepsilon_{56}|Z_5|} \qquad {\rm and} \qquad
\Im(Z_5^* Z_6^2)=0\,.
\eeq
Inserting this result for $Z_1$ back into \eqs{m2hH}{m2A} then yields:
\beqa
\mhl^2=\mha^2&=&Y_2+\half(Z_3+Z_4-|Z_5|)v^2\,,\label{mhma1}\\
\mhh^2&=&Y_2+\half(Z_3+Z_4+|Z_5|)v^2+\frac{|Z_6|^2 v^2}{|Z_5|}\,,
\qquad\qquad {\rm for}~~\varepsilon_{56}=+1\,,
\eeqa
and
\beqa
\mhl^2&=&Y_2+\half(Z_3+Z_4-|Z_5|)v^2-\frac{|Z_6|^2 v^2}{|Z_5|}\,,\\
\mhh^2=\mha^2&=&Y_2+\half(Z_3+Z_4+|Z_5|)v^2\,,
\qquad\qquad\qquad\qquad {\rm for}~~\varepsilon_{56}=-1\,.\label{mhma2}
\eeqa

In the mass-degenerate case, the mixing angle $\theta_{23}$ and the
corresponding invariant angle $\phi$
are no longer well-defined, as one can redefine the mixing
angles by rotating within the degenerate subspace.
Hence, the choice of $\sin 2\phi$ is arbitrary.  However, because
CP is conserved in the neutral Higgs sector, the structure
of the Higgs interactions  guarantees that
there exists one linear
combination of the mass-degenerate neutral Higgs states
that is CP-even and an orthogonal linear combination
that is CP-odd.  The latter defines the relevant mixing angle,
$\theta_{12}$ in Case I and $\theta_{13}$ in Case~II, respectively.
In particular, the identification
of eigenstates of definite CP quantum numbers
imposes the constraint $s_{13}=s_{12}c_{12}
=\sin 2\phi=0$, and the conditions of Cases I and~II
continue to hold.
A summary of the basis-independent conditions for CP-invariance,
under the assumption that $Z_6\neq 0$,
along with the identification of the CP quantum
numbers of the three neutral Higgs states can
be found in Table~\ref{Z6cond}.

In Section~\ref{invcust}, we determined the basis-independent conditions for
a custodial symmetric scalar potential.  In the case of
$Z_6\neq 0$, the relevant condition is $Z_4 =\epsilon_{56}|Z_5|$ [cf.~\eq{basindcust}].
Using \eqthree{equalmasses}{mhma1}{mhma2}, we conclude that when
\eq{z1special} holds, the custodial symmetric scalar potential
yields two neutral Higgs bosons, one CP-even and one CP-odd, that are both
degenerate in mass with the charged Higgs boson.

\subsection{The CP-conserving Limit: $\boldsymbol{Z_6 = 0}$}
The case where $Z_6=0$ (with $Z_5\neq 0$)
merits special attention.  In this case,
\eqst{tan13}{tan212} simplify to:
\beqa
c_{13}\Im(Z_5\,e^{-2i\theta_{23}}) &=& 0\,,\label{spcase1}\\[6pt]
(Z_1 v^2-A^2)s_{13}c_{13} &=& 0\,,\label{spcase2}\\[6pt]
\half s_{13}(c_{12}^2-s_{12}^2)\Im(Z_5\,e^{-2i\theta_{23}}) &=& s_{12}
c_{12}\left[\Re(Z_5\,e^{-2i\theta_{23}})-(Z_1-A^2/v^2)c_{13}^2\right]\,.
\label{spcase3}
\eeqa

First we consider cases in which
the three neutral Higgs masses are non-degenerate.  Then
in the CP-conserving limit, \eq{imz56} implies that
$s_{13}s_{12}c_{12}=0$.  If $c_{13}s_{13}\neq 0$, then \eqs{spcase1}{spcase2}
yield $\Im(Z_5\,e^{-2i\theta_{23}})=0$ and $Z_1 v^2=A^2$.  In this
case $\Re(Z_5\,e^{-2i\theta_{23}})=\pm |Z_5|$ and
$A^2=Y_2+\half v^2(Z_3+Z_4\mp|Z_5|)$, where both sign choices are
possible.  For either sign choice,
\eqst{mH1}{mH3} imply that two of the neutral
Higgs bosons are degenerate in mass, which contradicts our initial
assumption.  Thus, if the Higgs bosons are non-degenerate, then
\eq{spcase2} implies that either $s_{13}=0$ or $c_{13}=0$.
If $s_{13}=0$ then \eqs{spcase1}{spcase2} yield either
$s_{12}c_{12}=0$ or $\Re(Z_5\,e^{-2i\theta_{23}})=Z_1-A^2/v^2$.
However, in the latter case one again finds that two of the neutral
Higgs bosons are degenerate in mass, which again contradicts our initial
assumption.  Thus, in the case of non-degenerate neutral Higgs masses,
there are three cases to consider:
\vskip 0.1in

\textbf{Case I$\boldsymbol{^\prime}$:}
$\boldsymbol{s_{13}=s_{12}=\Im(Z_5\,e^{-2i\theta_{23}})=0.}$
This is a combination of the previous Cases I and IIa.
\vskip 0.1in

\textbf{Case II$\boldsymbol{^\prime}$:}
$\boldsymbol{s_{13}=c_{12}=\Im(Z_5\,e^{-2i\theta_{23}})=0.}$
This is a combination of the previous Cases I and IIb.
\vskip 0.1in


When $Z_6=0$ and CP is conserved,
a new third possibility arises in which $c_{13}=0$.
In this case, \eq{spcase3} yields
$\half s_{13}(c_{12}^2-s_{12}^2)\Im(Z_5\,e^{-2i\theta_{23}})=
s_{12}c_{12}\Re(Z_5\,e^{-2i\theta_{23}})$, where $s_{13}=\pm 1$.
Following the convention specified in Table~\ref{tabq}, we choose
$s_{13}=-1$. In this convention,
$\theta_{12}+\theta_{23}$ is indeterminate, and the quantity
\beq \label{t2312}
\overline{\theta}_{23}\equiv \theta_{23}-\theta_{12}\,,
\eeq
plays the role of $\theta_{23}$.  We designate this new case
\vskip 0.1in

\textbf{Case III$\boldsymbol{^\prime}$:}
$\boldsymbol{c_{13}=\Im(Z_5\,e^{-2i\overline\theta_{23}})=0.}$
\vskip 0.1in

%

To determine the CP-quantum numbers of the $h_k$, we first examine
the $G^+G^-h_k$ couplings when $Z_6=0$ [cf.~\eqst{ggh1}{ggh3}]:
\beqa
G^+ G^- h_1: &\qquad\quad & c_{12} c_{13} Z_1\,,\label{ggh10}\\
G^+ G^- h_2: &\qquad\quad & s_{12} c_{13} Z_1\,,\label{ggh20}\\
G^+ G^- h_3: &\qquad\quad & s_{13} Z_1\,,\label{ggh30}
\eeqa
One of these three couplings is non-vanishing in Cases I$^\prime$,
II$^\prime$ and III$^\prime$, which implies that the corresponding
neutral Higgs state ($h_k$ for $k=1,2$ or 3) is CP-even.  Moreover,
in this case, $q_{k1}=\pm 1$ and $q_{k2}=0$, which implies
that the couplings of the neutral Higgs field $\pm h_k$ are precisely
those of the Standard Model Higgs boson.
Henceforth, we identify the Standard Model--like CP-even neutral Higgs field
by $h_1^0$.  We then use \eq{hmassesinv} to obtain $m^2_{h_1^0}=Z_1 v^2$
[cf.~\eq{mH1}].
If we order the states $h_i$ such that $m_{h_1}<m_{h_2}<m_{h_3}$,
then the three cases I$^\prime$, II$^\prime$ and III$^\prime$
above correspond to the three possible mass orderings of $h_1^0$.

By examining the non-vanishing $Zh_i h_j$ couplings,
one immediately concludes that the
relative CP quantum number of the other two neutral Higgs bosons
is negative.   However, there is no unique assignment for the
individual CP quantum numbers if $Z_7=\rho^Q=0$.
For simplicity, we assume that
$Z_7\neq 0$.\footnote{If $Z_7=0$ but $\rho^Q\neq 0$,
then our analysis still goes through with $Z_7$ replaced by
$\rho^{Q\,*}$ ($Q=U$, $D$ or $E$).}
The following identity, analogous to \eq{imz56}, will also prove useful:
\beqa \label{imz57}
\Im(Z_5^* Z_7^2)&=& 2\,\Re(Z_5 e^{-2i\theta_{23}})\,
\Re(Z_7\,e^{-i\theta_{23}})\,\Im(Z_7\,e^{-i\theta_{23}}) \nonumber \\
&& \qquad -\,\Im(Z_5 e^{-2i\theta_{23}})\left\{
[\Re(Z_7\,e^{-i\theta_{23}})]^2-[\Im(Z_7\,e^{-i\theta_{23}})]^2\right\}\,.
\eeqa

Under the assumption of a
CP-conserving Higgs sector, we impose the condition
$\Im(Z_5^* Z_7^2)=0$, which
implies that $[\Re(Z_5^* Z_7^2)]^2=|Z_5|^2\,|Z_7|^4$, and leads to
two possible cases:
\beq \label{app:z5z7e}
\Re(Z_5^* Z_7^2)=\varepsilon_{57}|Z_5||Z_7|^2\,,\qquad
\varepsilon_{57}=\pm 1\,.
\eeq
 In Cases I$^{\prime}$ and
II$^{\prime}$,
$\Im(Z_5\,e^{-2i\theta_{23}})=0$, whereas
$\Im(Z_5\,e^{-2i\overline\theta_{23}})=0$ in Case III$^\prime$.  Then,
\eq{imz57} yields:
\beqa \label{zz7}
\hspace{-3.5in}
{\rm Case~I}^\prime, {\rm II}^\prime: & \qquad &
\Re(Z_7\,e^{-i\theta_{23}})\,\Im(Z_7\,e^{-i\theta_{23}})=0\,.\label{zzp}
\\
{\rm Case~III}^\prime: & \qquad &
\Re(Z_7\,e^{-i\overline\theta_{23}})\,
\Im(Z_7\,e^{-i\overline\theta_{23}})=0\,.\label{zzpp}
\eeqa

We can use the
$H^+H^- h_k$ couplings to determine the CP-quantum numbers of the
other two neutral Higgs bosons.  In cases I$^\prime$ and II$^\prime$,
\beqa
H^+ H^- h_1: &\qquad\quad & c_{12} c_{13} Z_3 -s_{12}\Re(Z_7\,e^{-i\theta_{23}})
+c_{12} s_{13}\Im(Z_7\,e^{-i\theta_{23}})\,,\label{hh1}\\
H^+ H^- h_2: &\qquad\quad & s_{12} c_{13} Z_3 +c_{12}\Re(Z_7\,e^{-i\theta_{23}})
+s_{12} s_{13}\Im(Z_7\,e^{-i\theta_{23}})\,,\label{hh2}\\
H^+ H^- h_3: &\qquad\quad & s_{13} Z_3 -c_{13}\Im(Z_7\,e^{-i\theta_{23}})\,.
\label{hh3}
\eeqa
In Case III$^\prime$, these couplings simplify to:
\beqa
H^+ H^- h_1: &\qquad\quad & -\Im(Z_7\,e^{-i\overline\theta_{23}})\,,\label{hh13}\\
H^+ H^- h_2: &\qquad\quad & \phm\Re(Z_7\,e^{-i\overline\theta_{23}})\,,\label{hh23} \\
H^+ H^- h_3: &\qquad\quad & -Z_3\,.
\label{hh33}
\eeqa

Since one of the three neutral states is CP-odd, its coupling to $H^+H^-$
must vanish.  Taking $Z_3$ and $Z_7$ as independent and non-vanishing,
we can identify the CP-odd Higgs boson.
Hence, using \eqst{hh1}{hh33},
\vspace{0.1in}

$\bullet$ if $h_1$ is CP-odd, then either
\beq
s_{13}=c_{12}=\Re(Z_7\,e^{-i\theta_{23}})=0\qquad \text{[Case II$^\prime$b]}\,
\qquad {\rm or}\qquad
c_{13}=\Im(Z_7\,e^{-i\overline\theta_{23}})=0\qquad \text{[Case III$^\prime$a]}
\,,\label{hhcp1}
\eeq

$\bullet$ if $h_2$ is CP-odd, then either
\beq
s_{13}=s_{12}={\rm Re}(Z_7\,e^{-i\theta_{23}})=0
\qquad \text{[Case I$^\prime$b]}\,
\qquad {\rm or}\qquad
c_{13}={\rm Re}(Z_7\,e^{-i\overline\theta_{23}})=0
\qquad \text{[Case III$^\prime$b]} \,,\label{hhcp2}
\eeq

$\bullet$ if $h_3$ is CP-odd, then either
\beq
s_{13}=s_{12}={\rm Im}(Z_7\,e^{-i\theta_{23}})=0
\qquad \text{[Case I$^\prime$a]}\,
\qquad {\rm or}\qquad
s_{13}=c_{12}={\rm Im}(Z_7\,e^{-i\theta_{23}})=0
\qquad \text{[Case II$^\prime$a]}\,.
\eeq
These correspond to six possible mass orderings of $h_1$, $h_2$ and
$h_3$ in Cases I$^\prime$, II$^\prime$ and III$^\prime$.

We previously identified the CP-even state $h_1^0$, whose
couplings coincide with those of the Standard Model Higgs boson.
Using \eq{hmassesinv}, the squared-masses of the remaining two
neutral Higgs bosons (a~CP-even state $h_2^0$ and a CP-odd
state~$\ha$) are given by \eqs{mH2}{mH3},
after making use of \eq{rez57} for Cases I$^\prime$ and
II$^\prime$ (and replacing $\theta_{23}$ with $\overline{\theta}_{23}$ for Case III$^\prime$).
We identify the states $h_1^0$ and
$h_2^0$ with $\hl$ and $\hh$ or vice versa, depending on the mass-ordering.
If $Z_7=0$, then $\varepsilon_{57}$ is not well-defined (since in the
real basis, the sign of $Z_5$ can be flipped by transforming $H_2\to iH_2$).  In this case, the
individual CP-quantum numbers of $h_2^0$ and $\ha$ are not fixed
by the interactions of the Higgs boson/gauge boson sector.
The corresponding masses are given in \eq{m2hh}, which can be
derived by directly solving the characteristic
equation of the neutral Higgs squared-mass matrix [cf.~\eq{charpoly}].

A summary of the basis-independent conditions for CP-invariance,
under the assumption that $Z_6=0$ and the Higgs masses
are non-degenerate, along with the identification of the CP quantum
numbers of the three neutral Higgs states can be found in Table~\ref{Z60cond}.

\subsection{Degenerate masses in the CP-conserving limit with $\boldsymbol{Z_6=0}$}
It is possible to have two mass-degenerate neutral Higgs
bosons in the 2HDM with $Z_6=0$, for special choices of $Z_1$.
If $Z_1$ satisfies
\beq \label{z1a}
Z_1 = Y_2/v^2+\half(Z_3+Z_4-\varepsilon_{57}|Z_5|)v^2\,,
\eeq
then \eqs{mH1}{mH3} yield
$m_{h^0_1}=\mha$ [this is the analogue of \eq{z1special}].
Likewise, if  $Z_1$ satisfies
\beq \label{z1b}
Z_1 = Y_2/v^2+\half(Z_3+Z_4+\varepsilon_{57}|Z_5|)v^2\,,
\eeq
then \eqs{mH1}{mH2} yield $m_{h_1^0}=m_{h_2^0}$ (this
has no analogue with any of the $Z_6\neq 0$ cases).
In the presence of mass degeneracies, one must reconsider the definition
of the mixing angles $\theta_{12}$ and $\theta_{13}$.   As in the
discussion below \eq{mhma2}, if two neutral scalar states of opposite
CP quantum number are mass degenerate, then the structure
of the Higgs interactions guarantees that
there exists one linear combination of the mass-degenerate neutral Higgs states
that is CP-even and an orthogonal linear combination
that is CP-odd.  If the two neutral mass-degenerate scalar states are
CP-even, then there exists one linear combination whose properties
coincide precisely with those of the Standard Model Higgs boson.
We designate this scalar field by $h_1^0$ and the orthogonal linear
combination by $h_2^0$.  In light of these remarks, the
results of Table~\ref{Z60cond} continue to hold even in the
mass-degenerate case.

However, there are three new cases that arise if two of the
neutral Higgs fields are mass-degenerate, which are not accounted for by
Cases I$^\prime$, II$^\prime$ and III$^\prime$.  These exceptional cases
correspond to the omitted cases described below \eq{spcase3}.
In particular,
mass-degeneracies arise for a special choice of $Z_1$ in following cases:
\vskip 0.1in

\textbf{Case IV$\boldsymbol{^\prime}$:}
$\boldsymbol{s_{13}={\rm Im}(Z_5 e^{-2i\theta_{23}})=0}$
{\bf and} $\boldsymbol{s_{12}c_{12}\neq 0.}$
In this case, \eq{spcase3} yields
\beq
Z_1=A^2/v^2+{\rm Re}(Z_5 e^{-2i\theta_{23}})
=Y_2/v^2+\half\left[Z_3+Z_4+{\rm Re}(Z_5 e^{-2i\theta_{23}})\right]\,,
\eeq
where we have used the definition of $A^2$ given in \eq{defA2}.
The quantity ${\rm Re}(Z_5 e^{-2i\theta_{23}})$ is fixed by \eq{rez57}.
\vskip 0.1in

\textbf{Case V$\boldsymbol{^\prime}$:}
$\boldsymbol{c_{12}={\rm Im}(Z_5 e^{-2i\theta_{23}})=0}$
{\bf and} $\boldsymbol{s_{13}c_{13}\neq 0.}$
In this case, \eq{spcase2} yields
\beq
Z_1=A^2/v^2
=Y_2/v^2+\half\left[Z_3+Z_4-{\rm Re}(Z_5 e^{-2i\theta_{23}})\right]\,.
\eeq
\vskip 0.1in

\textbf{Case VI$\boldsymbol{^\prime}$:}
$\boldsymbol{s_{12}={\rm Im}(Z_5 e^{-2i\theta_{23}})=0}$
{\bf and} $\boldsymbol{s_{13}c_{13}\neq 0.}$
In this case, \eq{spcase2} yields
\beq
Z_1=A^2/v^2
=Y_2/v^2+\half\left[Z_3+Z_4-{\rm Re}(Z_5 e^{-2i\theta_{23}})\right]\,.
\eeq
\vskip 0.1in



\noindent
Once again, we find that $m_{h_1^0}=m_{A^0}$ if \eq{z1a} is satisfied
and  $m_{h_1^0}=m_{h_2^0}$ if \eq{z1b} is satisfied.

To identify the CP quantum numbers of the neutral Higgs mass
eigenstates, we first examine the $G^+G^- h_k$ couplings given
in \eqst{ggh10}{ggh30} in order to identify the mass-degenerate
state $h_1^0$, which is
defined below \eq{z1b} to be the linear combination of
mass-degenerate neutral Higgs fields
whose interactions coincide with that of the Standard Model Higgs boson .
The CP quantum numbers of the orthogonal linear combination of
mass-degenerate neutral Higgs fields and the third non-degenerate state
can be obtained by examining the $H^+ H^-h_k$ couplings given
in \eqst{hh1}{hh3}.

For example, in Case IV$^\prime$, $s_{13}=0$
which yields
\beqa
H^+ H^- h_1: &\qquad & c_{12} Z_3 -s_{12}\Re(Z_7\,e^{-i\theta_{23}})
\,,\qquad m_{h_1}^2=Z_1 v^2\,,\label{hh1iv}\\
H^+ H^- h_2: &\qquad & s_{12} Z_3 +c_{12}\Re(Z_7\,e^{-i\theta_{23}})
\,,\qquad m_{h_2}^2=Z_1 v^2\,,\label{hh2iv}\\
H^+ H^- h_3: &\qquad & -\Im(Z_7\,e^{-i\theta_{23}})
\,,\qquad\qquad\qquad\, m^2_{h_3}=
\left[Z_1-{\rm Re}(Z_5 e^{-2i\theta_{23}})\right]v^2\,.
\label{hh3iv}
\eeqa
Since $h_1$ and $h_2$ are degenerate, we can redefine new linear
combinations to obtain:
\beqa
H^+ H^- (c_{12}h_1+s_{12}h_2): &\qquad &
\,\,\,\,c_{12} Z_3 \,,\label{hh1iva}\\
H^+ H^- (c_{12}h_2-s_{12}h_1): &\qquad & \phm\Re(Z_7\,e^{-i\theta_{23}})
\,,\label{hh2iva}\\
H^+ H^- h_3: &\quad & -\Im(Z_7\,e^{-i\theta_{23}})
\,.\label{hh3iva}
\eeqa
Likewise, the corresponding $G^+G^- h_k$ interactions are:
\beqa
G^+ G^- (c_{12}h_1+s_{12}h_2): &\qquad &
\!Z_1 \,,\\
G^+ G^- (c_{12}h_2-s_{12}h_1): &\qquad & 0
\,,\\
G^+ G^- h_3: &\quad & 0
\,.
\eeqa
Thus, one can immediately identify $h_1^0=c_{12}h_1+s_{12}h_2$, since
this linear combination possesses the Higgs couplings of the Standard
Model Higgs boson. The second CP-even Higgs
state is identified by its non-zero coupling to $H^+ H^-$ and
depends on whether $Z_7 e^{-i\theta_{23}}$ is purely real or
purely imaginary.  For example, if
${\rm Im}(Z_7 e^{-i\theta_{23}})=0$, then $c_{12}h_2-s_{12} h_1$
is CP-even and $h_3$ is CP-odd, and vice versa if
${\rm Re}(Z_7 e^{-i\theta_{23}})=0$.

Cases V$^\prime$ and VI$^\prime$ can be similarly treated.  In
particular,
\beqa
&& \phantom{\rm I}\text{Case V}^\prime:\quad m^2_{h_1}=
\left[Z_1+{\rm Re}(Z_5 e^{-2i\theta_{23}})\right]v^2\,,\qquad
m^2_{h_2}=m^2_{h_3}=Z_1 v^2\,,\\
&& \text{Case VI}^\prime:\quad m^2_{h_2}=
\left[Z_1+{\rm Re}(Z_5 e^{-2i\theta_{23}})\right]v^2\,,\qquad
m^2_{h_1}=m^2_{h_3}=Z_1 v^2\,.
\eeqa
If we impose the mass ordering $m_{h_1}\leq m_{h_2}\leq m_{h_3}$ in
order not to duplicate regions of the 2HDM parameter space, then
we can omit Case VI$^\prime$.
We summarize the exceptional mass-degenerate cases in Tables~\ref{Z60cond2},
\ref{d4}, \ref{d5} and \ref{d6}.
\begin{table}[ht!]
\centering
\caption{Basis-independent conditions
for a CP-conserving 2HDM scalar potential and vacuum
when $Z_6=0$ and $Z_5$, $Z_7\neq 0$, assuming
at least two degenerate neutral Higgs boson masses.
The cases below are exceptional, as they do not arise as
limits of Cases I$^\prime$, II$^\prime$ and III$^\prime$
[cf.~Table~\ref{Z60cond}].  If we impose
the mass-ordering $m_{h_1} \leq m_{h_2}\leq m_{h_3}$, then
Cases VI$^\prime$a and b can be eliminated.
The neutral Higgs mixing angles $\theta_{12}$ in Case IV$^\prime$
and $\theta_{13}$ in Cases  V$^\prime$ and VI$^\prime$
are defined such that the couplings of
$h_1^0$ (defined as the linear combination of mass-degenerate
neutral Higgs fields specified below) coincides precisely with those
of the Standard Model Higgs boson.
The phase factor $\eta^2$ that governs the CP transformation law
[cf.~\eq{CPgen}] is equal to $+1$ in cases IV$^\prime$a, V$^\prime$a,
and VI$^\prime$a, and $-1$ in cases IV$^\prime$b, V$^\prime$b, and
VI$^\prime$b.
Additional conditions in which $Z_7$ is replaced
by $\rho^{Q\,*}$ ($Q=U,D$ and $E$), respectively, must also hold due to
the phase correlations implicit in \eqs{correlation1}{correlation2}.
The squared-mass of the two mass-degenerate neutral Higgs states is equal to
$Z_1 v^2$, while the third non-degenerate neutral state
has a squared-mass equal to $(Z_1\pm
\epsilon_{57}|Z_5|)v^2$, where the plus sign is taken in
cases IV$^\prime$b, V$^\prime$a, and VI$^\prime$a, and the minus sign
is taken in cases IV$^\prime$a, V$^\prime$b, and VI$^\prime$b.
\label{Z60cond2}}
\vskip 0.2in
\begin{tabular}{|c||c|c|c|c|}\hline
\phaa Cases\phaa & \phaa conditions [in all cases below,
$\Im(Z_5\,e^{-2i\theta_{23}})=0$]
\phaa & \phaa $A^0$  \phaa & \phaa $h_1^0$ \phaa &
 \phaa $h_2^0$ \phaa  \\ \hline
IV$^\prime$a & $\,s_{13}
=\Im(Z_7\,e^{-i\theta_{23}})=0$\,,\,
$Z_1=Y_2/v^2+\half\left(Z_3+Z_4+\epsilon_{57}|Z_5|\right)\,$
& $h_3$ &  $c_{12}h_1+s_{12}h_2$
& $c_{12}h_2-s_{12}h_1$ \\
IV$^\prime$b & $\,s_{13}
=\Re(Z_7\,e^{-i\theta_{23}})=0$\,,\,
$Z_1=Y_2/v^2+\half\left(Z_3+Z_4-\epsilon_{57}|Z_5|\right)\,$
&  $c_{12}h_2-s_{12}h_1$ & $c_{12}h_1+s_{12}h_2$ & $h_3$ \\
V$^\prime$a & $\,c_{12}
=\Im(Z_7\,e^{-i\theta_{23}})=0$\,,\,
$Z_1=Y_2/v^2+\half\left(Z_3+Z_4-\epsilon_{57}|Z_5|\right)\,$
& $c_{13}h_3+s_{13}h_2$ & $c_{13}h_2-s_{13}h_3$  & $h_1$ \\
V$^\prime$b & $\,c_{12}
=\Re(Z_7\,e^{-i\theta_{23}})=0$\,,\,
$Z_1=Y_2/v^2+\half\left(Z_3+Z_4+\epsilon_{57}|Z_5|\right)\,$
& $h_1$ &  $c_{13}h_2-s_{13}h_3$ & $c_{13}h_3+s_{13}h_2$\\
VI$^\prime$a & $\,s_{12}
=\Im(Z_7\,e^{-i\theta_{23}})=0$\,,\,
$Z_1=Y_2/v^2+\half\left(Z_3+Z_4-\epsilon_{57}|Z_5|\right)\,$
& $c_{13}h_3-s_{13}h_1$ & $c_{13}h_1+s_{13}h_3$ & $h_2$ \\
VI$^\prime$b & $\,s_{12}
=\Re(Z_7\,e^{-i\theta_{23}})=0$\,,\,
$Z_1=Y_2/v^2+\half\left(Z_3+Z_4+\epsilon_{57}|Z_5|\right)\,$
& $h_2$& $c_{13}h_1+s_{13}h_3$ & $c_{13}h_3-s_{13}h_1$ \\
\hline
\end{tabular}
\end{table}
\begin{table}[ht!]
\begin{minipage}[t]{2.0in}
\centering
\parbox[t]{1.75in}{\caption{The U(2)-invariant quantities $q_{k\ell}$
for Cases IV$^\prime$a and IV$^\prime$b}\label{d4}} \\ \vspace{0.1in}
\begin{tabular}{|c||c|c|}\hline
$\phaa k\phaa $ &\phaa $q_{k1}\phaa $ & \phaa $q_{k2} \phaa $ \\ \hline
$1$ & $c_{12}$ & $-s_{12}$ \\
$2$ & $s_{12}$ & $\phm c_{12}$ \\
$3$ & $0$ & $i$ \\ \hline
\end{tabular}
\end{minipage}
\hfill
\begin{minipage}[t]{2.0in}
\centering
\parbox[t]{1.75in}{\caption{The U(2)-invariant quantities $q_{k\ell}$
for Cases V$^\prime$a and V$^\prime$b}\label{d5}} \\ \vspace{0.1in}
\begin{tabular}{|c||c|c|}\hline
$\phaa k\phaa $ &\phaa $q_{k1}\phaa $ & \phaa $q_{k2} \phaa $ \\ \hline
$1$ & $c_{13}$ & $-is_{13}$ \\
$2$ & $0$ & $1$ \\
$3$ & $s_{13}$ & $\phm ic_{13}$ \\ \hline
\end{tabular}
\end{minipage}
\hfill
\begin{minipage}[t]{2.0in}
\centering
\parbox[t]{1.75in}{\caption{The U(2)-invariant quantities $q_{k\ell}$
for Cases VI$^\prime$a and VI$^\prime$b}\label{d6}} \\ \vspace{0.1in}
\begin{tabular}{|c||c|c|}\hline
$\phaa k\phaa $ &\phaa $q_{k1}\phaa $ & \phaa $q_{k2} \phaa $ \\ \hline
$1$ & $\phm 0$ & $1$ \\
$2$ & $-c_{13}$ & $is_{13}$ \\
$3$ & $\phm s_{13}$ & $ic_{13}$ \\ \hline
\end{tabular}
\end{minipage}
\end{table}

In the analysis presented above, we assumed that $Z_5\neq 0$.
If $Z_5=Z_6=0$, then \eqs{mH2}{mH3} imply that $h_2^0$ and $A^0$
are mass-degenerate, independently of any special choice
for $Z_1$. In the special case of
$Z_1=Y_2/v^2+\half(Z_3+Z_4)$, all three neutral scalars are degenerate in
mass.  Moreover, the invariant form of the
Higgs squared-mass matrix given in \eq{mtilmatrix} is diagonal.
Thus, in this case it is simplest to take $\theta_{12}=\theta_{13}=
\theta_{23}=0$ (instead of imposing a mass ordering of the $h_k$ fields).
\Eq{ggh10} implies that $h_1$ is the CP-even neutral Higgs
field whose couplings coincide with that of the Standard Model Higgs
boson. \Eqs{hh2}{hh3} imply that $h_3$ is CP-odd and $h_2$ is
CP-even if $\Im(Z_7 e^{-i\theta_{23}})=0$ and vice versa if
$\Re(Z_7 e^{-i\theta_{23}})=0$.  In this case, $\theta_{23}$ simply
keeps track of the overall phase of $Z_7$.  Finally, in the special case
of $Z_5=Z_6=Z_7=0$ (cf.~Section~\ref{cp1}), the individual CP quantum
numbers of $h_2$ and $h_3$ can not be determined from the bosonic sector
alone.

In Section~\ref{invcust}, we determined the basis-independent conditions for
a custodial symmetric scalar potential.  In the case of
$Z_6=0$ and $Z_5$, $Z_7\neq 0$,
the relevant condition is $Z_4 =\epsilon_{57}|Z_5|$ [cf.~\eq{basindcust}],
which yields $m_{A^0} =\mc$  [cf.~\eq{equalmasses}].  If we apply this limit to Table~\ref{Z60cond2},
we discover that there are two possibilities: either $A^0$ is degenerate in mass with $H^\pm$
(cases IV$^\prime$a, V$^\prime$b and VI$^\prime$b), or there are two neutral fields,
one CP-even and one CP-odd, that are degenerate in mass
with  $H^\pm$ (cases IV$^\prime$b, V$^\prime$a, and VI$^\prime$a).
If $Z_5\neq 0$, $Z_6=Z_7=0$ and the Higgs--fermion interactions are CP-conserving, then as
shown in Section~\ref{twisted}, there are two possible conditions, $Z_4=\pm\epsilon_{5Q}|Z_5|$,
that yield a custodial symmetric scalar potential.
Table~\ref{Z60cond2} can again be used if
one replaces $\varepsilon_{57}$ replaced by $\varepsilon_{5Q}$.
As shown in \eq{equalmasses2}, the relation $Z_4=\epsilon_{5Q}|Z_5|$ yields $m_{A^0} =\mc$,
and one recovers the results given above for the possible Higgs mass degeneracies.
In contrast, the relation $Z_4=-\epsilon_{5Q}|Z_5|$
yields $m_{H^0}=\mc$.  In this case, there are again
two possibilities: either $H^0$ is degenerate in mass with $H^\pm$
(cases IV$^\prime$b, V$^\prime$a and VI$^\prime$a), or there are two neutral CP-even fields,
$h_1^0$ and $h_2^0$, that are degenerate in mass
with  $H^\pm$ (cases IV$^\prime$a, V$^\prime$b, and VI$^\prime$b).
If one now imposes one additional condition, $Z_4=Z_5=0$, then
all three neutral Higgs bosons are degenerate with the charged Higgs boson.
Hence, any permutation of possible neutral Higgs mass-degeneracies
with the charged Higgs boson is a possible
consequence of custodial symmetry, if one allows for
sufficiently restrictive conditions on the scalar
potential.

\section{Calculation of the 2HDM contributions to $\boldsymbol{S}$, $\boldsymbol{T}$ and $\boldsymbol{U}$}\label{app:one}

The one-loop corrections to the gauge boson two-point functions
contain three- and four-point interactions between gauge bosons and
the Higgs bosons, the form of which can be read off from
\eqs{VVH}{VVHH}.  The resulting Feynman rules in the
t'Hooft-Feynman gauge are exhibited in Table~\ref{tabFeyn}.  The 2HDM
contributions to $S$ are displayed in Tables~\ref{tabS1} and \ref{tabS2}.
The 2HDM contributions to
$T$ are displayed in Tables~\ref{tabT1} and \ref{tabT2}.
The 2HDM contributions to $U$ are displayed in Table~\ref{tabU}.
The reference Standard Model contributions, which are
subtracted out from the 2HDM contributions, are shown in
Table~\ref{tabSM}.

\begin{table}[ht!]
\centering
\caption{Feynman rules used in the calculation of the oblique parameters.
The four momentum $p_1$ points into the vertex, and the four-momentum
$p_2$ points out of the vertex.}
\label{tabFeyn}
\begin{tabular}{|c|c|}\hline
\feynrulefour{W_+}{h_k}{h_k}{\half ig^2 g^{\mu\nu}} {}&
\feynrulefour{Z}{h_k}{h_k}{\frac{ig^2}{2c_W^2} g^{\mu\nu} }{}\\
\feynrulefour{W_+}{H^+}{H^+}{\half ig^2}{}&
\feynrulefour{Z}{H^+,G^+}{H^+,G^+}{\frac{ig^2}{2c_W^2}(c_{W}^2-s_W^2)}{}\\
\feynruleVV{W_+}{W_+}{h_k}{ig q_{ik}m_W  }  & \feynruleVV{Z}{Z}{h_i}{\frac{ig}{c_W}q_{i1}m_Z }
\\
\feynrule{W_+}{H^-}{h_k}{-\half ig q_{k2}}&
\feynrule{Z}{h_3,G^0}{h_2,h_1}{\frac{g}{2 c_W} q_{11}}\\
\feynrule{W_+}{G^-}{h_k}{-\half ig q_{k1}}&
\feynrule{Z}{h_1,G^0}{h_3,h_2}{\frac{g}{2 c_W}q_{21}}
\\
\feynrule{W_+}{G^-}{\phi}{-\half ig}&\feynrule{Z}{h_2,G^0}{h_1,h_3}{\frac{g}{2 c_W}q_{31}}\\
\feynrule{\gamma}{H^+,G^+}{H^+,G^+}{i g s_W}&\feynrule{Z}{H^+}{H^+}{\frac{ig}{2 c_W}(c_W^2-s_W^2)}\\
\feynruleVV{W_+}{W_+}{\phi}{ig m_W }& \feynruleVV{Z}{Z}{\phi}{\frac{ig}{c_W} m_Z  } \\
& \feynrule{Z}{\phi}{G^0}{- \frac{g}{2c_W}}\\\hline
\end{tabular}
\vspace{0.2in}
\end{table}

The loop integrals are defined and evaluated
following ref.~\cite{langacker}:
\beqa
\int \frac{d^4k}{(2\pi)^4} \frac{1}{(k^2-m^2)} &=&
\frac{i}{16\pi^2}A_0(m^2)\,,
\eeqa
\beqa
\int \frac{d^4k}{(2\pi)^4} \frac{1}{(k^2-m_1^2)[(k+q)^2-m_2^2]} &=& \frac{i}{16\pi^2}B_0(q^2;m_1^2,m_2^2)\,,\\
\int \frac{d^4k}{(2\pi)^4} \frac{k^\mu k^\nu}{(k^2-m_1^2)[(k+q)^2-m_2^2]} &=& \frac{i}{16\pi^2}g^{\mu\nu}B_{22}(q^2;m_1^2,m_2^2)\,.
\eeqa
The following two relations are noteworthy:
\beqa
B_{0}(0;m_1^2,m_2^2) &=& \frac{A_0(m_1^2)-A_0(m_2^2)}{m_1^2-m_2^2}\,,\label{bzerodef} \\
4 B_{22}(0;m_1^2,m_2^2) &=& \mathcal{F}(m_1^2,m_2^2)+A_0(m_1^2)+A_0(m_2^2)\,,\label{BtoF}
\eeqa
where
\beq \label{calfdef}
\mathcal{F}(m_1^2,m_2^2) \equiv \half (m_1^2+m_2^2)-\frac{m_1^2m_2^2}{m_1^2-m_2^2}\ln\left(\frac{m_1^2}{m_2^2}\right)\,.
\eeq

The contributions to $S$ from the
diagrams in Table~\ref{tabS1}, \ref{tabS2} and Table~\ref{tabSM} are
evaluated by employing \eq{piA} and \eq{Sdef}, with the following
result:
\footnote{The $2H$ superscript indicates the 2HDM contributions
and the $SM$ superscript indicates the contributions from
the Standard Model with a reference Higgs mass $m_\phi$ that is
subtracted off from the 2HDM result.  This subtraction procedure is
necessary in order to get a finite result.}
\beqa
 S &\equiv& \frac{16\pi c_W^2}{g^2}\left\{F_{ZZ}^{2H}(m_Z^2)
- F_{ZZ}^{SM}(m_Z^2) -F_{\gamma\gamma}^{2H}(m_Z^2)
+ F_{\gamma\gamma}^{SM}(m_Z^2)
-\frac{c_{2W}}{s_W c_W}\left[F_{Z\gamma}^{2H}(m_Z^2)-
F_{Z\gamma}^{SM}(m_Z^2)\right]\right\}\nonumber\\
&=&
\frac{1}{\pi m_Z^2} \Biggl\{\sum_{k=1}^3 q_{k1}^2\biggl[
\mathcal{B}_{22}(m_Z^2;m_Z^2,m_k^2)- m_Z^2
\mathcal{B}_{0}(m_Z^2;m_Z^2,m_k^2)\biggr]
+q_{11}^2 \mathcal{B}_{22}(m_Z^2;m_2^2,m_3^2)
+q_{21}^2\mathcal{B}_{22}(m_Z^2;m_1^2,m_3^2)\nonumber \\
& & \qquad\qquad +q_{31}^2\mathcal{B}_{22}(m_Z^2;m_1^2,m_2^2)
- \mathcal{B}_{22}(m_Z^2;{m^2_{H^\pm}},{m^2_{H^\pm}})
-\mathcal{B}_{22}(m_Z^2;m_Z^2,m_\phi^2)
+ m_Z^2\mathcal{B}_{0}(m_Z^2;m_Z^2,m_\phi^2)\Biggr\}\,,\nonumber
\\
\phantom{line}
\eeqa
where the $F_{ij}(m_V^2)$ are defined in \eq{piA}.

%
\begin{table}[ht!]
\centering
\caption{Diagrams representing the 2HDM contributions to $S$, part 1.}
\label{tabS1}
\begin{tabular}{|c|}\hline
Contributions to $\Pi_{ZZ}^{2H}(m_Z^2)$ \T \Bot\\ \hline
\Gaugeprop{Z}{Z}{Z}{\hi}{=-\frac{g^2M_Z^2}{16 \pi^2 c_W^2}q_{i1}^2B_0(m_Z^2;m_Z^2,m_i^2)}\\
\Loopgraph{Z}{Z}{G^0}{\hi}{=\gc q_{i1}^2B_{22}(m_Z^2;m_Z^2,m_i^2)}{}\\
\Loopgraph{Z}{Z}{h_3}{h_1}{=\frac{g^2}{16 \pi^2 c_W^2}q_{21}^2B_{22}(m_Z^2;m_1^2,m_3^2)}{}\\
\Loopgraph{Z}{Z}{h_3}{h_2}{=\frac{g^2}{16 \pi^2 c_W^2}q_{11}^2B_{22}(m_Z^2;m_2^2,m_3^2)}{}\\
\Loopgraph{Z}{Z}{h_1}{h_2}{=\frac{g^2}{16 \pi^2 c_W^2}q_{31}^2B_{22}(m_Z^2;m_1^2,m_2^2)}{}\\
\Loopgraph{Z}{Z}{H^+}{H^+}{=\gc c_{2W}^2 B_{22}(m_Z^2;{m^2_{H^\pm}},{m^2_{H^\pm}})}{}\\ \hline
\end{tabular}
\end{table}

\begin{table}[ht!]
\centering
\caption{Diagrams representing the 2HDM contributions to $S$, part 2.}
\label{tabS2}
\begin{tabular}{|c|}\hline
Contributions to $\Pi_{\gamma\gamma}^{2H}(m_Z^2)$ and $\Pi_{Z\gamma}^{2H}(m_Z^2)$\T \Bot\\ \hline
 \Loopgraph{\gamma}{\gamma}{H^+}{H^+}{=4\gtwo s_W^2 B_{22}(m_Z^2;{m^2_{H^\pm}},{m^2_{H^\pm}})}{}\\
\Loopgraph{Z}{\gamma}{H^+}{H^+}{= \frac{2 g^2}{16\pi^2c_W} s_W  c_{2W} B_{22}(m_Z^2;{m^2_{H^\pm}},{m^2_{H^\pm}})}{} \\ \hline
\end{tabular}
\end{table}

The parameter $T$ can be calculated in a similar manner, where
\eqst{bzerodef}{calfdef} are especially useful.
Adding the contributions to $T$ from all the diagrams shown in Tables~\ref{tabT1},~\ref{tabT2} and~\ref{tabSM} yields
\beqa
\alpha T&\equiv&\frac{A^{2H}_{WW}(0)}{m_W^2}- \frac{A^{2H}_{ZZ}(0)}{m_Z^2}-\left[\frac{ A^{SM}_{WW}(0)}{m_W^2}- \frac{A^{SM}_{ZZ}(0)}{m_Z^2}\right]\nonumber\\
&=&\frac{g^2}{16\pi^2m_W^2}\Biggl\{\sum_{k=1}^3 |q_{k2}|^2
B_{22}(0;{m^2_{H^\pm}},m_k^2)-q_{11}^2 B_{22}(0;m_2^2,m_3^2)
-q_{21}^2 B_{22}(0;m_1^2,m_3^2)-q_{31}^2 B_{22}(0;m_1^2,m_2^2)\nonumber\\
& &\qquad\qquad
+\sum_{k=1}^3 q_{k1}^2\biggl[B_{22}(0;m_W^2,m_k^2)-B_{22}(0;m_Z^2,m_k^2)
-m_W^2 B_0(0;m_W^2,m_k^2)+m_Z^2 B_0(0;m_Z^2,m_k^2)\biggr]\nonumber \\
& & \qquad -\tfrac{1}{2}A_0({m^2_{H^\pm}})
-B_{22}(0;m_W^2,m_\phi^2)+B_{22}(0;m_Z^2,m_\phi^2)
+m_W^2 B_0(0;m_W^2,m_\phi^2)-m_Z^2 B_0(0;m_Z^2,m_\phi^2)\Biggr\},
\nonumber\\
\phantom{line}
\eeqa
where the $A_{ij}(0)$ are defined in \eq{piA}.
Using $\alpha =g^2 s_W^2/(4\pi)$ to isolate $T$ and simplifying the result
by employing \eq{BtoF}, we end up with:

\beqa \label{generalformT}
T &=& \frac{1}{16\pi m_W^2s_W^2} \Biggl\{\,\sum_{k=1}^3|q_{k2}|^2
\mathcal{F}({m^2_{H^\pm}},m_k^2)-q_{11}^2 \mathcal{F}(m_2^2,m_3^2)
-q_{21}^2 \mathcal{F}(m_1^2,m_3^2)-q_{31}^2 \mathcal{F}(m_1^2,m_2^2)\nonumber\\
& &+\sum_{k=1}^3 q_{k1}^2\biggl[\mathcal{F}(m_W^2,m_k^2)-\mathcal{F}(m_Z^2,m_k^2)
-4m_W^2 B_0(0;m_W^2,m_k^2)+4m_Z^2 B_0(0;m_Z^2,m_k^2)\bigr]\biggr]\nonumber\\
&&+\mathcal{F}(m_Z^2,m_\phi^2) -\mathcal{F}(m_W^2,m_\phi^2)
+4m_W^2 B_0(0;m_W^2,m_\phi^2)-4m_Z^2 B_0(0;m_Z^2,m_\phi^2)\Biggr\}\,.
\eeqa
\begin{table}[ht!]
\addtolength{\tabcolsep}{5pt}
\centering
\caption{Diagrams representing the 2HDM contributions to $T$, part 1.}
\label{tabT1}
\begin{tabular}{|c|}\hline
\qquad \quad \qquad \quad Contributions to $A_{WW}^{2H}(0)$ \T \Bot \qquad \quad \qquad \quad\\ \hline
\gaugeprop{W^+}{W^+}{W^+}{\hi}{=-\frac{g^2m_W^2}{16 \pi^2}q_{i1}^2B_0(0;m_W^2,m_i^2)}\\
\loopgraph{W^+}{W^+}{G^+}{\hi}{=\gtwo q_{i1}^2 B_{22}(0;m_W^2,m_i^2)}{}\\
\loopgraph{W^+}{W^+}{h_1}{H^+}{=\gtwo |q_{12}|^2B_{22}(0;{m^2_{H^\pm}},m_1^2)}{}\\
\loopgraph{W^+}{W^+}{h_2}{H^+}{=\gtwo |q_{22}|^2B_{22}(0;{m^2_{H^\pm}},m_2^2)}{}\\
\loopgraph{W^+}{W^+}{h_3}{H^+}{=\gtwo |q_{32}|^2B_{22}(0;{m^2_{H^\pm}},m_3^2)}{}\\
\fourpoint{W^+}{W^+}{\hi}{=-\frac{1}{2}\gtwo A_0(m_i^2)}\\
\fourpoint{W^+}{W^+}{H^+}{=-\frac{1}{2}\gtwo A_0({m^2_{H^\pm}})}\\ \hline
\end{tabular}
\end{table}
\clearpage
\begin{table}[ht!]
\addtolength{\tabcolsep}{20pt}
\centering
\caption{Diagrams representing the 2HDM contributions to $T$, part 2.}
\label{tabT2}
\begin{tabular}{|c|}\hline
\qquad \quad \qquad \quad Contributions to $A_{ZZ}^{2H}(0)$\qquad \quad \qquad \quad\\ \hline
\gaugeprop{Z}{Z}{Z}{\hi}{=-\frac{g^2m_Z^2}{16 \pi^2 c_W^2}q_{i1}^2B_0(0;m_Z^2,m_i^2)}\\
\loopgraph{Z}{Z}{G^0}{\hi}{=\gc q_{i1}^2B_{22}(0;m_Z^2,m_i^2)}{}\\
 \loopgraph{Z}{Z}{h_3}{h_1}{=\frac{g^2}{16 \pi^2 c_W^2}q_{21}^2B_{22}(0;m_1^2,m_3^2)}{}\\
\loopgraph{Z}{Z}{h_3}{h_2}{=\frac{g^2}{16 \pi^2 c_W^2}q_{11}^2B_{22}(0;m_2^2,m_3^2)}{}\\
\loopgraph{Z}{Z}{h_1}{h_2}{=\frac{g^2}{16 \pi^2 c_W^2}q_{31}^2B_{22}(0;m_1^2,m_2^2)}{}\\
\fourpoint{Z}{Z}{\hi}{=-\frac{1}{2}\gc A_0(m_i^2)}\\
\fourpoint{Z}{Z}{H^+}{=-\frac{1}{2}\gc c_{2W}^2 A_0({m^2_{H^\pm}})}\\
\loopgraph{Z}{Z}{H^+}{H^+}{=\gc c_{2W}^2 B_{22}(0;{m^2_{H^\pm}},{m^2_{H^\pm}})}{=\frac{1}{2}\gc c_{2W}^2 A_0({m^2_{H^\pm}})}\\ \hline
\end{tabular}
\end{table}

Lastly, adding all of the contributions to $S+U$ in Tables~\ref{tabU} and~\ref{tabSM} gives the following:
\beqa
S+U &=& \frac{16\pi}{g^2}\left[F_{WW}(m_W^2)
- F_{\gamma\gamma}(m_W^2)-\frac{c_W}{s_W}F_{Z\gamma}(m_W^2)\right]\nonumber\\[8pt]
 &=&\frac{1}{\pi m_W^2} \Biggl\{-\sum_{k=1}^3 q_{k1}^2 m_W^2
\mathcal{B}_{0}(m_W^2;m_W^2,m_k^2)
+ m_W^2 \mathcal{B}_{0}(m_W^2;m_W^2,m_\phi^2)
- \mathcal{B}_{22}(m_W^2;m_W^2,m_\phi^2) \nonumber\\[6pt]
& & + \sum_{k=1}^3\biggl[q_{k1}^2\mathcal{B}_{22}(m_W^2;m_W^2,m_k^2)
+|q_{k2}|^2 \mathcal{B}_{22}(m_W^2;{m^2_{H^\pm}},m_k^2)\biggr]
-2 \mathcal{B}_{22}(m_W^2;{m^2_{H^\pm}},{m^2_{H^\pm}})\Biggr\}\,.
\eeqa

The formulae for the
oblique parameters in a general extended-Higgs sector model with an
arbitrary number of scalar singlets and doublets has been presented in
\Ref{lavoura}.  In contrast to our above results, the
treatment of \Ref{lavoura} employs a basis-dependent
scalar-mixing matrix.  In particular, our expressions for
the oblique parameters depend only on the masses of the physical
Higgs fields and the basis-invariant functions~$q_{k\ell}$.

\begin{table}[hb!]
\addtolength{\tabcolsep}{20pt}
\centering
\caption{Diagrams representing the 2HDM contributions to $S+U$.}
\label{tabU}
\begin{tabular}{|c|}\hline
Contributions to $\Pi_{WW}^{2H}(m_W^2)$ \T \Bot\\ \hline
\gaugeprop{W^+}{W^+}{W^+}{\hi}{=-\frac{g^2m_W^2}{16 \pi^2}q_{i1}^2B_0(m_W^2;m_W^2,m_i^2)}\\
\loopgraph{W^+}{W^+}{G^+}{\hi}{=\gtwo q_{i1}^2 B_{22}(m_W^2;m_W^2,m_i^2)}{} \\
\loopgraph{W^+}{W^+}{H^+}{\hi}{=\gtwo |q_{i2}|^2B_{22}(m_W^2;{m^2_{H^\pm}},m_i^2)}{}\\ \hline
\qquad \quad Contributions to $\Pi_{\gamma\gamma}^{2H}(m_W^2)$ and $\Pi_{Z\gamma}^{2H}(m_W^2)$ \qquad\quad\\ \hline
\loopgraph{\gamma}{\gamma}{H^+}{H^+}{=4\gtwo s_W^2 B_{22}(m_W^2;{m^2_{H^\pm}},{m^2_{H^\pm}})}{}\\
\loopgraph{Z}{\gamma}{H^+}{H^+}{= 2\frac{g^2s_W c_{2W} }{16\pi^2c_W} B_{22}(m_W^2;{m^2_{H^\pm}},{m^2_{H^\pm}})}{}\\ \hline
\end{tabular}
\end{table}
\begin{table}[ht!]
\addtolength{\tabcolsep}{5pt}
\small
\centering
\caption{Standard Model contributions to the oblique parameters.}
\label{tabSM}
\begin{tabular}{|c|}\hline
 Contributions to $\Pi_{WW}^{SM}(m_W^2)$ and $\Pi_{ZZ}^{SM}(m_Z^2)$\T \Bot \\ \hline
\gaugeprop{W^+}{W^+}{W^+}{\phi}{=-\frac{g^2m_W^2}{16\pi^2}B_0(m_W^2;m_W^2,m_1^2)}\\
\loopgraph{W^+}{W^+}{G^+}{\phi}{=\gtwo B_{22}(m_W^2;m_W^2,m_1^2)}{}\\
\gaugeprop{Z}{Z}{Z}{\phi}{=-\frac{g^2m_Z^2}{16 \pi^2 c_W^2}B_0(m_Z^2;m_Z^2,m_1^2)}\\
\loopgraph{Z}{Z}{G^0}{\phi}{= \gc B_{22}(m_Z^2;m_Z^2,m_1^2)}{}\\ \hline
Contributions to $A_{WW}^{SM}(0)$ and $A_{ZZ}^{SM}(0)$\T \Bot\\ \hline
\gaugeprop{W^+}{W^+}{W^+}{\phi}{=-\frac{g^2m_W^2}{16\pi^2}B_0(0;m_W^2,m_1^2)}\\
\loopgraph{W^+}{W^+}{G^+}{\phi}{=\gtwo B_{22}(0;m_W^2,m_1^2)}{}\\
\gaugeprop{Z}{Z}{Z}{\phi}{=-\frac{g^2m_Z^2}{16 \pi^2 c_W^2}B_0(0;m_Z^2,m_1^2)}\\
\loopgraph{Z}{Z}{G^0}{\phi}{= \gc B_{22}(0;m_Z^2,m_1^2)}{}\\ \hline
\end{tabular}
\end{table}

\clearpage
\section{Higgs masses and mixing angles in the decoupling limit}\label{app:two}

In the decoupling limit of the 2HDM~\cite{decoupling}, one
neutral Higgs boson is kept
light, with mass $\lsim\mathcal{O}(m_Z)$, and the other Higgs boson
masses are taken large compared to $m_Z$.  In this case, one can
formally integrate out the heavy Higgs states, and the effective
low-energy theory consists of a one-Higgs-doublet model.  In the
decoupling limit, the properties of the light neutral Higgs boson must
approach those of the Standard Model Higgs boson.
It is simplest to characterize the decoupling limit in the Higgs basis
as follows:
\beqa
{\rm (i)}~~|Z_i|&\lsim& \mathcal{O}(1)\,,\label{dec1}\\
{\rm (ii)}~~~Y_2 &\gg& v^2\,.\label{dec2}
\eeqa
We shall define $\Lambda$ to be the mass scale that characterizes the
heavy Higgs states, i.e. $Y_2\sim \mathcal{O}(\Lambda)$.
In light of \eq{mc}, these
two requirements imply that $m_{H^\pm}\gg v$.

It is convenient to work in the basis of neutral Higgs mass-eigenstate
$h_1$, $h_2$ and $h_3$, in which the corresponding squared-masses are
given by \eq{hmassesinv}.  Assuming
\eqs{dec1}{dec2}, the squared-masses are given by:
\beqa
m_1^2&=&(s_{12}^2+c_{12}^2 s_{13}^2)Y_2+\mathcal{O}(v^2)\,,\\
m_2^2&=&(c_{12}^2+s_{12}^2 s_{13}^2)Y_2+\mathcal{O}(v^2)\,,\\
m_3^2&=&c_{13}^2Y_2+\mathcal{O}(v^2)\,,
\eeqa
after employing the $q_{k2}$ given in Table~\ref{tabq}.
In the decoupling limit, precisely
two of the three neutral Higgs masses are of $\mathcal{O}(Y_2)$
whereas the third neutral Higgs boson mass is of order
$\mathcal{O}(v^2)$.  Moreover, to preserve consistency of \eqs{tan213}{tan212},
we required that terms of $\mathcal{O}(Y_2/v^2)$ cancel in these two equations,
which yield the conditions:
\beq
Y_2 s_{13}c_{13}\lsim \mathcal{O}(v^2)\,,
\qquad\qquad
Y_2 c_{13}^2 s_{12} c_{12}\lsim \mathcal{O}(v^2)\,.
\eeq
The above requirements lead to three possible cases:
\vspace{0.1in}

\hspace{1in}
\textbf{Case I$^{\boldsymbol\prime}$:} \phantom{aa}\,\,
$\boldsymbol{|s_{12}|\sim |s_{13}|\lsim\mathcal{O}(v^2/Y_2)}$
\quad $\Longrightarrow$ \quad $m_2$, $m_3\gg m_1$,
\vspace{0.1in}

\hspace{1in}
\textbf{Case II$^{\boldsymbol\prime}$:} \,\! \phantom{a}
$\boldsymbol{|c_{12}|\sim |s_{13}|\lsim\mathcal{O}(v^2/Y_2)}$
\quad $\Longrightarrow$ \quad $m_1$, $m_3\gg m_2$,
\vspace{0.1in}

\hspace{1in}
\textbf{Case III$^{\boldsymbol\prime}$:} \phantom{a}
$\boldsymbol{|c_{13}|\lsim\mathcal{O}(v^2/Y_2)}$\phantom{$=s_{13}$}\phantom{W}\,
\,\quad $\Longrightarrow$ \quad $m_1$, $m_2\gg m_3$.
\vspace{0.1in}

The nomenclature for these three cases follows that of Appendix C.3,
although we do \textit{not} assume that $Z_6=0$ in the present
discussion.  In all three cases, we can now obtain expressions for the
corresponding neutral Higgs squared-masses.
\vspace{0.1in}

In Case I$^\prime$,
\beqa
m_1^2&\simeq& Z_1 v^2\,,\label{mass1}\\
m_2^2&\simeq & Y_2+\half\left[Z_3+Z_4+\Re(Z_5
  e^{-2i\theta_{23}})\right]v^2\,,\\
m_3^2&\simeq & Y_2+\half\left[Z_3+Z_4-\Re(Z_5
  e^{-2i\theta_{23}})\right]v^2\,.
\eeqa

In Case II$^\prime$,
\beqa
m_1^2&\simeq & Y_2+\half\left[Z_3+Z_4+\Re(Z_5
  e^{-2i\theta_{23}})\right]v^2\,,\\
m_2^2&\simeq& Z_1 v^2\,,\\
m_3^2&\simeq & Y_2+\half\left[Z_3+Z_4-\Re(Z_5
  e^{-2i\theta_{23}})\right]v^2\,.
\eeqa

In Case III$^\prime$,
\beqa
m_1^2&\simeq & Y_2+\half\left[Z_3+Z_4-\Re(Z_5
  e^{-2i\overline\theta_{23}})\right]v^2\,,\\
m_2^2&\simeq & Y_2+\half\left[Z_3+Z_4+\Re(Z_5
  e^{-2i\overline\theta_{23}})\right]v^2\,,\\
m_3^2&\simeq& Z_1 v^2\,,\label{mass9}
\eeqa
where $\overline\theta_{23}$ is defined in \eq{t2312} [cf.~the comments
that precede this equation].  In all cases above, we omit terms of
$\mathcal{O}\left(v^4/Y_2\right)$.

Despite appearances, the above mass formulae are consistent.
In Cases I$^\prime$ and II$^\prime$, \eq{tan13} implies that
$\Im(Z_5 e^{-2i\theta_{23}})\lsim\mathcal{O}\left(v^2/Y_2\right)$.
It follows that
$\Re(Z_5 e^{-2i\theta_{23}})=\varepsilon|Z_5|+
\mathcal{O}\left(v^2/Y_2\right)$,
where
\beq \label{realsign}
\varepsilon\equiv
\rm{sgn}\left[\Re(Z_5 e^{-2i\theta_{23}})\right]\,.
\eeq
Hence, in Cases I$^\prime$ and II$^\prime$, the
squared-masses of the two heavy states are given by
\beq \label{heavy}
Y_2+\half\left(Z_3+Z_4\pm |Z_5|\right)+
\mathcal{O}\left(\frac{v^2}{Y_2}\right)\,.
\eeq
In Case III$^\prime$, \eq{tan212} implies that
$\Im(Z_5 e^{-2i\overline\theta_{23}})\lsim\mathcal{O}\left(v^2/Y_2\right)$.
In this case, in then follows that
 $\Re(Z_5 e^{-2i\overline\theta_{23}})=\overline\varepsilon|Z_5|+
\mathcal{O}\left(v^2/Y_2\right)$, where
$\overline\varepsilon=\rm{sgn}[\Re(Z_5 e^{-2i\overline\theta_{23}})]$.
Once again,
the squared-masses of the two heavy states again reduce to \eq{heavy}.

In the analysis above, no assumption was made for the value of $Z_6$.
If $Z_6=0$ then \eqst{spcase1}{spcase3} imply that
$\Im(Z_5 e^{2i\theta_{23}})=0$ in Cases I$^\prime$ and II$^\prime$ and
$\Im(Z_5 e^{2i\overline\theta_{23}})=0$ in Case III$^\prime$.
In this case, one can diagonalize
the neutral Higgs squared-mass matrix exactly [cf.~\eq{mtilmatrix}].
In particular, in the case of $Z_6=0$, the squared-mass
formulae given in \eqst{mass1}{mass9} [and in \eq{heavy}] are exact, with no
$\mathcal{O}(v^2/Y_2)$ corrections.

In the case of a CP-conserving scalar potential and $Z_6\neq 0$,
we note that Case I of Table~\ref{Z6cond} is consistent with Case
I$^\prime$ above, and the decoupling limit
is specified by $|s_{12}|\sim\mathcal{O}(v^2/Y_2)$.
Case IIb of Table~\ref{Z6cond} is consistent with Case
II$^\prime$ above, and the decoupling limit is specified by
$|s_{13}|\sim\mathcal{O}(v^2/Y_2)$.  Finally, Case IIa of
Table~\ref{Z6cond} is consistent with either Cases I$^\prime$ or III$^\prime$.
The corresponding decoupling limit is $|s_{13}|\sim\mathcal{O}(v^2/Y_2)$
in Case I$^\prime$ and $|c_{13}|\sim\mathcal{O}(v^2/Y_2)$ in Case III$^\prime$.

It is convenient to adopt a convention where $m_1<m_2$, $m_3$.  In this
convention, only Case I$^\prime$ is relevant in the decoupling limit.
Henceforth, we shall assume that $h_1$ is the lightest neutral Higgs
boson in the decoupling limit.  No mass-ordering of $h_2$ and $h_3$,
which depends on the sign $\varepsilon$, will be assumed.

For completeness (and as a check of the above results), we provide an
alternate derivation of the neutral Higgs squared-masses and mixing in
the decoupling limit.   We may
compute the eigenvalues of \eq{mtilmatrix} directly
by setting $Z_6=0$ and treating
$Z_6$ as a small perturbation.  In first approximation,
\beqa
m_1^2&\simeq &Z_1 v^2\,,\label{m11approx}\\
m_{2,3}^2&\simeq &m_{H^\pm}^2+\half(Z_4\pm |Z_5|)v^2\,,\label{m23approx}
\eeqa
where we have used \eq{mc}.  To determine which squared masses in \eq{m23approx} correspond to $h_2$ and $h_3$,
one can also treat the off-diagonal 23 and 32 elements of \eq{mtilmatrix} perturbatively, in which case
one finds:
\beq
m_2^2-m_3^2\simeq\begin{cases} \phm|Z_5|v^2\,,&\quad \text{for}~\Re(Z_5 e^{-2i\theta_{23}})\geq 0\,,\\
-|Z_5|v^2\,,&\quad \text{for}~\Re(Z_5 e^{-2i\theta_{23}})\leq 0\,.\end{cases}
\eeq
That is, all the heavy scalar squared-masses can be written in terms of a single
large squared-mass parameter $\Lambda^2$ as follows:
\beqa
m_3^2 &\equiv& \LL\,,\label{massrelations1} \\
m_2^2 &=& \LL + \varepsilon |Z_5|v^2\,, \label{massrelations2}\\
{m^2_{H^\pm}} &=& \LL - \half\left[Z_4-\varepsilon|Z_5|\right]v^2\,, \label{massrelations3}
\eeqa
where $\varepsilon$ is defined in \eq{realsign}.

Corrections proportional to $Z_6$ enter at second-order in perturbation theory and
contribute terms that are parametrically smaller than the results displayed
in \eqs{m11approx}{m23approx}.  In particular,
\beq
m_1^2\simeq Z_1v^2 -\frac{|Z_6|^2 v^4}{\Lambda^2}\,.\label{secondorder}
\eeq

The invariant neutral Higgs mixing angles in the decoupling limit
can be determined directly from eqs.~(C21) and (C25) of
\Ref{haberoneil}, which we reproduce below:
\beqa
s_{13}^2&=&\frac{(Z_1 v^2-m_1^2)(Z_1 v^2-m_2^2)+|Z_6|^2 v^4}{(m_3^2-m_1^2)(m_3^2-m_2^2)}\,,\label{f1}\\
c_{13}^2s_{12}^2&=&\frac{(Z_1 v^2-m_1^2)(m_3^2-Z_1 v^2)-|Z_6|^2 v^4}{(m_2^2-m_1^2)(m_3^2-m_2^2)}\label{f2}\,.
\eeqa
These expressions are exact.
Assuming that $Z_5\neq 0$, it then immediately follows from \eqst{massrelations1}{secondorder}
that:\footnote{If $Z_6=0$ then $m_1^2=Z_1 v^2$ is exact, in which case $s_{12}=s_{13}=0$ with no
additional corrections.}
\beq \label{angleD}
s^2_{13}\sim s^2_{12}\sim\mathcal{O}\left(\frac{v^4}{\Lambda^4}\right)\,,
\eeq
since the numerators of \eqs{f1}{f2} are of order $v^6/\Lambda^2$, whereas the denominators
are of order $\Lambda^2 v^2$.
Some care is required to treat the case of $Z_5=0$ [since in this case $m_2^2-m_3^2\lsim
\mathcal{O}(v^4/\Lambda^2)$].  Nevertheless, our original analysis above
confirms that \eq{angleD} still holds.  Hence, in the decoupling limit,\footnote{By
convention, we take $-\half\pi\leq\theta_{12}\,,\,\theta_{13}<\half\pi$, in which case
$c_{12}$, $c_{13}\geq 0$.}
\beq
|s_{12}|\sim |s_{13}|  \simeq \mathcal{O}\left(\frac{v^2}{\LL}\right), \qquad\qquad
c_{12} \sim  c_{13} \simeq 1\,, \label{sandc}
\eeq
in a convention where $h_1$ is defined to be the lightest neutral
Higgs boson.


In the CP-conserving limit, the neutral Higgs masses in the decoupling
limit can be obtained directly from \eqs{m2hH}{m2A} by assuming
that $Y_2\gg Z_1 v^2$.  In the case of $Z_6\neq 0$,
\beqa
m^2_{h^0}&\simeq & Z_1 v^2\,,\label{m1approx}\\
m^2_{H^0}&\simeq &m_A^2+\varepsilon_{56}|Z_5|v^2\,,\label{m2approx}\\
m^2_{A^0}&\simeq &m_{H^\pm}^2+\half(Z_4-\epsilon_{56}|Z_5|)v^2\,,\label{m3approx}
\eeqa
as one approaches the decoupling limit, in agreement with eqs.~(\ref{m11approx})
and (\ref{massrelations1})--(\ref{massrelations3}).
In particular, referring to Table~\ref{Z6cond}, we identify $h_1=h^0$ and
\beqa
h_2&=&H^0\,,\quad h_3=A^0\,,\qquad \text{and}~\Re(Z_5 e^{-2i\theta_{23}})=\varepsilon_{56}|Z_5|\,,
\qquad\phm (\text{Case I}) \label{identity1}\\
h_2&=&A^0\,,\quad h_3=H^0\,,\qquad \text{and}~\Re(Z_5 e^{-2i\theta_{23}})=-\varepsilon_{56}|Z_5|
\,,\qquad (\text{Case IIa})
\label{identity2}
\eeqa
from which it follows that
\beq \label{var}
\varepsilon=\eta^2\epsilon_{56}\,.
\eeq

In the case of $Z_6=0$, \eq{m1approx} is an exact result.  In addition,
\eqst{m2approx}{var} apply with $\varepsilon_{56}$ replaced by
$\varepsilon_{57}$ and/or $\varepsilon_{5Q}$ as appropriate,
with the new versions of \eqs{identity1}{identity2} applying
in Cases I$^\prime$a and I$^\prime$b of Table~\ref{Z60cond}, respectively.

Finally, we note that in the limit of custodial symmetry,
the 2HDM potential and vacuum are CP-conserving and
\eq{z45eta} is satisfied.   That is,
\beq
Z_4=\eta^2\varepsilon|Z_5|\,,
\eeq
in the case of a generic scalar potential.  Consider first the
case of $Z_6\neq 0$.  Then, using \eqst{m1approx}{m3approx},
\beqa
m_{A^0}^2-m_{H^\pm}^2&=&\half|Z_5|(\eta^2\varepsilon-\epsilon_{56})v^2\,,\\
m_{H^0}^2-m_{H^\pm}^2&=&\half|Z_5|(\eta^2\varepsilon+\epsilon_{56})v^2\,.
\eeqa
Using \eq{var}, it follows that $m_{H^\pm}^2=m^2_{A^0}$, as expected.  For
$Z_6=0$ and $Z_7\neq 0$, simply replace $\varepsilon_{56}$ with
$\varepsilon_{57}$ and the same conclusions follow.
In the special case of $Z_6=Z_7=0$ (and assuming CP-conserving Higgs--fermion
Yukawa interactions), it is also possible to have a custodial symmetric scalar
potential with $Z_4=-\eta^2\varepsilon|Z_5|$ [cf.~\eq{z45etaalt} and
Section~\ref{twisted}].  Replacing $\varepsilon_{56}$ above with
$\varepsilon_{5Q}$, it then follows that $m_{H^\pm}^2=m^2_{H^0}$.  Of
course, both these Higgs mass-degeneracies are enforced by the custodial
symmetry independently of the decoupling limit.


\section{Derivation of tree-level unitarity limits}\label{app:three}

The assumption of tree-level unitarity in scattering processes implies an upper bound on
the magnitudes of the $Z_i$ parameters.  This places an upper limit on the masses of
the heavy Higgs states in parameter regimes in which the decoupling limit does not apply.
The implications of unitarity for the 2HDM has been studied in the context of the scattering
of gauge bosons and the physical scalars in refs.~\cite{Kanemura,akeroyd}.  By placing an upper limit
on the amplitude for a process $\varphi_A \varphi_B \rightarrow \varphi_C \varphi_D$, one can quantify
the constraints from tree-level unitarity as follows:
\beq \left|g_{ABCD} \right| < 8\pi.\eeq
For tree-level scattering processes, only the quartic bosonic couplings are relevant, namely
$W^+ W^-W^+W^-$, $W^+W^-H^+H^-$, $(H^+e^{i\theta_{23}})(H^+e^{i\theta_{23}})W^-W^-+\rm{h.c.}$,
$Z^0Z^0Z^0h_m$, $G^0 h_mG^-(H^+e^{i\theta_{23}}) +\rm{h.c.}$, $Z^0Z^0H^+H^-$, and $ Z^0 Z^0W^-(H^+e^{i\theta_{23}}).$

The equivalence theorem~\cite{equivalence} allows one equate a high energy scattering amplitude involving
gauge bosons to the analogous amplitude involving Goldstone bosons, up to an unimportant
overall sign, by making the replacements
$W^{\pm}\rightarrow G^\pm$ and $Z^0 \rightarrow G^0$.  Thus, one can translate limits on
the gauge boson/Higgs couplings into limits on the Goldstone/Higgs couplings. The
resulting constraints on $Z_1$, $Z_3$, $Z_3 + Z_4$, $\Re(\zfive)$, and $\Re(\zsix)$
can be read off directly from eqs.~(\ref{scalpot}), as shown in Table~\ref{unitarity}.

\begin{table}[ht!]
\centering
\caption{Calculation of tree-level unitarity limits on the CP-conserving quartic couplings. Combinatorial factors are included to take into account identical particles. }
\label{unitarity}
\begin{tabular}{|c|c|c|}\hline
Relevant term in the scalar potential& Amplitude \T \Bot&Resulting unitarity bound\\ \hline
$\half Z_1 G^+ G^- G^+ G^- $ \T \Bot &$ \frac{1}{16\pi}(\half Z_1)\cdot 4 $&$ |Z_1|<4\pi $\\
$\half Z_3 G^0 G^0 H^+H^-$\T \Bot& $\frac{1}{16\pi}(\half Z_3)\cdot 2 $&$ |Z_3|<8\pi $\\
$(Z_3+Z_4)G^+ G^- H^+ H^- $\T \Bot& $\frac{1}{16\pi}(Z_3+Z_4) $&$ |Z_3+Z_4|<8\pi $\\
$\half \zfive (H^+ e^{i\theta_{23}} )(H^+ e^{i\theta_{23}} )G^-G^-+{\rm h.c.}$\T \Bot& $\frac{1}{16\pi}\Re(\zfive)\cdot 4 $& $|\Re(\zfive)|<2\pi $\\
$\zsix G^0 G^0 G^- (H^+ e^{i\theta_{23}} ) +{\rm h.c.}$\T \Bot&$ \frac{1}{16\pi}\Re(\zsix)\cdot 4 $&$ |\Re(\zsix)|<2\pi$ \\ \hline
\end{tabular}
\end{table}

The CP-violating parameters $\Im(\zfive)$ and $\Im(\zsix)$ appear in a more complicated form in the quartic scalar potential.  From the interaction $\half \Im(q_{m2}Z_6\,e^{-i\theta_{23}})\,G^0 G^0 G^0 h_m$ and Table~\ref{tabq}, one can write Feynman rules for $m = 1,2$:
\beqa
g_{G^0G^0G^0h_1}&=&3\bigl\{-s_{12} \Im[\zsix] - c_{12} s_{13} \Re[\zsix])\bigr\},\nonumber\\
g_{G^0G^0G^0h_2}&=&3\bigl\{c_{12} \Im[\zsix] - s_{12} s_{13} \Re[\zsix]\bigr\},\label{g}
\eeqa
after including an overall symmetry factor $3!$ corresponding to three identical particles
at the vertex.
Unitarity requires $|g_{G^0G^0G^0h_m}|< 8\pi$.  It is convenient to combine the two limits in quadrature to isolate $\Im(\zsix)$.
That is, $|g_{G^0G^0G^0h_1}|^2 + |g_{G^0G^0G^0h_2}|^2< 64\pi^2$, which yields
\beq
\left[\Im(\zsix)\right]^2 + s_{13}^2 \left[\Re(\zsix)\right]^2< \frac{64\pi^2}{9}.
\eeq
Since $s_{13}^2 \left[\Re(\zsix)\right]^2$ is real and non-negative, it must be true that $|\Im(\zsix)|< 8\pi/3$.

Similarly, one can use the term $\half i\,G^0 h_m \biggl\{G^-H^+e^{i\theta_{23}}\left[q^*_{m2}Z_4-q_{m2}Z_5 e^{-2i\theta_{23}}\right]+{\rm h.c.}\biggr\}$ with $m=1,2$ to derive
\beqa
g_{G^0G^-(H^+ e^{i\theta_{23}})h_1}&=&-c_{12}s_{13} Z_4 - s_{12}\Im(\zfive) - c_{12} s_{13} \Re(\zfive),\nonumber\\
g_{G^0G^-(H^+ e^{i\theta_{23}})h_2}&=&-s_{12}s_{13} Z_4 + c_{12}\Im(\zfive) - s_{12} s_{13} \Re(\zfive).\label{gg}
\eeqa
Adding the contributions of the two couplings above in quadrature and applying the unitarity bound yields after some simplification:
\beq s_{13}^2 \left[ Z_4 + \Re(\zfive)\right]^2 + \left[\Im(\zfive)\right]^2 < 64 \pi^2.\eeq
In particular, one must satisfy $|\Im(\zfive)|< 8\pi$.



\begin{thebibliography}{99}
\bibitem{davidson}
  S.~Davidson and H.~E.~Haber,
  Phys.\ Rev.\  D {\bf 72}, 035004 (2005); D {\bf 72}, 099902(E) (2005)
  [arXiv:hep-ph/0504050].



\bibitem{haberoneil}
  H.~E.~Haber and D.~O'Neil,
  Phys.\ Rev.\ D {\bf 74}, 015018 (2006);
  D {\bf 74}, 059905(E) (2006)
[arXiv:hep-ph/0602242].
  
\bibitem{Maniatis:2006fs}
  M.~Maniatis, A.~von Manteuffel, O.~Nachtmann and F.~Nagel,
 Eur.\ Phys.\ J.\  C {\bf 48}, 805 (2006)
 [arXiv:hep-ph/0605184].

\bibitem{Nishi:2006tg}
  C.C.~Nishi,
Phys.\ Rev.\  D {\bf 74}, 036003 (2006); D {\bf 76}, 119901(E) (2007)
[arXiv:hep-ph/0605153].

\bibitem{Ivanov:2006yq}
  I.P.~Ivanov,
Phys.\ Rev.\  D {\bf 75}, 035001 (2007) D {\bf 76}, 039902(E) (2007)
[arXiv:hep-ph/0609018].

\bibitem{Maniatis:2007vn}
  M.~Maniatis, A.~von Manteuffel and O.~Nachtmann,
Eur.\ Phys.\ J.\  C {\bf 57}, 719 (2008)
[arXiv:0707.3344 [hep-ph]].

\bibitem{pomarol}
  A.~Pomarol and R.~Vega,
  Nucl.\ Phys.\ B {\bf 413}, 3 (1994)
  [arXiv:hep-ph/9305272].

\bibitem{peskin}
  M.~E.~Peskin and T.~Takeuchi,
  Phys.\ Rev.\ Lett.\  {\bf 65}, 964 (1990);
  Phys.\ Rev.\ D {\bf 46}, 381 (1992).

\bibitem{branco}
  G.~C.~Branco, L.~Lavoura and J.~P.~Silva,
  \textit{CP Violation} (Oxford University Press, Oxford, UK, 1999).

\bibitem{cpx2}
F.~J.~Botella and J.~P.~Silva,
Phys.\ Rev.\ {\bf D51}, 3870 (1995)
[arXiv:hep-ph/9411288].

\bibitem{cpbasis}
  J.~F.~Gunion and H.~E.~Haber,
  Phys.\ Rev.\  {\bf D72}, 095002 (2005)
  [arXiv:hep-ph/0506227].

\bibitem{hhg}
J.~F.~Gunion, H.~E.~Haber, G.~Kane and S.~Dawson,
\textit{The Higgs Hunter's Guide}  (Westview Press, Boulder, CO, 2000).


\bibitem{inert}
 R.~Barbieri, L.~J.~Hall and V.~S.~Rychkov,
  Phys.\ Rev.\  D {\bf 74}, 015007 (2006)
  [arXiv:hep-ph/0603188].


\bibitem{Ferreira:2009wh}
  P.~M.~Ferreira, H~E.~Haber and J.~P.~Silva,
  Phys.\ Rev.\  D {\bf 79}, 116004 (2009)
  [arXiv:0902.1537 [hep-ph]].


\bibitem{Ferreira:2010yh}
  P~M.~Ferreira, H.~E.~Haber, M.~Maniatis, O.~Nachtmann and J.~P.~Silva,
  arXiv:1010.0935 [hep-ph].

\bibitem{herquet}
  J.~M.~Gerard and M.~Herquet,
  Phys.\ Rev.\ Lett.\  {\bf 98}, 251802 (2007)
  [arXiv:hep-ph/0703051];

 \bibitem{herquet2}
 S.~de Visscher, J.~M.~Gerard, M.~Herquet, V.~Lemaitre and F.~Maltoni,
  JHEP {\bf 0908}, 042 (2009)
  [arXiv:0904.0705 [hep-ph]].

\bibitem{lepewwg}
The ALEPH, CDF, D0, DELPHI, L3, OPAL, SLD Collaborations,
the LEP Electroweak Working Group, the Tevatron Electroweak Working
Group, and the SLD Electroweak and Heavy Flavor groups,
CERN-PH-EP/2009-023 (13 November, 2009); the most recent updated
results can be found at
\texttt{http://lepewwg.web.cern.ch/LEPEWWG/}

\bibitem{erlerPDG}
J.~Erler and P.~Langacker, \textit{Electroweak Model and Constraints
  on New Physics},
in K. Nakamura et al. [Particle Data Group],
J. Phys. G {\bf 37}, 075021 (2010).

\bibitem{gfitter}
 H.~Flacher, M.~Goebel, J.~Haller, A.~Hocker, K.~Monig and J.~Stelzer,
  Eur.\ Phys.\ J.\  C {\bf 60}, 543 (2009)
  [arXiv:0811.0009 [hep-ph]];
the most recent updated results can be found at
\texttt{http://gfitter.desy.de/}.

\bibitem{lepbound}
R.~Barate {\it et al.}  [The ALEPH, DELPHI, L3, OPAL Collaborations
and The LEP Working Group for Higgs Boson Searches]
  Phys.\ Lett.\  B {\bf 565}, 61 (2003)
  [arXiv:hep-ex/0306033].

\bibitem{haberpomarol}
  H.~E.~Haber and A.~Pomarol,
  Phys.\ Lett.\  B {\bf 302}, 435 (1993)
  [arXiv:hep-ph/9207267].

\bibitem{veltman}
  M.~J.~G.~Veltman,
  Nucl.\ Phys.\  B {\bf 123}, 89 (1977);
 M.~S.~Chanowitz, M.~A.~Furman and I.~Hinchliffe,
  Phys.\ Lett.\  B {\bf 78}, 285 (1978);
M.~B.~Einhorn, D.~R.~T.~Jones and M.~J.~G.~Veltman,
  Nucl.\ Phys.\  B {\bf 191}, 146 (1981).

\bibitem{kennedy}
  D.~C.~Kennedy,
``Renormalization of electroweak gauge interactions,''
in \textit{Perspectives in the standard model},  Proceedings of the
1991 Theoretical Advanced
Studies Institute, Boulder, CO, Jun 2--28, 1991, edited by
R.K.~Ellis, C.T.~Hill and J.D.~Lykken.
(World Scientific Publishing, Singapore, 1992) pp.~163--282.


\bibitem{habertasi}
  H.~E.~Haber,
  ``Introductory Low-Energy Supersymmetry,'' in
\textit{Recent directions in particle theory: from superstrings and black holes to the standard model}, Proceedings of
the Theoretical Advanced Study Institute (TASI 92), Boulder, CO, 1--26
June 1992, edited by J.~Harvey and J.~Polchinski (World Scientific
Publishing, Singapore, 1993) pp.~589--688.


\bibitem{passario}
  G.~Passarino and M.~J.~G.~Veltman,
  Nucl.\ Phys.\ B {\bf 160}, 151 (1979).

\bibitem{Froggatt}
  C.~D.~Froggatt, R.~G.~Moorhouse and I.~G.~Knowles,
  Phys.\ Rev.\  D {\bf 45}, 2471 (1992);
  Nucl.\ Phys.\  B {\bf 386}, 63 (1992).


\bibitem{decoupling}
 H.~E.~Haber and Y.~Nir,
  Nucl.\ Phys.\  B {\bf 335}, 363 (1990);
J.~F.~Gunion and H.~E.~Haber,
  Phys.\ Rev.\  D {\bf 67}, 075019 (2003)
  [arXiv:hep-ph/0207010].

\bibitem{susybasis}
  P.~M.~Ferreira, H.~E.~Haber, J.~P.~Silva,
  Phys.\ Rev.\  {\bf D82}, 016001 (2010)
  [arXiv:1004.3292 [hep-ph]].



\bibitem{kaffas}
 A.~Wahab El Kaffas, P.~Osland and O.~M.~Ogreid,
  Phys.\ Rev.\  D {\bf 76}, 095001 (2007)
  [arXiv:0706.2997 [hep-ph]].

\bibitem{deva}
 D.~O'Neil,
  ``Phenomenology of the Basis-Independent CP-Violating Two-Higgs Doublet Model,''
  Ph.D. thesis (University of California, Santa Cruz, 2009)
  arXiv:0908.1363 [hep-ph].

\bibitem{gunetal}
P.~H.~Chankowski, T.~Farris, B.~Grzadkowski, J.~F.~Gunion, J.~Kalinowski and M.~Krawczyk,
  Phys.\ Lett.\  B {\bf 496}, 195 (2000)
  [arXiv:hep-ph/0009271].

\bibitem{gmw}
B.~Grzadkowski, M.~Maniatis and J.~Wudka,
  arXiv:1011.5228 [hep-ph].


\bibitem{cpx}
L.~Lavoura and J.~P.~Silva,
Phys.\ Rev.\ {\bf D50}, 4619 (1994)
[arXiv:hep-ph/9404276].

\bibitem{abramowitz}
M.~Abramowitz and I.~A.~Stegun,
\textit{Handbook of Mathematical Functions}\,
(Dover Publications Inc., New York, 1972).


\bibitem{langacker}
W.~Hollik, ``Renormalization of the Standard Model,'' in
P.~Langacker,
  \textit{Precision Tests of the Standard Electroweak Model}, edited
by P.~Langacker (World Scientific Publishing, Singapore, 1995) pp.~37--116.

\bibitem{lavoura}
  W.~Grimus, L.~Lavoura, O.~M.~Ogreid and P.~Osland,
  Nucl.\ Phys.\  B {\bf 801}, 81 (2008)
  [arXiv:0802.4353 [hep-ph]].

\bibitem{Kanemura}
  S.~Kanemura, T.~Kubota and E.~Takasugi,
  Phys.\ Lett.\  B {\bf 313}, 155 (1993)
  [arXiv:hep-ph/9303263],

  S.~Kanemura, Y.~Okada, E.~Senaha and C.~P.~Yuan,
  Phys.\ Rev.\  D {\bf 70}, 115002 (2004)
  [arXiv:hep-ph/0408364],
  H.~Huffel and G.~Pocsik,
  Z.\ Phys.\  C {\bf 8}, 13 (1981).
  H.~A.~Weldon,
  Phys.\ Rev.\  D {\bf 30}, 1547 (1984).

\bibitem{akeroyd}
  A.~G.~Akeroyd, A.~Arhrib and E.~M.~Naimi,
  Phys.\ Lett.\  B {\bf 490}, 119 (2000)
  [arXiv:hep-ph/0006035];
  I.~F.~Ginzburg and I.~P.~Ivanov,
  arXiv:hep-ph/0312374.

\bibitem{equivalence}
See e.g.,
 J.~Bagger and C.~Schmidt,
  Phys.\ Rev.\  D {\bf 41}, 264 (1990);
 H.~G.~J.~Veltman,
  Phys.\ Rev.\  D {\bf 41}, 2294 (1990);
and references therein.


\end{thebibliography}
\end{document}